\documentclass[12pt,a4paper]{article}

\usepackage{amsmath,amssymb}

\usepackage[bookmarks=true,hyperfigures=true]{hyperref}
\usepackage[final]{graphicx}
\usepackage[nosort]{cite}
\usepackage[bulletsep]{collref}
\usepackage{slashed}
\usepackage{tocloft}
\usepackage{array}
\usepackage{pifont}

\ifx\hypersetup\sadfkjashdfkxja\else
\hypersetup{pdftitle={Symmetries of Scattering Amplitudes in N=6 Superconformal Chern--Simons Theory}}
\hypersetup{pdfauthor={Till Bargheer, Florian Loebbert, Carlo Meneghelli}}
\hypersetup{pdfsubject={}}
\hypersetup{pdfkeywords={ABJM, Scattering Amplitude, N=6 Supersymmetric Chern--Simons Theory, Tree Level, Yangian Symmetry}}
\hypersetup{plainpages=false}
\hypersetup{pdfpagemode=UseNone}
\hypersetup{bookmarksnumbered=true}
\hypersetup{pdfstartview=FitH}
\hypersetup{colorlinks=false}
\hypersetup{citebordercolor={.6 .9 .6}}
\hypersetup{urlbordercolor={.7 .8 1}}
\hypersetup{linkbordercolor={1 .7 .7}}
\hypersetup{pdfborder={0 0 .5}}
\fi

\makeatletter
\newlength{\apb@width}
\newcommand{\autoparbox}[2][c]{\settowidth{\apb@width}{#2}\parbox[#1]{\apb@width}{#2}}
\newcommand{\includegraphicsbox}[2][]{\autoparbox{\includegraphics[#1]{#2}}}
\makeatother

\usepackage[a4paper,text={450pt,650pt},centering]{geometry}

\allowdisplaybreaks[3]

\numberwithin{equation}{section}

\usepackage[font=small,labelfont=bf,width=0.85\textwidth]{caption}

\usepackage{color}

\makeatletter
\let\old@startsection=\@startsection
\renewcommand{\@startsection}[6]{\old@startsection{#1}{#2}{#3}{#4}{#5}{#6\mathversion{bold}}}
\let\oldl@section=\l@section
\renewcommand{\l@section}[2]{%
\oldl@section{\mathversion{bold}#1}{#2}}
\makeatother

\let\oldPhi=\Phi
\let\oldPsi=\Psi
\let\oldGamma=\Gamma
\let\oldDelta=\Delta
\let\oldSigma=\Sigma
\let\oldLambda=\Lambda
\let\oldTheta=\Theta
\let\oldPi=\Pi
\let\oldXi=\Xi
\renewcommand{\Phi}{\mathnormal{\oldPhi}}
\renewcommand{\Psi}{\mathnormal{\oldPsi}}
\renewcommand{\Gamma}{\mathnormal{\oldGamma}}
\renewcommand{\Sigma}{\mathnormal{\oldSigma}}
\renewcommand{\Delta}{\mathnormal{\oldDelta}}
\renewcommand{\Theta}{\mathnormal{\oldTheta}}
\renewcommand{\Lambda}{\mathnormal{\oldLambda}}
\renewcommand{\Pi}{\mathnormal{\oldPi}}
\renewcommand{\Xi}{\mathnormal{\oldXi}}

\newcommand{\op}[1]{\mathcal{#1}}
\newcommand{\gen}[1]{\mathfrak{#1}}

\newcommand{\alg}[1]{\mathfrak{#1}}
\newcommand{\grp}[1]{\mathrm{#1}}
\newcommand{\superN}{\mathcal{N}}
\newcommand{\sym}{$\superN=4$ SYM}
\newcommand{\scs}{$\superN=6$ SCS}
\newcommand{\psu}{$\alg{psu}(2,2|4)$}
\newcommand{\osp}{$\alg{osp}(6|4)$}
\newcommand{\amp}{\mathcal{A}}
\newcommand{\deltaP}{\ensuremath{\delta^3(P)}}
\newcommand{\deltaQ}{\ensuremath{\delta^6(Q)}}
\newcommand{\deltaPQ}{\ensuremath{\deltaP\,\deltaQ}}

\newcommand{\rep}[1]{\mathbf{#1}}
\newcommand{\fnum}{\mathcal{F}}

\newcommand{\grade}[1]{|#1|}

\newcommand{\sTr}{\mathop{\mathrm{sTr}}}

\newcommand{\tr}{\mathop{\mathrm{Tr}}}
\newcommand{\sign}{\mathop{\mathrm{sgn}}}

\newcommand{\Res}{\mathrm{Res}}
\renewcommand{\mod}[1]{(\mathrm{mod}\:#1)}

\newcommand{\Complex}{\mathbb{C}}
\newcommand{\Reals}{\mathbb{R}}
\newcommand{\cp}{\grp{\Complex P}}

\newcommand{\ff}{f\kern-5pt f}

\newcommand{\dd}{d}
\newcommand{\eps}{\varepsilon}

\newcommand{\defeq}{\mathrel{:=}}

\ifx\genfrac\sdflkaj\else\fi
\newcommand{\sfrac}[2]{{\textstyle\frac{#1}{#2}}}
\newcommand{\half}{\sfrac{1}{2}}

\newcommand{\third}{\sfrac{1}{3}}
\newcommand{\twothird}{\sfrac{2}{3}}

\newcommand{\pardel}[1]{\frac{\partial}{\partial #1}}

\newcommand{\indup}[1]{_{\mathrm{#1}}}

\newcommand{\supup}[1]{^{\mathrm{#1}}}

\newcommand{\matr}[2]{\left(\begin{array}{@{\,}#1@{\,}}#2\end{array}\right)}

\newcommand{\brk}[1]{(#1)}
\newcommand{\lrbrk}[1]{\left(#1\right)}
\newcommand{\bigbrk}[1]{\bigl(#1\bigr)}
\newcommand{\biggbrk}[1]{\biggl(#1\biggr)}
\newcommand{\Bigbrk}[1]{\Bigl(#1\Bigr)}

\newcommand{\biggsbrk}[1]{\biggl[#1\biggr]}
\newcommand{\Bigsbrk}[1]{\Bigl[#1\Bigr]}
\newcommand{\brc}[1]{\{#1\}}

\newcommand{\bigbrc}[1]{\bigl\{#1\bigr\}}

\newcommand{\comm}[2]{[#1,#2]}
\newcommand{\scomm}[2]{[#1,#2\}}

\newcommand{\acomm}[2]{\{#1,#2\}}

\newcommand{\abs}[1]{|#1|}

\newcommand{\spaa}[1]{\langle#1\rangle}

\newcommand{\ket}[1]{|#1\rangle}

\newcommand{\nn}{\nonumber}

\newcommand{\beq}{\begin{equation}}
\newcommand{\eeq}{\end{equation}}

\def\[{\begin{equation}}
\def\]{\end{equation}}
\def\<{\begin{eqnarray}}
\def\>{\end{eqnarray}}

\newdimen\yysquaresize
\newdimen\yyrsquaresize
\newdimen\yythickness
\newdimen\yyskip
\def\yysquare#1{%
\setlength{\yyrsquaresize}{\yysquaresize}%
\addtolength{\yyrsquaresize}{-2\yythickness}%
\vrule width \yythickness%
\vbox to \yysquaresize{%
  \hrule height \yythickness\vss%
  \hbox to \yyrsquaresize{\hss#1\hss}%
  \vss\hrule height \yythickness}%
\vrule width \yythickness}

\def\yyyoung#1{\vtop{\baselineskip0pt\lineskip-\yythickness\halign{\tabskip-\yythickness&\yysquare{##}\cr #1}}}

\newcommand{\young}[1]{\hskip\yyskip\mbox{\yyyoung{#1}}\hskip\yyskip}

\makeatletter
\def\mr@ignsp#1 {\ifx\:#1\@empty\else #1\expandafter\mr@ignsp\fi}%
\newcommand{\multiref}[1]{\begingroup
\xdef\mr@no@sparg{\expandafter\mr@ignsp#1 \: }%
\def\mr@comma{}%
\@for\mr@refs:=\mr@no@sparg\do{\mr@comma\def\mr@comma{,}\ref{\mr@refs}}%
\endgroup}
\makeatother

\newcommand{\hypref}[2]{\ifx\href\asklfhas #2\else\href{#1}{#2}\fi}
\newcommand{\Secref}[1]{Section~\multiref{#1}}
\newcommand{\Secsref}[1]{Sections~\multiref{#1}}

\newcommand{\Appref}[1]{Appendix~\multiref{#1}}

\newcommand{\Tabref}[1]{Table~\multiref{#1}}

\newcommand{\Figref}[1]{Figure~\multiref{#1}}
\newcommand{\Figsref}[1]{Figures~\multiref{#1}}

\renewcommand{\eqref}[1]{(\multiref{#1})}

\ifx\href\asklfhas\newcommand{\href}[2]{#2}\fi
\newcommand{\arxivlink}[1]{\href{http://arxiv.org/abs/#1}{arxiv:#1}}

\begin{document}

\thispagestyle{empty}
\begin{flushright}\footnotesize
\texttt{\arxivlink{1003.6120}}\\
\texttt{AEI-2010-049}%
\end{flushright}
\vspace*{1cm}

\begin{center}%
{\Large\textbf{\mathversion{bold}%
Symmetries of Tree-level Scattering Amplitudes\\
in $\superN=6$ Superconformal Chern--Simons Theory
}\par}
\vspace{1cm}%

\textsc{Till Bargheer, Florian Loebbert, Carlo Meneghelli}\vspace{5mm}%

\textit{Max-Planck-Institut f\"ur Gravitationsphysik\\%
Albert-Einstein-Institut\\%
Am M\"uhlenberg 1, 14476 Potsdam, Germany}\vspace{3mm}%

\{\texttt{till,loebbert,carlo}\}\texttt{@aei.mpg.de}
\par\vspace{1cm}

\vspace{1cm}

\textbf{Abstract}\vspace{7mm}

\begin{minipage}{12.7cm}
Constraints of the \osp\ symmetry on tree-level scattering
amplitudes in $\superN=6$ superconformal Chern--Simons theory are
derived. Supplemented by Feynman diagram calculations, solutions to
these constraints, namely the four- and six-point superamplitudes, are
presented and shown to be invariant under Yangian symmetry.
This introduces integrability into the amplitude sector of the theory.
\end{minipage}

\end{center}

\newpage

\setcounter{tocdepth}{2}
\hrule height 0.75pt
\tableofcontents
\vspace{0.8cm}
\hrule height 0.75pt
\vspace{1cm}

\setcounter{tocdepth}{2}

\section{Introduction and Overview}

While the prime example of the AdS/CFT correspondence is the duality
between four-dimensional $\superN=4$ super Yang--Mills theory (SYM)
and type IIB superstring theory on $\grp{AdS}_5\times\grp{S}^5$
\cite{Maldacena:1998re,Gubser:1998bc,Witten:1998qj},
another
remarkable instance equates $\superN=6$ superconformal Chern--Simons
theory in three dimensions (SCS) and type IIA strings on
$\grp{AdS}_4\times\cp^3$
\cite{Aharony:2008ug}.
In the study of the spectrum on both sides of these two
correspondences, the discovery of integrability
\newcommand{\intsymweak}{Minahan:2002ve,Beisert:2003tq}
\newcommand{\intsymstrong}{Bena:2003wd,Kazakov:2004qf,Beisert:2005bm}
\newcommand{\intscsweak}{Minahan:2008hf,Bak:2008cp,Zwiebel:2009vb,Minahan:2009te,Bak:2009mq,Bak:2009tq}
\newcommand{\intscsstrong}{Arutyunov:2008if,Stefanski:2008ik,Gromov:2008bz,Gomis:2008jt,Grassi:2009yj}
\newcommand{\intscsws}{Zarembo:2009au,Kalousios:2009ey,Sundin:2009zu}
\nocollect{Bena:2003wd}
\nocollect{Minahan:2008hf}
\nocollect{Arutyunov:2008if}
\nocollect{Zarembo:2009au}
\cite{\intsymweak,\intsymstrong,\intscsweak,\intscsstrong,\intscsws}
in the planar limit has been of crucial importance, and has lead to
the belief that the planar theories might be exactly solvable.

Exact solvability would suggest that integrability also
manifests itself in the scattering amplitudes of the above theories.
For the
$\grp{AdS}_5$/$\mathrm{CFT}_4$ correspondence, this is indeed
the case. Motivated by a duality between Wilson loops and scattering
amplitudes in \sym\ theory
\cite{Alday:2007hr,Brandhuber:2007yx,Alday:2007he,Drummond:2008aq},
a dual superconformal symmetry of scattering amplitudes was found at weak coupling
\cite{Drummond:2006rz,Drummond:2007aua,Drummond:2008vq,Brandhuber:2008pf}.
This dual symmetry can be traced back to a T-self-duality of the
$\grp{AdS}_5\times\grp{S}^5$ string background
\cite{Berkovits:2008ic,Beisert:2008iq}
(see also
\cite{Beisert:2009cs}
for a review).
In addition to the standard superconformal symmetry, the
dual realization
acts on dual momentum variables leaving all \sym\ tree-level
amplitudes invariant
\cite{Drummond:2008cr}.
Integrability at weak coupling then arises as the
closure of standard and dual superconformal symmetry into a Yangian
symmetry algebra for tree-level scattering amplitudes
\cite{Drummond:2009fd}.
In fact, \sym\ tree-level amplitudes seem to be uniquely
determined by a modified Yangian representation that takes into
account the peculiarities of collinear configurations due to
conformal symmetry
\cite{Bargheer:2009qu,Korchemsky:2009hm,Beisert:2010gn}.
There has also been remarkable progress on the application of
integrable methods to the strong-coupling regime of scattering amplitudes
in \sym\ theory
\cite{Alday:2009yn,Alday:2009dv,Alday:2010vh}.

On the other hand, little is known about
scattering amplitudes in the $\grp{AdS}_4$/$\mathrm{CFT}_3$
correspondence. For \scs, so far only four-point amplitudes have
been computed
\cite{Agarwal:2008pu}.
In particular, while some possibilities for T-self-duality have been
explored
\cite{Adam:2009kt},
no direct analog of dual superconformal symmetry was found for
this theory.

Given the perturbative integrability of the spectral problem of \scs\ theory
paralleling the discoveries in the $\grp{AdS}_5$/$\mathrm{CFT}_4$ case,
and the recent findings on scattering amplitudes in the latter,
it seems reasonable to search for integrable structures (alias Yangian
symmetry) in \scs\ scattering amplitudes. In the absence of a dual symmetry, a
straightforward generalization of the developments in \sym\ appears to be obscured.
Even without a dual symmetry, however, a procedure to consistently promote
certain standard Lie algebra
representations to Yangian representations
is well-known
\cite{Drinfeld:1985rx,Dolan:2004ps,Drummond:2009fd}.
That is, Yangian
generators that act on scattering amplitudes in a similar way as in
\sym\ can be constructed directly.
However, a priori it is not true that invariants of the standard Lie algebra representation
are also invariant under the Yangian algebra.
Invariance of scattering amplitudes under the Yangian generators would be a
manifestation of integrability.

The standard \osp\ symmetry of \scs\ is
realized on the tree-level amplitudes $\amp_n\supup{tree}$ as a sum of
the action of the free generators $\gen{J}^{(0)}_{\alpha,k}$ on the
individual legs $k$,
\begin{equation}
 \gen{J}^{(0)}_{\alpha}\amp_n\supup{tree}
=\sum_{k=1}^n\gen{J}^{(0)}_{\alpha,k}\amp_n\supup{tree}
=0\,.
\end{equation}
For scattering amplitudes in \sym, as well as for local
gauge invariant operators both in \sym\ and in \scs,
the Yangian generators $\gen{J}^{(1)}_\alpha$ at tree-level are realized
according to the construction of
\cite{Drinfeld:1985rx,Dolan:2004ps}:
They act as
bilocal compositions of standard symmetry generators,
\begin{equation}
\gen{J}^{(1)}_\alpha
\sim
f_\alpha{}^{\beta\gamma}\sum_{j<k}\gen{J}^{(0)}_{\beta,j}\,\gen{J}^{(0)}_{\gamma,k}\,.
\label{eq:introyang}
\end{equation}
Hence these are also natural candidates for Yangian symmetry generators for
\scs\ scattering amplitudes.

In this paper, the constraints of the \osp\ (level-zero)
symmetry algebra on $n$-point scattering amplitudes are analyzed. The
four- and six-point superamplitudes of \scs\ theory are
given as solutions to these constraints, and are shown to be invariant
under the Yangian (level-one) algebra constructed as described above.
This introduces integrability into the amplitude sector of \scs\ theory.

\subsection*{Outline}
The paper is structured as follows:
In \Secref{sec:kin}, the kinematics for three-dimensional field
theories are discussed, and momentum spinors are introduced.
An on-shell superspace and the corresponding superfields for
\scs\ are presented in \Secref{sec:superfield}, where also
color-ordering is discussed.
The realization of the symmetry algebra \osp\ in terms of
the superspace variables is exhibited in \Secref{sec:rep}.
In \Secref{sec:level0invariants} the invariants of this realization
are studied.
The four- and six-point tree-level superamplitudes are presented in
\Secref{sec:sixpoint}.
In \Secref{sec:yanginv}, the realization of the \osp\ Yangian algebra
is analyzed and shown to be consistent by means of the Serre
relations. Yangian invariance of the four- and six-point amplitudes is
shown.
Finally, our conventions as well as several technical details,
including the computation of two six-point component amplitudes from Feynman
diagrams, are presented in the appendix.

\section{Three-Dimensional Kinematics}
\label{sec:kin}

\paragraph{Momentum Spinors.}

The Lorentz algebra in three dimensions is given by $\alg{so}(2,1) $ being isomorphic to $\alg{sl}(2;\mathbb R)$.
Thanks to this isomorphism,  an $\alg{so}(2,1)$  vector equivalently is an $\alg{sl}(2;\mathbb R)$
bispinor. More explicitly, three-dimensional vectors can be expanded
in a basis of symmetric matrices $\sigma^\mu$,
\begin{equation}
p^{ab}
=(\sigma^\mu)^{ab}p_\mu
=\matr{cc}{p^0-p^1&p^2\\p^2&p^0+p^1},
\label{eq:vecbispin}
\end{equation}
and any symmetric $2\times2$ matrix $p^{ab}$ can be written as
\begin{equation}
p^{ab}
=\lambda^{(a} \mu^{b)}\,.
\label{eq:vecbispin2}
\end{equation}
By means of the identifications \eqref{eq:vecbispin,eq:vecbispin2},
the square norm of the vector $p^\mu$ equals the
determinant of the corresponding matrix:
\begin{equation}
p^\mu p_\mu
=-\det(p^{ab})
=-\left(\lambda^a\varepsilon_{ab}\mu^b\right)^2\,.
\label{eq:normdet}
\end{equation}
In particular, this means that the masslessness condition $p^2=0$ can be
explicitly solved
\begin{equation}
p^{ab}=\lambda^a\lambda^b\,.
\label{eq:vecbispinligh}
\end{equation}
Given a massless momentum, the choice of $\lambda^a$ in
\eqref{eq:vecbispinligh} is unique up to a sign being the
manifestation of the fact that the group $\grp{SL}(2;\mathbb R)$
is the double cover of $\grp{SO}(2,1)$. That the sign is the only
freedom in the choice of $\lambda^a$ is due to the fact that
the little group of massless particles%
\footnote{$\grp{SO}(d-2)$ in $d$ dimensions.}
is discrete in three dimensions.
For massive momenta on the other hand,
the choice of $\lambda^a, \mu^a$ in \eqref{eq:vecbispin2}
has an $\Reals^+\times\grp{U}(1)$ freedom
\begin{equation}
\lambda^a \rightarrow c \lambda^a\,,
\qquad
\mu^a \rightarrow \mu^a/c\,,
\qquad
c\in\Complex_{\setminus\brc{0}}\,.
\end{equation}
In particular this contains the little group $\grp{U}(1)$ of \emph{massive} particles%
\footnote{$\grp{SO}(d-1)$ in $d$ dimensions.}
in three dimensions.

Some comments on reality conditions for $\lambda^a$ are in order.
Physical momenta are real; this means that $\lambda^a$ can be either
purely real or purely imaginary.
For positive-energy momenta ($p^0>0$), $\lambda^a$ is purely real,
while it is purely imaginary for negative-energy momenta.
Even for complex momenta, $p^{ab}$ is expressed in terms of a single
complex $\lambda$ as in \eqref{eq:vecbispinligh}.
This seems very different to the four-dimensional case, where momenta can be written as
\begin{equation}
p_{d=4}^{a \dot b}=\lambda^a \tilde \lambda^{\dot b}\, ,
\end{equation}
and $\lambda^a$ and $\tilde \lambda^{\dot b}$ are independent
in complexified kinematics. In Minkowski signature, $\lambda^a$ and
$\tilde \lambda^{\dot b}$ are actually complex conjugate to each
other. This is the origin of the holomorphic anomaly
\cite{Cachazo:2004by,Cachazo:2004dr,Britto:2004nj}.
Looking at \eqref{eq:vecbispinligh}, nothing similar appears to happen
in  three dimensions if one imposes the correct reality conditions.

\begin{table}
\centering
\renewcommand{\arraystretch}{1.3}
\begin{tabular}{|l
                |>{\centering }m{2.1cm}
                |>{\centering }m{1.9cm}
                |>{\centering }m{3.2cm}
                |>{\centering }m{1.3cm}
                |>{\centering\arraybackslash }m{3.7cm}|}
\hline
&Lorentz\linebreak $\grp{SO}(d-1,1)$
&Conformal\linebreak $\grp{SO}(d,2)$
&Lightlike\linebreak Momentum
&Little\linebreak Group
&Superconformal\linebreak Group
\\\hline
d=3
&$\grp{SL}(2;\mathbb{R})$
&$\grp{SP}(4;\mathbb{R})$
&$p^{ab}=\lambda^a\lambda^b$
&$\mathrm{Z}_2$
&$\grp{OSP}(\superN_{\leq 8}|4)$
\\
d=4
&$\grp{SL}(2;\mathbb{C})$
&$\grp{SU}(2,2)$
&$p^{a\dot b}=\lambda^a\bar\lambda^{\dot b}$
&$\grp{U}(1)$
&$\grp{(P)SU}(2,2|\superN_{\leq 4})$
\\
d=6
&$\grp{SL}(2;\mathbb{H})$\linebreak$\simeq\grp{SU}^*(4)$
&$\grp{SO}^*(8)$
&$p^{[AB]}=\eps_{ab}\lambda^{Aa}\lambda^{Bb}$
 \linebreak $p_{[AB]}=\eps_{\dot a\dot b}\tilde\lambda^{\dot a}_A\tilde\lambda^{\dot b}_B$
&$\grp{SU}(2)^2$
&$\grp{OSP}(8|2)$, $\grp{OSP}(8|4)$
\\\hline
\end{tabular}
\renewcommand{\arraystretch}{1.0}
\caption{Spinor-helicity formalism and superconformal symmetry in various dimensions.}
\label{tab:spinsup}
\end{table}

It is worth noting that the existence of a spinor-helicity framework in a certain
dimension is intimately connected to the existence of superconformal
symmetry in that dimension, cf.~\Tabref{tab:spinsup}.
For the six-dimensional case the spinor-helicity formalism has been recently applied
to scattering amplitudes in \cite{Cheung:2009dc,Dennen:2009vk}.

\paragraph{Kinematical Invariants.}

In terms of momentum spinors, two-particle Lorentz invariants can
be conveniently expressed as
\begin{equation}
\eta_{\mu\nu}p_1^\mu p_2^\nu
=-\half\spaa{12}^2\,,
\qquad
\spaa{jk}\defeq\lambda_j^a\varepsilon_{ab}\lambda_k^b\,.
\label{eq:spaa}
\end{equation}
It is easy to count the number of (independent) Poincar\'e invariants
that can be built out of $n$ massless three-dimensional momenta.
Every spinor carries two degrees of freedom resulting in $2n$
variables for $n$ massless momenta.
The number of two-particle Lorentz invariants one can build from these is
$2n-3$, where $3$ is the number of Lorentz generators.
This can be explicitly done using Schouten's identity
\begin{equation}
\spaa{kl}\spaa{ij}+\spaa{ki}\spaa{jl}+\spaa{kj}\spaa{li}=0.
\end{equation}
Finally, total momentum conservation imposes three further constraints,
such that the number of Poincar\'e invariants is $2n-6$.
Note that for $n=3$ there is no Poincar\'e invariant, even in complex
kinematics.

\paragraph{One-Particle States.}

One-particle states are solutions of the linearized equation of motion.
This equation is an irreducibility condition for the
representation of the Poincar\'e group.
For massless particles, these Poincar\'e representations are lifted
to representations of the conformal group $\grp{SO}(d,2)$.
Once again, the existence of the spinor formulation in three dimensions
makes it possible to explicitly solve the irreducibility condition.

For scalars, the irreducibility condition is trivially satisfied by an
arbitrary function of the massless momentum:
\begin{equation}
p^2 \phi(p^{ab})=0 \quad\Rightarrow \quad \phi(p^{ab})= \phi(\lambda^a \lambda^b)\,.
\end{equation}
For fermions, the irreducibility condition is given by the Dirac
equation, which forces the fermionic state $\Psi_a$ to be proportional
to $\varepsilon_{ab}\lambda^b$,
\begin{equation}
p^{ab}\Psi_b(p^{cd})=0
\quad\Rightarrow \quad  \Psi_a(p^{cd})= \varepsilon_{ab} \lambda^b \psi(\lambda^c \lambda^d)\,.
\label{eq:dirac}
\end{equation}
Thus when $\lambda^a$ changes its sign, the scalar state is invariant, while
the fermionic state picks up a minus sign. Once again,
this just corresponds to the fact
that fermions are representations of
$\grp{Spin}(2,1)\sim\grp{SL}(2;\Reals)$,
which is the double cover of $\grp{SO}(2,1)$.
Put differently,
\begin{equation}
\exp\Bigbrk{i\pi\lambda^a\pardel{\lambda^a}}\ket{\text{State}}
=(-1)^\fnum\ket{\text{State}}\,,
\label{eq:fermionoperator}
\end{equation}
where $\fnum$ denotes the fermion number operator.

It is worth mentioning that these representations of the conformal group
$\grp{SO}(3,2) \sim \grp{Sp}(4, \mathbb R)$ have a long history.
They go back to Dirac \cite{Dirac:1963ta} and were particularly studied by
Flato and Fronsdal in an ancestor form of the AdS/CFT correspondence \cite{Flato:1978qz}.

\section{Superfields and Color Ordering}
\label{sec:superfield}

\paragraph{Field Content.}

The matter fields of $\superN=6$ superconformal Chern--Simons theory
comprise eight scalar fields and eight fermion fields that form four
fundamental multiplets of the internal $\alg{su}(4)$ symmetry:
\begin{equation}
\phi^A(\lambda)\,,
\quad
\bar\phi_A(\lambda)\,,
\quad
\psi_A(\lambda)\,,
\quad
\bar\psi^A(\lambda)\,,
\quad
A\in\brc{1,2,3,4}\,.
\end{equation}
The fields $\phi^A$ and $\psi_A$ transform in the
$(\rep{N},\rep{\bar N})$
representation, while $\bar\phi_A$, $\bar\psi^A$ transform in the
$(\rep{\bar N},\rep{N})$ representation of the gauge group
$\grp{U}(N)\times\grp{U}(N)$.%
\footnote{$\rep{N}$: Fundamental representation of $\grp{U}(N)$,
$\rep{\bar N}$: Antifundamental representation of $\grp{U}(N)$.}
The former shall be called ``particles'', the latter
``antiparticles''.
In addition, the theory contains gauge fields
$A_\mu$, $\hat A_\mu$ that transform in $(\rep{ad},\rep{1})$,
$(\rep{1},\rep{ad})$ representations of the gauge group. The gauge fields
however cannot appear as external fields in scattering amplitudes, as
their free equations of motion
$\partial_{[\mu}A_{\nu]}=0=\partial_{[\mu}\hat A_{\nu]}$ do not allow
for excitations.

\paragraph{Superfields.}

For the construction of scattering amplitudes, it is convenient to
employ a superspace formalism, in which the fundamental fields of
$\superN=6$ superconformal Chern--Simons theory combine into
superfields and supersymmetry becomes manifest.
In \sym, the fields (gluons, fermions, scalars) transform in different
representations of the internal symmetry group.
Thus in the superfield of \sym, the
fields can be multiplied by different powers of the fermionic
coordinates $\eta^A$ according to their different representation.
Internal symmetry, realized as
$\gen{R}^A{}_B\sim\eta^A\partial/\partial\eta^B$, is then manifest.
All particles in \scs\ form
(anti)fundamental multiplets of the internal
$\alg{su}(4)$ symmetry.
Thus an analogous superfield construction, i.e.\ one in which R-symmetry only acts on
the fermionic variables, seems obstructed for this theory.
Nevertheless, by breaking manifest R-symmetry, one can employ
$\superN=3$ superspace, in which the
fundamental fields combine into one bosonic and one
fermionic superfield with the help of an
$\alg{su}(3)$ Gra{\ss}mann spinor $\eta^A$,
\begin{align}
\Phi(\Lambda)
&= \phi^4(\lambda)
  +\eta^A\psi_A(\lambda)
  +\half\varepsilon_{ABC}\eta^A\eta^B\phi^C(\lambda)
  +\sfrac{1}{3!}\varepsilon_{ABC}\eta^A\eta^B\eta^C\psi_4(\lambda)\,,\nn\\
\bar\Phi(\Lambda)
&= \bar\psi^4(\lambda)
  +\eta^A\bar\phi_A(\lambda)
  +\half\varepsilon_{ABC}\eta^A\eta^B\bar\psi^C(\lambda)
  +\sfrac{1}{3!}\varepsilon_{ABC}\eta^A\eta^B\eta^C\bar\phi_4(\lambda)\,.
\end{align}
Here and in the following, $\Lambda$ is used as a shorthand notation
for the pair of variables $(\lambda,\eta)$. Introducing these
superfields amounts to splitting the internal $\alg{su}(4)$ symmetry
into a manifest $\alg{u}(3)$, realized as
$\gen{R}^A{}_B\sim\eta^A\partial/\partial\eta^B$, plus a
non-manifest remainder, realized as multiplication and second-order
derivative operators.
For the complete representation of
the symmetry group on the superfields, see the following \Secref{sec:rep}.

Using the superfields, scattering amplitudes conveniently combine into
\emph{superamplitudes}
\begin{equation}
\hat\amp_n=\hat\amp_n(\Phi_1,\bar\Phi_2,\Phi_3,\ldots,\bar\Phi_n)\,,
\quad
\Phi_k\defeq\Phi(\Lambda_k)\,.
\end{equation}
Component amplitudes for all possible configurations of fields then
appear as coefficients of $\hat\amp_n$ in the fermionic variables
$\eta_1^A,\ldots,\eta_n^A$.

\paragraph{Color Ordering.}

In all tree-level Feynman diagrams, each external particle
(antiparticle) is connected to one antiparticle (particle) by a
fundamental color line and to another antiparticle (particle) by an
antifundamental color line.%
\footnote{This implies in particular that only scattering processes
involving the same number of particles and antiparticles are
non-vanishing.}
Tree-level scattering amplitudes can
therefore conveniently be expanded in their color factors:
\begin{equation}
\hat\amp_n\bigbrk{\Phi_1{}^{A_1}_{\bar A_1},
                  \bar\Phi_2{}^{\bar B_2}_{B_2},
                  \Phi_3{}^{A_3}_{\bar A_3},
                  \ldots,
                  \bar\Phi_n{}^{\bar B_n}_{B_n}}
=\sum_{\makebox[5ex]{\scriptsize$\sigma\in\brk{\grp{S}_{n/2}\times\grp{S}_{n/2}}/\grp{C}_{n/2}$}}
 \amp_n\bigbrk{\Lambda_{\sigma_1},\ldots,\Lambda_{\sigma_n}}
 \delta^{A_{\sigma_1}}_{B_{\sigma_2}}
 \delta^{\bar B_{\sigma_2}}_{\bar A_{\sigma_3}}
 \delta^{A_{\sigma_3}}_{B_{\sigma_4}}
 \cdots
 \delta^{\bar B_{\sigma_n}}_{\bar A_{\sigma_1}}\,.
\label{eq:colord}
\end{equation}
Here, the sum extends over permutations $\sigma$ of $n$ sites that only mix
even and odd sites among themselves, modulo cyclic permutations by two
sites. By definition, the color-ordered amplitudes $\amp_n$ do not
depend on the color indices of the external superfields. The
total amplitude $\hat\amp_n$ is invariant up to a fermionic sign under
all permutations of its arguments. Therefore the color-ordered
amplitudes $\amp_n$ are invariant under cyclic
permutations of their arguments by two sites,
\begin{equation}
\amp_n(\Lambda_3,\ldots,\Lambda_n,\Lambda_1,\Lambda_2)
=(-1)^{(n-2)/2}\amp_n(\Lambda_1,\ldots,\Lambda_n)\,,
\end{equation}
where the sign is due to the fact that $\Phi$ is bosonic and
$\bar\Phi$ is fermionic. While the color-ordered \emph{component} amplitudes
can at most change by a sign under shifts of the arguments by one
site,%
\footnote{A single-site shift amounts to exchanging the fundamental
with the antifundamental gauge group, which equals a parity
transformation in \scs\ \cite{Aharony:2008ug}.}
the superamplitude $\amp_n$ might transform non-trivially under
single-site shifts, as the definition of $\amp_n(\Lambda_1,\ldots,\Lambda_n)$ in
\eqref{eq:colord} implies that $\Lambda\indup{odd/even}$
belong to bosonic/fermionic superfields.

For the color-ordered amplitudes $\amp_n$, the superanalog of
the condition \eqref{eq:fermionoperator} takes the form
\begin{equation}\label{eq:centralthing}
\exp i\pi\Bigbrk{\lambda_k^a\pardel{\lambda_k^a}+\eta_k^A\pardel{\eta_k^A}}\amp_n
=(-1)^k\amp_n\,.
\end{equation}
Note that this local constraint looks similar to the (local) central charge condition
in four dimensions. Moreover,
$\exp i\pi\bigbrk{\lambda_k^a\pardel{\lambda_k^a}+\eta_k^A\pardel{\eta_k^A}}$
is central for the \osp\ realization given in the next
\Secref{sec:rep}.

Note that the above color
structure \eqref{eq:colord} is very similar to the structure of quark-antiquark
scattering in QCD, see e.g.\ \cite{Mangano:1988kk}.

\section{Singleton Realization of \texorpdfstring{\osp}{osp(6|4)}}
\label{sec:rep}

\begin{figure}\centering
\includegraphics[scale=1.2]{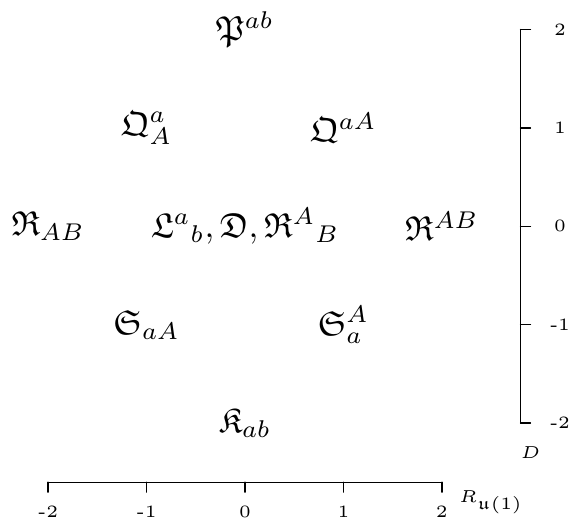}
\caption{The generators of \osp\ can be arranged according to their dilatation charge and their $\alg{u}(1)$ charge under $\gen{R}^C{}_C$.}
\label{fig:ospgen}
\end{figure}

The \osp\ algebra is spanned by the $\alg{sp}(4)$
generators of translations $\gen{P}^{ab}$, Lorentz transformations
$\gen{L}^a{}_b$, special conformal transformations $\gen{K}_{ab}$ and
dilatations $\gen{D}$, by the $\alg{so}(6)$ R-symmetries $\gen{R}^{AB}$,
$\gen{R}^A{}_B$ and $\gen{R}_{AB}$ as well as 24 supercharges
$\gen{Q}^{aA}$, $\gen{Q}^a{}_A$, $\gen{S}_a{}^A$ and $\gen{S}_{aA}$.
Here we use $\alg{sl}(2)$ indices $a, b,\ldots=1,2$ and $\alg{su}(3)$
indices $A,B,\ldots =1,2,3$. As mentioned above, the internal $\alg{so}(6)$ symmetry is
not manifest in this realization of the algebra. The generators
$\gen{R}_{AB}$ and $\gen{R}^{AB}$ are antisymmetric in their indices,
while $\gen{R}^A{}_B$ does contain a non-vanishing trace and thus
generates $\alg{su}(3)+\alg{u}(1)$. Hence, in total we have $15$
independent R-symmetry generators corresponding to
$\alg{so}(6)\sim\alg{su}(4)$, cf.\ also \Figref{fig:ospgen}.

\paragraph{Commutators.}

The generators of \osp\ obey the following commutation
relations: Lorentz and internal rotations read
\begin{align}
\comm{\gen{L}^a{}_b}{\gen{J}^c}&=+\delta^c_b\gen{J}^a-\half\delta^a_b\gen{J}^c\,,&
\comm{\gen{L}^a{}_b}{\gen{J}_c}&=-\delta^a_c\gen{J}_b+\half\delta^a_b\gen{J}_c\,,\\
\comm{\gen{R}^A{}_B}{\gen{J}^C}&=+\delta^C_B\gen{J}^A\,,&
\comm{\gen{R}^A{}_B}{\gen{J}_C}&=-\delta^A_C\gen{J}_B\,,\\
\comm{\gen{R}_{AB}}{\gen{J}^C}&=\delta^C_B\gen{J}_A-\delta^C_A\gen{J}_B\,,&
\comm{\gen{R}^{AB}}{\gen{J}_C}&=\delta^B_C\gen{J}^A-\delta^A_C\gen{J}^B\,.\label{eq:commRuudd}
\end{align}
Commutators including translations and special conformal
transformations take the form
\begin{equation}
\comm{\gen{K}_{ab}}{\gen{P}^{cd}}
	=\delta^d_b\gen{L}^c{}_a
	+\delta^c_b\gen{L}^d{}_a
	+\delta^d_a\gen{L}^c{}_b
	+\delta^c_a\gen{L}^d{}_b
	+2\delta^d_b\delta^c_a\gen{D}
	+2\delta^c_b\delta^d_a\gen{D}\,,
\end{equation}
\begin{align}
 \comm{\gen{P}^{ab}}{\gen{S}_c^A}
&=
-\delta_c^a \gen{Q}^{bA}-\delta_c^b\gen{Q}^{aA}\,,
&\comm{\gen{K}_{ab}}{\gen{Q}^{cA}}
&=\delta_b^c\gen{S}_a^A+\delta_a^c\gen{S}^A_b\,,
\\
\comm{\gen{P}^{ab}}{\gen{S}_{cA}}
&=-\delta^a_c\gen{Q}^b_A-\delta^b_c\gen{Q}^a_A\,,
&\comm{\gen{K}_{ab}}{\gen{Q}^c_A}
&=\delta_a^c\gen{S}_{bA}+\delta_b^c\gen{S}_{aA}\,,
\end{align}
while the supercharges commute into translations and rotations:
\begin{equation}
\acomm{\gen{Q}^{aA}}{\gen{Q}^b_B}=\delta^A_B\gen{P}^{ab}\,,\qquad
\acomm{\gen{S}_{aA}}{\gen{S}_b^B}=\delta^B_A\gen{K}_{ab}\,,
\end{equation}
\begin{align}
\acomm{\gen{Q}^{aA}}{\gen{S}_{bB}}&=\delta^A_B\gen{L}^a{}_b-\delta^a_b\gen{R}^A{}_B+\delta^A_B\delta^a_b\gen{D}\,,
&\acomm{\gen{Q}^{aA}}{\gen{S}_b^B}&=-\delta^a_b\gen{R}^{AB}\,,
\\
\acomm{\gen{Q}^a_A}{\gen{S}_b^B}&=\delta^B_A\gen{L}^a{}_b+\delta^a_b\gen{R}^B{}_A+\delta^A_B\delta^a_b\gen{D}\,,
&\acomm{\gen{Q}^a_A}{\gen{S}_{bB}}&=-\delta^a_b\gen{R}_{AB}\,.
\end{align}
Furthermore the non-vanishing dilatation weights are given by
\begin{align}
\comm{\gen{D}}{\gen{P}^{ab}}&=+\gen{P}^{ab}\,,&
\comm{\gen{D}}{\gen{Q}^{aA}}&=+\half\gen{Q}^{aA}\,,&
\comm{\gen{D}}{\gen{Q}^a_A} &=+\half\gen{Q}^a_A\,,
\\
\comm{\gen{D}}{\gen{K}_{ab}}&=-\gen{K}_{ab}\,,&
\comm{\gen{D}}{\gen{S}_a^A} &=-\half\gen{S}_a^A\,,&
\comm{\gen{D}}{\gen{S}_{aA}}&=-\half\gen{S}_{aA}\,.
\end{align}
All other commutators vanish. Note that in contrast to the
\psu\ symmetry algebra of \sym\ theory,
all fermionic generators are connected by commutation relations with bosonic generators.

\paragraph{Singleton Realization.}

The above algebra \osp\ can be realized in terms of the
bosonic and fermionic spinor variables $\lambda^a$ and $\eta^A$
introduced in \Secsref{sec:kin,sec:superfield}. Acting on one-particle states the
representation takes the form (cf.\ also \cite{Gunaydin:1984vz}, used
in the present context in \cite{Zwiebel:2009vb,Minahan:2009te,Papathanasiou:2009en}):
\begin{alignat*}{2}
\gen{L}^a{}_b&=\lambda^a\partial_b-\half\delta^a_b\lambda^c\partial_c\,,&\qquad
\gen{P}^{ab}&=\lambda^a\lambda^b\,,\\
\gen{D}&=\half\lambda^a\partial_a+\half\,,&
\gen{K}_{ab}&=\partial_a\partial_b\,,
\end{alignat*}
\begin{equation*}
\gen{R}^{AB}=\eta^A\eta^B\,,\qquad
\gen{R}^A{}_B
=\eta^A\partial_B-\half\delta^A_B\,,\qquad
\gen{R}_{AB}=\partial_A\partial_B\,,
\end{equation*}
\begin{alignat}{2}
\gen{Q}^{aA}&=\lambda^a\eta^A\,,&\qquad
\gen{S}_a^A&=\eta^A\partial_a\,,\nn\\
\gen{Q}^a_A&=\lambda^a\partial_A\,,&
\gen{S}_{aA}&=\partial_a\partial_A\,.
\label{eq:genone}
\end{alignat}
For a general discussion of representations of this type, cf.~\Appref{app:clifford}.
The multi-particle generalization of these generators at tree-level is given by a
sum over single-particle generators \eqref{eq:genone}
acting on each individual particle $k$, i.e.\
\begin{equation}
\gen{J}_\alpha^{\mathrm{multi}}=\sum_{k=1}^n \gen{J}_{\alpha, k}^{\mathrm{single}},
\qquad
\gen{J}_\alpha\in\alg{osp}(6|4).
\label{eq:jmulti}
\end{equation}
As opposed to \psu, the symmetry algebra of \sym, the algebra \osp\ cannot
be enhanced by a central and/or a hyper-charge.
Since in \sym\ theory the hyper-charge of \psu\
measures the helicity, this can be considered the algebraic
manifestation of the lack of helicity in
three dimensions. Still we can define some central element like in \eqref{eq:centralthing}.

\section{Constraints on Symmetry Invariants}
\label{sec:level0invariants}

We are interested in the determination of tree-level scattering
amplitudes of $n$ particles in \scs\ theory. These should be
functions of the superspace coordinates introduced in
\Secref{sec:superfield} and be invariant under the symmetry algebra \osp\ of
the \scs\ Lagrangian. In order to approach this problem, this section is concerned with the symmetry constraints imposed on generic functions of $n$ bosonic and
$n$ fermionic variables $\lambda_i^a$ and $\eta_i^A$, respectively.
That is, we study the form of invariants $I_n(\lambda_i,\eta_i)$ under
the above representation of \osp.
It is demonstrated that requiring invariance under the symmetry
reduces to finding $\alg{so}(6)$ singlets plus solving a set of
first-order partial differential equations; the latter following from
invariance under the superconformal generator $\gen{S}$. Invariance under all other generators will then be manifest in our construction.

Due to the color decomposition discussed in \Secref{sec:superfield},
scattering amplitudes are expected to be invariant under two-site
cyclic shifts. Since the generators given in the previous section are
invariant under arbitrary permutations of the particle sites, they do
not impose any cyclicity constraints. Those constraints as well as
analyticity conditions are important ingredients for the determination
of amplitudes, but are not studied in the following.
Note that apart from assuming a specific realization of the symmetry
algebra, the investigations in this section are completely general.
In \Secref{sec:sixpoint}, we will specialize to four and six
particles and give explicit solutions to the constraints. The
following discussion will be rather technical. For convenience, the
main results are summarized at the end of this section.

\paragraph{Invariance under \texorpdfstring{$\alg{sp}(4)$}{sp(4)}.}

The subalgebra $\alg{sp}(4)$ of \osp\ is spanned by the
generators of translations $\gen{P}^{ab}$, Lorentz transformations
$\gen{L}^a{}_b$, special conformal transformations $\gen{K}_{ab}$ and
dilatations $\gen{D}$. Invariance under the multiplication operator
$\gen{P}^{ab}=\lambda^a\lambda^b$ constrains an
invariant of $\alg{sp}(4)$ to be of the form
\begin{equation}
 I_n(\lambda_i,\eta_i)=\delta^3(P) G(\lambda_i,\eta_i),
\end{equation}
where $P^{ab}=\sum_{i=1}^n\lambda_i^a\lambda_i^b$ is the overall
momentum and $G(\lambda_i,\eta_i)$ some function to be determined. The momentum delta function is Lorentz invariant on its own
so that $G(\lambda_i,\eta_i)$ has to be invariant under
$\gen{L}^a{}_b$ as well. As $\deltaP$ has weight $-3$ in $P$,
dilatation invariance furthermore requires that
$\sum_{k=1}^n\lambda_k^a\partial_{ka} G=(6-n)G$.
We will not specify any invariance condition for the conformal boost
here, since invariance under $\gen{K}_{ab}$ will follow from invariance
under the superconformal generators $\gen{S}_{aA}$, $\gen{S}_b^B$ using the algebra.

\paragraph{Invariance under \texorpdfstring{$\gen{Q}$}{Q} and \texorpdfstring{$\gen{R}$}{R}.}
Invariance under the multiplicative supermomentum
$\gen{Q}^{aA}$ requires the
invariant $I_n$ to be proportional to a corresponding supermomentum
delta function:
\begin{equation}
I_n(\lambda_i,\eta_i)=\delta^3(P)\delta^6(Q) F(\lambda_i,\eta_i),
\label{eq:ninvPQ}
\end{equation}
where one way to define the delta function is given by
\begin{equation}
\delta^6(Q)=\prod_{\substack{a=1,2\\A=1,2,3}}Q^{aA}, \qquad Q^{aA}=\sum_{i=1}^n \lambda_i^a\eta_i^A.
\label{eq:deltaQdef}
\end{equation}
Again, the function $F(\lambda_i,\eta_i)$ should be Lorentz
invariant, and dilatation invariance implies
\begin{equation}
 \sum_{k=1}^n\lambda_k^a\partial_{ka}F=-n\, F.
\label{eq:DinvF}
\end{equation}
Invariance under the second momentum supercharge $\gen{Q}^a_A$ will
follow from R-symmetry, but will also be discussed in \eqref{eq:Q2inv}.

In order to construct a singlet under the multiplicative R-symmetry
generator $\gen{R}^{AB}=\sum_{i=1}^n\eta_i^A\eta_i^B$ one might want to
add another delta function ``$\delta(R)$'' to our invariant. Things, however, turn out
to be not as straightforward as for the generators $\gen{P}$ and
$\gen{Q}$. As a function of the bosonic object $\gen{R}^{AB}$ made out of fermionic quantities, ``$\delta(R)$'' is not well defined.

We first of all note that invariance under the $\alg{u}(1)$ R-symmetry
generator
\begin{equation}
\gen{R}^C{}_C=\eta^C\partial_C-\sfrac{3}{2}n
\label{eq:Rgammacounter}
\end{equation}
fixes the power $m$ of Gra{\ss}mann parameters $\eta$ in the $n$-leg
invariant $I_n$ to
\begin{equation}
m=\sfrac{3}{2}n.
\label{eq:gammaweight}
\end{equation}
Hence, increasing the number of legs of the invariant by $2$ increases
the Gra{\ss}mann degree of the invariant by $3$ (remember that
amplitudes with an odd number of external particles vanish). This is a
crucial difference to scattering amplitudes in \sym\ theory.
As a consequence, the complexity of amplitudes in \scs\
automatically increases with the number of legs. There are no simple
MHV-type amplitudes for all numbers of external particles as in the
four-dimensional counterpart. Rather, the $n$-point amplitude
resembles the $\mathrm{N}^{(n-4)/2}\mathrm{MHV}$ amplitude in \sym\
theory (being the most complicated).

We can ask ourselves what happens to the R-symmetry generators in the
presence of $\delta^3(P)\delta^6(Q)$. To approach this problem, we introduce a new basis for the fermionic parameters $\eta^A_i$:
\begin{equation}\label{eq:newbasis}
 \eta_i^A, \quad i=1,\dots,n\quad\to\quad \alpha_J^A,\, \beta_J^A,\, Q^{aA},\, Y^{aA}, \quad J=1,\dots ,\sfrac{n-4}{2}.
\end{equation}
That is we trade $n$ anticommuting parameters $\eta^A$ for
$n=2\times(n-4)/2+4$ new fermionic variables. The new
quantities are defined as
\begin{align}
\alpha_J^A\defeq x^+_J\cdot \eta^A=\sum_{i=1}^nx^+_{Ji} \eta_i^A,
&\qquad\qquad
\beta^A_J\defeq x^-_J\cdot \eta^A=\sum_{i=1}^nx^-_{Ji} \eta_i^A,\label{eq:defalphabeta}
\\
Y^{aA}&\defeq y^a\cdot \eta^A=\sum_{i=1}^ny^a_i \eta_i^A,\label{eq:defY}
\end{align}
where the coordinate vectors $x_{Ji}^\pm(\lambda_k)$ and
$y_i^a(\lambda_k)$ express the new variables $\alpha_J$, $\beta_J$ and
$Y^a$ in terms of the old variables $\eta_i$. At first sight,
introducing this new set of variables might seem unnatural. It will,
however, be very convenient for treating invariants of
\osp\ and appears to be a natural basis for scattering
amplitudes in \scs\ theory.

In order for the new set of Gra{\ss}mann variables \eqref{eq:newbasis} to
provide $n$ independent parameters, the coordinates have to satisfy
some independence conditions. Since the two variables $Q^a$ are given
by the coordinate $n$-vectors $\lambda_i^a$ for $a=1,2$, a
natural choice are the orthogonality conditions
\begin{equation}
x^\pm_J\cdot \lambda^b
=0,
\qquad
y^a\cdot \lambda^b
=\eps^{ab},
\qquad
y^a\cdot x^\pm_J
=0,
\label{eq:xpmdefrel}
\end{equation}
where the dot $\cdot$ represents the contraction of two $n$-vectors as
in \eqref{eq:defalphabeta,eq:defY}. For convenience we furthermore
choose the following normalizations
\begin{equation}\label{eq:xpmnormal}
x^\pm_I \cdot x^\pm_J \,
=0,\qquad
x^+_I\cdot x^-_J\,
=\delta_{IJ},\qquad
y^a\cdot y^b\,
=0.
\end{equation}
Given $\lambda^a_i$ such that $\lambda^a \cdot \lambda^b=0$,
\eqref{eq:xpmdefrel} and  \eqref{eq:xpmnormal} do not fix $x^\pm_I$
and $y^a$ uniquely. The leftover freedom can be split into an
\emph{irrelevant} part and a \emph{relevant} one.
The \emph{irrelevant} freedom is
\begin{equation}
\label{eq:irrelevantfree}
x^\pm_{Ii}\rightarrow x^\pm_{Ii} + v^\pm_{aI} \lambda^a_i,\qquad
y^a \rightarrow y^a + w\lambda^a \, ,
\end{equation}
where $v^\pm_{aI}$ and $w$ are functions of $\lambda$.
The freedom expressed in  \eqref{eq:irrelevantfree}
is nothing but the freedom of shifting the fermionic variables
defined in \eqref{eq:defalphabeta,eq:defY}
by terms proportional to $Q^{aB}$. In the presence of $\deltaQ$
this freedom is obviously irrelevant.
The \emph{relevant} freedom corresponds to $\lambda$-dependent $\grp{O}(n-4)$
rotations of $x^\pm_I$, see \Appref{app:SO6inv} for more
details.

We can now explicitly
express $\eta_i$ in terms of the new parameters,
\begin{equation}
 \eta_i^A=\sum_{M=1}^{(n-4)/2}x^-_{Mi}\alpha_M^A+\sum_{M=1}^{(n-4)/2}x^+_{Mi}\beta_M^A-\eps_{ab}y_i^aQ^{bA}+\eps_{ab}\lambda_i^aY^{bA}.
\label{eq:gammaexp}
\end{equation}
Since for general momentum spinors $\lambda_i^a$ obeying overall momentum conservation the two operators
\begin{equation}
 A_{ij}=\sum_{J=1}^{(n-4)/2}x_{J(i}^+x^-_{Jj)},\qquad\quad B_{ij}=\eps_{ab}\lambda_{(i}^a y_{j)}^b,
\end{equation}
define projectors on the $x^\pm$ and $\lambda$-$y$ subspace, respectively,
the statement that the new variables span the whole space of Gra{\ss}mann
parameters can be rephrased as
\begin{equation}\label{eq:Rcoord}
 \delta_{ij}=\sum_{J=1}^{(n-4)/2}x_{J(i}^+x^-_{Jj)}+\eps_{ab}\lambda_{(i}^a y_{j)}^b.
\end{equation}
Here ${(..)}$ denotes symmetrization in the indices, whereas ${[..]}$ will be used for antisymmetrization in the following. Equation \eqref{eq:Rcoord}, however, only represents the coordinate version of rewriting the
multiplicative R-symmetry generator in terms of the new parameters:
\begin{equation}\label{eq:Rnewpar}
\delta^3(P)\gen{R}^{AB}
=\delta^3(P)\sum_{i=1}^n\eta_i^A\eta_i^B
=\delta^3(P)\Bigg(\sum_{J=1}^{(n-4)/2}\alpha_J^{[A}\beta_J^{B]}+\eps_{ab}Q^{a[A}Y^{aB]}\Bigg).
\end{equation}
Introducing the new set of variables $\{\alpha,\beta,Q,Y\}$ was
originally motivated by this rewriting. In particular, we now find
that the R-symmetry generators further simplify under the supermomentum delta function
\begin{equation}
\delta^3(P)\delta^6(Q)\,\gen{R}^{AB}=\delta^3(P)\delta^6(Q)\sum_{J=1}^{(n-4)/2}\alpha_J^{[A}\beta_J^{B]}.
\end{equation}

In order to investigate the properties of the unknown function $F$ in
\eqref{eq:ninvPQ}
\begin{equation}\label{eq:InG}
  I_n(\lambda_i,\alpha,\beta,Y,Q)=\delta^3(P)\delta^6(Q)\, F(\lambda_i,\alpha,\beta,Y,Q)
\end{equation}
in terms of the new fermionic variables, we act with $\gen{Q}^a_A$ on
the invariant and use the properties of $x^\pm$, $y^a$ and $\lambda^a$
under the momentum delta function to obtain
\begin{equation}
 \gen{Q}^a_A I_n=-\delta^3(P)\delta^6(Q)\eps^{ab} \frac{\partial F}{\partial Y^{bA}}.
 \label{eq:Q2inv}
\end{equation}
Since $\gen{Q}^a_A$ invariance forces this to vanish, the $Y$-dependence of $F$ is constrained to
\begin{equation}
 \frac{\partial F}{\partial Y}\sim  Q.
\end{equation}
All terms of $F$ proportional to $Q$ vanish in \eqref{eq:InG} such that under $\delta^6(Q)$ we find
\begin{equation}
F=F(\lambda_i,\alpha,\beta).
\end{equation}
This guarantees invariance under $\gen{Q}^a_A$. Hence, introducing the
new set of fermionic variables and making use of $\gen{Q}^{aA}$ and
$\gen{Q}^a_A$ invariance, we fixed the dependence of the invariant on
$12$ of the Gra{\ss}mann variables.
Rewriting the R-symmetries in terms of the new variables we obtain the conditions
\begin{align}
\gen{R}^{AB}I_n&= \delta^3(P)\delta^6(Q)\sum_{J=1}^{(n-4)/2}\alpha_J^{[A}\beta_J^{B]}F(\lambda_i,\alpha,\beta)\stackrel{!}{=}0,\nonumber\\
\gen{R}_{AB}I_n&= \delta^3(P)\delta^6(Q)\sum_{J=1}^{(n-4)/2}\frac{\partial}{\partial\alpha_J^{[A}}\frac{\partial}{\partial\beta_J^{B]}}F(\lambda_i,\alpha,\beta)\stackrel{!}{=}0.
\label{eq:Rinv}
\end{align}
Note that since $\alpha$, $\beta$ are independent of $Q$, these
equations equivalently have to hold in the absence of the supermomentum delta function.
%
Solutions to these equations for $n=6$ will be given in \Secref{sec:sixpoint}. Invariance under
\begin{equation}
\gen{R}^A{}_B=\sum_{J=1}^{p}\lrbrk{\alpha_J^A\pardel{\alpha_J^B}+\beta_J^A\pardel{\beta_J^B}-\delta^A_B}
\end{equation}
follows from \eqref{eq:Rinv} using the algebra relations
\eqref{eq:commRuudd}. For more details on the solutions to
these equations, see
\Appref{app:SO6inv} (cf.\ also \cite{Gunaydin:1984vz}).

The analysis up to here concerns only the super-Poincar\'e and
R-symmetry part of \osp. Since this part of the symmetry is believed
to not receive quantum corrections, the considerations up to now
are valid at the full quantum level.

\paragraph{Invariance under \texorpdfstring{$\gen{S}$}{S}.}

In this paragraph we consider the implications of S-invariance
on the function $I_n$. This is the most involved part of the
invariance conditions in this section and will imply invariance under
the conformal boost $\gen{K}_{ab}$ by means of the algebra relation
$\acomm{\gen{S}_{aA}}{\gen{S}^B_b}=\delta^B_A\gen{K}_{ab}$. We apply
the generator $\gen{S}_a^A$ to the invariant $I_n$ after imposing
invariance under $\gen{P}$, $\gen{L}$, $\gen{D}$, $\gen{Q}$ and
$\gen{R}$ as above:
\begin{equation}
\gen{S}_a^A I_n(\lambda,\alpha,\beta,Q)=\delta^3(P)\Big[\frac{\partial \delta^6(Q)}{\partial Q^{aB}}\gen{R}^{AB}F+\delta^6(Q)\gen{S}_a^AF\Big].
\end{equation}
Expressing the R-symmetry generator in terms of the parameters $\alpha$ and $\beta$
\begin{equation}
 \gen{S}_a^A I_n=\delta^3(P)\Bigg[\frac{\partial\delta^6(Q)}{\partial Q^{aB}}\sum_{J=1}^{(n-4)/2}\alpha_J^{[A}\beta_J^{B]}F+\eps_{bc}Y^{c[B}FQ^{bA]}\frac{\partial\delta^6(Q)}{\partial Q^{aB}}+\delta^6(Q)\gen{S}_a^A F\Bigg],
\end{equation}
the first term vanishes by means of \eqref{eq:Rinv}. Using $Q^{bA}\partial\delta^6(Q)/\partial
Q^{aB}=\delta^b_a\delta^A_B\delta^6(Q)$, we can rewrite this as
\begin{equation}
 \gen{S}_a^A I_n=\delta^3(P)\delta^6(Q)\big(2 \eps_{ca}Y^{cA}+\gen{S}_a^A \big)F,
 \label{eq:Srewrite}
\end{equation}
and express the second term in this sum in the form of
\begin{equation}\label{eq:secondexpand}
 \gen{S}_a^A F=\sum_{j,k=1}^n\sum_{J=1}^{(n-4)/2}\eta_k^A\eta_j^B\left(\frac{\partial x_{Jj}^+}{\partial\lambda_k^a}\frac{\partial}{\partial \alpha_J^B}+\frac{\partial x_{Jj}^-}{\partial \lambda_k^a}\frac{\partial}{\partial\beta_J^B}\right)F+\eta^A\cdot F_a.
\end{equation}
Here we have defined the partial derivative of $F$ as
\begin{equation}
 F_{ai}=\left.\frac{\partial F(\lambda,\alpha,\beta)}{\partial \lambda_i^a}\right|_{\alpha,\beta=\mathrm{const}}.
\end{equation}
If we now expand $\eta_i$ in \eqref{eq:secondexpand} in terms of the
new fermionic basis \eqref{eq:gammaexp} and use the conditions
\eqref{eq:xpmdefrel,eq:xpmnormal}, the first term in
\eqref{eq:Srewrite} cancels and the invariance
condition for the S-symmetry takes the form of a differential equation for
the unknown function $F$:
\begin{align}
\label{eq:Sinv}
\gen{S}_a^A I_n&=\delta^3(P)\delta^6(Q)\sum_{J=1}^{(n-4)/2}\Bigg\{\sum_{M,N=1}^{(n-4)/2}\Big[\left(\alpha_M^A Z_{MNJa}^{--+}-\beta_M^A Z_{MNJa}^{++-}\right)
\alpha_N^B\frac{\partial}{\partial\alpha_J^B}\\
&\quad+(\beta_M^A Z_{MNJa}^{+--}+\alpha_M^A Z_{MNJa}^{---})\alpha_N^B\frac{\partial}{\partial \beta_J^B}
\Big]F
+\left(x_J^-\cdot F_a\right)\alpha_J^A
+\brc{(\alpha,+)\leftrightarrow(\beta,-)}\Bigg\}.\nn
\end{align}
Here we have defined for convenience
\begin{equation}
Z_{MNJa}^{\pm\pm\pm}=\sum_{j,k=1}^n x_{Mk}^\pm x_{Nj}^\pm\frac{\partial x_{Jj}^\pm}{\partial \lambda_k^a}.
\label{eq:Zdef}
\end{equation}
Once the differential equation \eqref{eq:Sinv} is satisfied,
invariance under $\gen{S}_{aA}$ follows from the commutation relations
of \osp. While this equation is trivially satisfied for $n=4$, we will
give explicit solutions to it for $n=6$ in \Secref{sec:sixpoint}.

\paragraph{Summary.}

To summarize the previous analysis, a
general $n$-point invariant $I_n$ of the superalgebra can be expanded in a
basis of R-symmetry invariants $F_{n,k}$,%
\footnote{More precisely, the quantities $F_{n,k}$ have to be
multiplied by $\deltaQ$ in order to be actual R-symmetry invariants.
In a slight abuse of notation, we refer to the $F_{n,k}$ themselves
as R-symmetry invariants.}
\begin{equation}
I_{n}=\deltaPQ\sum_{k=1}^Kf_{n,k}(\lambda)F_{n,k}\,,
\label{eq:detinv}
\end{equation}
where a priori some $f_{n,k}(\lambda)$ could be zero.
The number $K$ of basis elements
$F_{n,k}$ is given by the number of singlets in the representation
$(\rep{4}\oplus\rep{\bar4})^{\otimes(n-4)}$, cf.~\Appref{app:SO6inv}.
We have introduced a new basis $\{\alpha_I,\beta_I,Y,Q\}$
for the fermionic superspace coordinates. Using invariance under $\gen{Q}^{aA}$
and $\gen{Q}^a_A$ these are very helpful to fix the dependence of the
invariant on $12$ of the Gra{\ss}mann variables:
The basis elements $F_{n,k}$ are
functions only of the $n-4$ Gra{\ss}mann spinors
$\alpha_1^A,\beta_1^A,\ldots,\alpha_{(n-4)/2}^A,\beta_{(n-4)/2}^A$, multiplied
by the supermomentum delta-function $\deltaQ$.
They have to satisfy the invariance conditions
\eqref{eq:Rinv}. In particular this implies, via the $\alg{u}(1)$
R-charge \eqref{eq:Rgammacounter}, that they have to be homogeneous polynomials
of degree $3(n-4)/2$ in the $\{\alpha_I,\beta_I\}$
variables. This is very different than in \sym, where the $n$-point
amplitude is inhomogeneous in the fermionic variables, and the
coefficients of the lowest and highest powers (MHV amplitudes) have
the simplest form. Here, the $n$-point amplitude rather resembles the most
complicated (N$^{(n-4)/2}$MHV) part of the \sym\ amplitude.
When expanding a general invariant in
the basis $\brc{F_{n,k}}$, the momentum-dependent coefficients must be
Lorentz-invariant and are further constrained by the
S-invariance equation \eqref{eq:Sinv}.
The analysis of that equation for general $n$ is beyond the scope of the present paper.
One would have to analyze whether and how the basic R-symmetry invariants $F_{n,k}$
mix under \eqref{eq:Sinv}. Moreover, the invariants $F_{n,k}$ transform into each other
under a change in the choice of $\{\alpha_I,\beta_I\}$ (for more
details see \Appref{app:SO6inv}).
One nice thing of \eqref{eq:Sinv} is that it
expands into a set
of purely bosonic \emph{first-order} differential equations.

\section{Amplitudes for Four and Six Points}
\label{sec:sixpoint}

After the general analysis of \osp\ $n$-point invariants in
\Secref{sec:level0invariants}, the simplest cases $n=4$ and $n=6$ are
discussed in this section.

\paragraph{Four-Point Amplitude.}

After imposing (super)momentum conservation via the factor $\deltaPQ$,
invariance under the $\alg{u}(1)$ R-charge
\eqref{eq:Rgammacounter} already requires the four-point
superamplitude to be of the form
\begin{equation}
\amp_4=\deltaPQ\,f(\lambda)\,,
\end{equation}
where $f(\lambda)$ is a Lorentz-invariant function of the $\lambda_k$
with weight $-4$. $\amp_4$ then trivially satisfies the
R- and S-invariance conditions
\eqref{eq:Rinv,eq:Sinv} and as a consequence is \osp\
invariant. A field-theory computation \cite{Agarwal:2008pu} shows
that indeed the superamplitude is given by%
\footnote{The two expressions are equal due to the identity
$0=\deltaP\spaa{2|P|4}=\deltaP\bigbrk{\spaa{21}\spaa{14}+\spaa{23}\spaa{34}}$.
Note that we could also write
$\amp_4=i\sign(\spaa{12}\spaa{14})\deltaPQ/\sqrt{\spaa{12}\spaa{23}\spaa{34}\spaa{41}}$, which seems more natural comparing to MHV amplitudes in \sym\ theory. Then, however, one has to deal with the sign factor such that we decided not to use this square root form of the four-point amplitude. }
\begin{equation}
\amp_4
=\frac{\deltaPQ}{\spaa{21}\spaa{14}}
=\frac{\deltaPQ}{-\spaa{23}\spaa{34}}\,,
\label{eq:4point}
\end{equation}
where we neglect an overall constant.
For later reference, we state the component amplitudes for four
fermions and for four scalars:
\begin{equation}
A_{4\psi}\defeq A_4(\psi_4,\bar\psi^4,\psi_4,\bar\psi^4)=\frac{\deltaP\spaa{13}^3}{\spaa{21}\spaa{14}}\,,
\quad
A_{4\phi}\defeq A_4(\phi^4,\bar\phi_4,\phi^4,\bar\phi_4)=\frac{\deltaP\spaa{24}^3}{\spaa{21}\spaa{14}}\,.
\label{eq:A4A4}
\end{equation}
%

\paragraph{Six-Point Invariants.}

In the case of six points, there is only one pair of fermionic
variables $\alpha,\beta$. The space of R-symmetry invariants
in these variables is spanned by the two elements (cf.~\Appref{app:SO6inv})
\begin{equation}
\delta^3(\alpha)=\sfrac{1}{3!}\varepsilon_{ABC}\alpha^A\alpha^B\alpha^C=\alpha^1\alpha^2\alpha^3\,,
\quad
\delta^3(\beta)=\sfrac{1}{3!}\varepsilon_{ABC}\beta^A\beta^B\beta^C=\beta^1\beta^2\beta^3\,.
\end{equation}
Thus the most general six-point function that is \osp\
invariant is given by
\begin{equation}
I_6=\deltaPQ\bigbrk{f^+(\lambda)\,\delta^3(\alpha)+f^-(\lambda)\,\delta^3(\beta)}\,,
\label{eq:Inv6}
\end{equation}
where $\alpha=x^+\cdot\eta$, $\beta=x^-\cdot\eta$ and $x^\pm$
satisfy \eqref{eq:xpmdefrel,eq:xpmnormal}.
In order to be Lorentz-invariant, the
functions $f^\pm(\lambda)$ must only depend on the spinor brackets
\eqref{eq:spaa}. For being invariant under the dilatation generator
\eqref{eq:DinvF}, they furthermore must have weight $-6$ in the
$\lambda_k$'s. Finally, they have to be chosen such that invariance under $\gen{S}^A_a$ is
satisfied. As there is only one pair of $x^\pm$ in the case of six
particles, many of the quantities $Z^{\pm\pm\pm}_a$ defined in
\eqref{eq:Zdef} vanish. Namely, $0=Z^{+\pm\pm}_a=Z^{-\pm\pm}_a$, as can
be seen by acting with $x^\pm\cdot\partial/\partial\lambda^a$ on
$0=x^\pm\cdot x^\pm$ \eqref{eq:xpmnormal}. The $\gen{S}^A_a$ invariance equation
\eqref{eq:Sinv} thus reduces to
\begin{equation}
\gen{S}^A_aI_6
=\deltaPQ\biggbrk{
	 \Bigbrk{x^+\cdot\frac{\partial f^+}{\partial\lambda^a}-3Z^{++-}_af^+}\beta^A\delta^3(\alpha)
	+\brc{(\alpha,^+)\leftrightarrow(\beta,^-)}}\,.
\label{eq:Sinv6pre}
\end{equation}
Invariance under $\gen{S}^A_a$ is therefore equivalent to
\begin{equation}
0=\sum_{k=1}^6x^\pm_k\Bigbrk{
	\frac{1}{f^\pm}\frac{\partial f^\pm}{\partial\lambda_k^a}
	-3\sum_{j=1}^6x^\pm_j\frac{\partial x^\mp_j}{\partial\lambda_k^a}}\,.
\label{eq:Sinv6}
\end{equation}
For given $x^\pm$, this eliminates one functional degree of
freedom of $f^\pm$, which generically depends on $2n-6|_{n=6}=6$ kinematical
invariants (cf.~\Secref{sec:kin}).

\paragraph{Six-Point Amplitude.}

It appears very hard to find a solution to \eqref{eq:Sinv6} directly. Moreover, a
solution would not fix the relative constant between the two terms of
\eqref{eq:Inv6}. In order to obtain the six-point superamplitude, one
thus has to calculate at least one component amplitude from Feynman
diagrams. With two component amplitudes, the invariant \eqref{eq:Inv6} can be fixed
uniquely, without having to solve \eqref{eq:Sinv6}.%
\footnote{This was noted already in \cite{Agarwal:2008pu}.}
The latter can
then be used as a cross-check on the result.
It is reasonable to compute the amplitudes
$A_{6\psi}=A_6(\psi_4,\bar\psi^4,\psi_4,\bar\psi^4,\psi_4,\bar\psi^4)$
and
$A_{6\phi}=A_6(\phi^4,\bar\phi_4,\phi^4,\bar\phi_4,\phi^4,\bar\phi_4)$,
as these have
relatively few contributing diagrams.

To obtain the component amplitudes $A_{6\psi}$ and $A_{6\phi}$ from
the superamplitude $\amp_6$, one has to
extract the coefficients of $\eta^3_1 \eta^3_3 \eta^3_5$ and
$\eta^3_2 \eta^3_4 \eta^3_6$, respectively, in the
expansion of \eqref{eq:Inv6}. The Gra{\ss}mann quantities  $\eta^A_i $
appear in expressions of the form
\begin{equation}
\label{eq:gammaexpan}
\delta^{9}(\eta^A \cdot t^\alpha) \equiv \deltaQ \delta^3(\alpha) \, ,
\end{equation}
where we introduce $t^\alpha_i \equiv (\lambda_i^a, x^+_i)$ (so
$\alpha=1,2,3$).
The $\eta^3_i \eta^3_j \eta^3_k$ term in \eqref{eq:gammaexpan}
is proportional to
\begin{equation}
\det
\begin{pmatrix}
t^1_i & t^2_i & t^3_i \\
t^1_j & t^2_j & t^3_j \\
t^1_k & t^2_k & t^3_k  \\
\end{pmatrix}^3 =
\det
\begin{pmatrix}
\lambda^1_i & \lambda^2_i & x^+_i \\
\lambda^1_j & \lambda^2_j & x^+_j \\
\lambda^1_k & \lambda^2_k & x^+_k  \\
\end{pmatrix}^3 =
 \bigbrk{\spaa{ij}x^+_k+\spaa{jk}x^+_i+\spaa{ki}x^+_j}^3  \, .
\end{equation}
In this way one can extract from \eqref{eq:Inv6} rather simple
expressions for the component amplitudes in terms of $f^\pm$, $x^\pm$:
\begin{align}
A_{6\psi}
&= \bigbrk{\spaa{13}x^+_5+\spaa{35}x^+_1+\spaa{51}x^+_3}^3f^+
  +\bigbrk{\spaa{13}x^-_5+\spaa{35}x^-_1+\spaa{51}x^-_3}^3f^-\,,\nn\\
A_{6\phi}
&= \bigbrk{\spaa{24}x^+_6+\spaa{46}x^+_2+\spaa{62}x^+_4}^3f^+
  +\bigbrk{\spaa{24}x^-_6+\spaa{46}x^-_2+\spaa{62}x^-_4}^3f^-\,.
\label{eq:A6A6}
\end{align}
As shown explicitly in \Appref{app:detthing}, the equations
\eqref{eq:A6A6} indeed determine $f^\pm$ and can be rewritten as
\begin{align}
\frac{A_{6\psi}}{\left(-(p_1+p_3+p_5)^2/2\right)^{3/2}}
&= z f^+ + z^{-1} f^-\,,\nn\\
\frac{is \, A_{6\phi}}{\left(-(p_1+p_3+p_5)^2/2\right)^{3/2}}
&=  z f^+ - z^{-1} f^-\,.
\label{eq:zshit}
\end{align}
where $s$ is an undetermined sign and both $s$ and $z$ are
functions of $\lambda$.
The functions $s$, $z$ parametrize the \emph{relevant} $\grp{O}(2)$
freedom in the choice of $x^\pm$ mentioned below
\eqref{eq:irrelevantfree} and discussed in \Appref{app:SO6inv}. $z$ can obviously be reabsorbed in the
definition of $f^\pm$, the sign $s$ corresponds to the interchange of $f^+$ with
$f^-$.

Using the explicit form of $A_{6\psi}$ and $A_{6\phi}$ obtained from a
Feynman diagram computation in
\Appref{app:feynman}, the equations \eqref{eq:A6A6} determine $f^\pm(\lambda)$ and thereby the whole six-point superamplitude:
\begin{equation}
\amp_6=\deltaPQ\bigbrk{f^+(\lambda)\,\delta^3(\alpha)+f^-(\lambda)\,\delta^3(\beta)}\,.
\end{equation}
We do not state $f^\pm(\lambda)$ here, as their form is not very
illuminating.
Note that an explicit six-point solution of \eqref{eq:xpmdefrel,eq:xpmnormal} for $x^\pm$ is given by
\begin{align}
 x^\pm_{i}&=\frac{1}{2\sqrt{2}} \eps_{ijk}\frac{\spaa{jk}}{\sqrt{\spaa{13}^2+\spaa{35}^2+\spaa{51}^2}},\qquad i,j,k \quad \text{odd},
\nonumber\\
 x^\pm_{i}&=\frac{\pm i}{2\sqrt{2}} \eps_{ijk}\frac{\spaa{jk}}{\sqrt{\spaa{24}^2+\spaa{46}^2+\spaa{62}^2}},\qquad i,j,k \quad \text{even}.
\end{align}
That
the resulting superamplitude indeed satisfies the invariance condition
\eqref{eq:Sinv6} can be seen by symbolically evaluating the latter
and plugging random numerical momentum-spinors
$\lambda_k$ on the support of $\delta(P)$ into the result. In fact,
as can be seen already in \eqref{eq:Sinv6pre}, invariance implies that
the two terms
\begin{equation}
\deltaPQ f^+(\lambda)\,\delta^3(\alpha)\,,
\qquad
\deltaPQ f^-(\lambda)\,\delta^3(\beta)
\end{equation}
are separately S-invariant.

\paragraph{Factorization and Collinear Limits.}

There is a general factorization property (see
e.g.~\cite{Bern:2007dw}) that any color-ordered tree-level
scattering amplitude has to satisfy as an intermediate momentum
$P_{1k}= p_1 + \dots + p_k$ goes on-shell:%
\footnote{Since we are dealing with cyclically invariant amplitudes,
there is no loss of generality in this choice of momenta.}
\begin{equation}
\label{eq:facprop}
\tilde A_n(1,\dots, n) \stackrel{P_{1k}^2 \rightarrow 0}{\rightarrow} \sum_{\text{int. part. $p$}} (\pm 1)^{\fnum_p}\frac{1}{P_{1k}^{2}}
\tilde A_{k+1}(1,\dots, k, \hat \lambda) \tilde A_{n-k+1}(\pm i \hat \lambda,k+1,\dots,n).
\end{equation}
Here $A_n= \tilde A_n \deltaP$ and $\hat \lambda^a$ is defined by the equation
$\hat \lambda^a\hat \lambda^b = P^{ab}_{1k}$, while $\fnum_p$ denotes the fermion number of particle $p$. The freedom in the choice of the sign of
$\hat \lambda^a$ is compensated by the term $ (\pm 1)^{\fnum_p}$. We sum over
all internal particles such that the amplitudes on the right hand side of \eqref{eq:facprop} are non-vanishing.
Finally, the power $2$ of  $1/P_{1k}$ in \eqref{eq:facprop} follows from dimensional analysis, keeping in mind that
\begin{equation}
[\tilde A_n]_{\text{mass dim.}}= 3 -\frac{n}{2}.
\end{equation}
The purpose of this paragraph is to consider \eqref{eq:facprop} using the explicit
expressions for the component amplitudes $A_{4\phi}$, $A_{4\psi}$
\eqref{eq:A4A4} and $A_{6\phi}$, $A_{6\psi}$ \eqref{eq:A6A6} and check for consistency.
In particular, since in the theory under study only amplitudes
with an even number of legs are non-vanishing, $A_{2n}$
should be finite in the generic factorization limit of an even number of legs, i.e.\ have no pole in $P^2_{1,2k}$.

For the four-point amplitude we can distinguish two cases for the two-particle factorization ($=$ collinear) limit.
Using momentum conservation we have $(p_1+p_2+p_3)^2=p_4^2=0$. If we take $P_{12}^2\to0$, i.e.\ $\lambda_1^a=x \lambda_2^a$ for some constant $x$, this gives
\begin{equation}
 0=(p_1+p_2+p_3)^2=2(1+x^2) p_2\cdot p_3.
\label{eq:twocases}
\end{equation}
For generic $x$, this equation implies $p_2\cdot p_3=0$,
yielding that all momenta are collinear and therefore
all kinematical invariants vanish, $\spaa{jk}\sim\spaa{12}$, i.e.\
%
\begin{equation}
\label{eq:coll4}
\tilde A_4 \sim \spaa{12}\quad\text{for}\quad \spaa{12} \rightarrow 0.
\end{equation}
On the other hand \eqref{eq:twocases} is satisfied if $x=\pm i$ or in other words%
\footnote{We thank Yu-tin Huang for pointing our attention to this second case.}%
\begin{equation}
 p_1^\mu+p_2^\mu=0,\qquad p_3^\mu+p_4^\mu=0.
\end{equation}
For this special momentum configuration $\tilde A_4$ does not vanish
in the two-particle collinear limit, but is singular. 

For the six-point amplitudes
there are two different limits to be considered:%
\footnote{For the two six-point amplitudes we computed \eqref{eq:A6A6}, there is no sum over internal particles.}
\begin{itemize}
\item $k=3$: $(p_1+p_2+p_3)^2 \rightarrow 0$. In this case \eqref{eq:facprop} reads:
\begin{equation}
\label{eq:fac1}
\tilde A_6 \rightarrow \frac{1}{P_{13}^2} \tilde A_4 \tilde A_{4}+ \text{finite}
\end{equation}
\item $k=2$: $(p_1+p_2)^2=2 p_1 p_2 \rightarrow 0$, $p_1+p_2\neq0$. In this case \eqref{eq:facprop} reads:
\begin{equation}
\label{eq:fac2}
\tilde A_6 \rightarrow \frac{1}{P_{12}^2} \tilde A_3 \tilde A_{5}+ \text{finite}= \text{finite}.
\end{equation}
\end{itemize}
The latter case is supposed to give a finite result since amplitudes with an
odd number of legs vanish. We checked that \eqref{eq:fac1,eq:fac2} are indeed
satisfied for the amplitudes $A_{6\phi}$ and $A_{6\psi}$ given in
\eqref{eq:A6A6}.
\begin{figure}
\centering\includegraphics[scale=1.2]{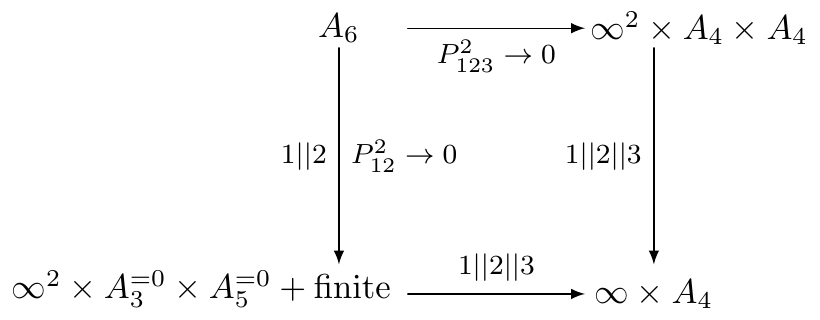}
\caption{Generic collinear ($||$) and factorization ($P^2\to0$) limits of the six point amplitude.}
\label{fig:limits}
\end{figure}


What are the implications of the pole structure of
$A_{6\psi}$, $A_{6\phi}$ on the functions $f^\pm(\lambda)$? First note
that \eqref{eq:A4A4}
\begin{equation}
A_{4\psi}(\lambda_1,\dots,\lambda_4)
=\frac{\spaa{24}}{\spaa{13}}A_{4\phi}(\lambda_1,\ldots,\lambda_4)
=\pm A_{4\phi}(\lambda_1,\ldots,\lambda_4)\,,
\end{equation}
because $p_1+p_3=-p_2-p_4$ and thus $\spaa{13}^2=\spaa{24}^2$,
therefore the sign depends on $\lambda_k$.
This implies that in the three-particle factorization limit
\begin{equation}
\Res_{P_{13}^2=0}\tilde A_{6\psi}=\pm\Res_{P_{13}^2=0}\tilde A_{6\phi}\,.
\end{equation}
Comparing this to \eqref{eq:zshit} shows that either $f^+(\lambda)$ or
$f^-(\lambda)$ does not contribute to the factorization limit. Note
that this is consistent with \Appref{app:super}, where the superanalog
of \eqref{eq:facprop} is worked out. In the three-particle factorization limit, only
one of the basic R-symmetry invariants $\delta^3(\alpha)$,
$\delta^3(\beta)$ survives.

To finish this section, we comment on the limit of three momenta becoming collinear.
This kinematical configuration is nothing but the intersection of the two limits considered
above.  If we first take the sum of three momenta to be on-shell and further restrict
to the configuration where these three momenta become collinear
we obtain
\begin{equation}
\tilde A_6(1,2,3,4,5,6)\,\, \stackrel{P_{13}^2\to0}{\longrightarrow}
\,\,\frac{\tilde A_4(1,2,3,\hat\lambda) \tilde A_{4}(\hat\lambda,4,5,6)}{\spaa{12}^2+\spaa{23}^2+\spaa{13}^2}
\,\,\stackrel{1||2||3}{\longrightarrow}\,\,\frac{\spaa{12}}{\spaa{12}^2} \tilde A_4(4,5,6,\hat\lambda),
\end{equation}
where $P_{13}^2\sim\spaa{12}^2+\spaa{23}^2+\spaa{13}^2$.
Hence, the  $\spaa{12}^{-2}$ divergence in \eqref{eq:fac1} becomes  $\spaa{12}^{-1}$
because $\tilde A_4(1,2,3,\hat \lambda)$ goes to zero as in \eqref{eq:coll4}.
On the other hand we could start from the two-particle collinear limit \eqref{eq:fac2} and see that the finite part on
the right hand side diverges as $\spaa{12}^{-1}$ if the third particle becomes collinear to the (already collinear)
first two particles (cf.~\Figref{fig:limits}).

Note in particular the difference to \sym\ theory, where the
two-particle factorization limit already results in a pole
proportional to a non-vanishing lower-point scattering amplitude.
Furthermore the two-particle factorization and the two-particle
collinear limit are equivalent as opposed to the limits for three
particles relevant for \scs\ theory.

\section{Integrability alias Yangian Invariance}
\label{sec:yanginv}

In this section, we show that the four- and six-point scattering
amplitudes of \scs\ theory given above are invariant under a
Yangian symmetry. In the following, we will refer to the local Lie
algebra representation of \osp\ given in \Secref{sec:rep} as the
level-zero symmetry with generators $\gen{J}^{(0)}_\alpha$, e.g.\
$\gen{P}\to \gen{P}^{(0)}$. Based on this level-zero symmetry, we will
construct a level-one symmetry with generators $\gen{J}^{(1)}_\alpha$
using a method due to Drinfel'd \cite{Drinfeld:1985rx}: We bilocally
compose two level-zero generators forming a level-one generator and
neglect possible additional local contributions. This results in the
bilocal structure of the level-one generators that also appear
in the context of \sym\ theory, see e.g.\
\cite{Dolan:2003uh,Dolan:2004ps}. Up to additional constraints in form
of the Serre relations, the closure of level-zero and level-one
generators then forms the Yangian algebra. Note in particular that,
while the dual superconformal symmetry in \sym\ theory was very
helpful for identifying the Yangian symmetry on scattering amplitudes
\cite{Drummond:2009fd}, it is not a necessary
ingredient for constructing a Yangian.

To be precise, a Yangian superalgebra is given by a set of level-zero and
level-one generators $\gen{J}^{(0)}_\alpha$ and $\gen{J}^{(1)}_\beta$
obeying the (graded) commutation relations
\begin{equation}
\scomm{\gen{J}^{(0)}_\alpha}{\gen{J}^{(0)}_\beta}=f_{\alpha\beta}{}^\gamma \gen{J}^{(0)}_\gamma,
\qquad
\scomm{\gen{J}^{(0)}_\alpha}{\gen{J}^{(1)}_\beta}=f_{\alpha\beta}{}^\gamma \gen{J}^{(1)}_\gamma,
\label{eq:Yangalg}
\end{equation}
as well as the Serre relations%
\footnote{Note that there is a second set of Serre relations that for
finite-dimensional semi-simple Lie algebras follows from
\eqref{eq:serre}, see \cite{MacKay:2004tc}.}
\begin{multline}
\scomm{\gen{J}^{(1)}_\alpha}{\scomm{\gen{J}^{(1)}_\beta}{\gen{J}^{(0)}_\gamma}}
+(-1)^{\abs{\alpha}(\abs{\beta}+\abs{\gamma})}\scomm{\gen{J}^{(1)}_\beta}{\scomm{\gen{J}^{(1)}_\gamma}{\gen{J}^{(0)}_\alpha}}
+(-1)^{\abs{\gamma}(\abs{\alpha}+\abs{\beta})}\scomm{\gen{J}^{(1)}_\gamma}{\scomm{\gen{J}^{(1)}_\alpha}{\gen{J}^{(0)}_\beta}}\\
=\frac{h^2}{24}\,
 (-1)^{\abs{\rho}\abs{\mu}+\abs{\tau}\abs{\nu}}
 f_{\alpha \rho}{}^\lambda
 f_{\beta\sigma}{}^\mu
 f_{\gamma\tau}{}^{\nu}
 f^{\rho\sigma\tau}
 \{\gen{J}_\lambda,\gen{J}_\mu,\gen{J}_\nu].
\label{eq:serre}
\end{multline}
Here, $h$ is a convention dependent constant corresponding to the
quantum deformation (in the sense of quantum groups) of the level-zero
algebra. The symbol $\abs{\alpha}$ denotes the Gra{\ss}mann degree of
the generator $\gen{J}_{\alpha}$ and $\{.\,,.\,,.]$ represents the
graded totally symmetric product of three generators. Given invariance
under $\mathfrak{J}^{(0)}_\alpha$ and $\mathfrak{J}^{(1)}_\alpha$,
successive commutation of the level-zero and level-one generators then
implies an infinite set of generators.

In the case at hand the level-zero generators $\gen{J}^{(0)}_{\alpha}$ can be identified with the standard \osp\ generators defined in \Secref{sec:rep}, where indices $\alpha, \beta, \dots$ label the different generators. We define the level-one generators by the bilocal composition
\begin{equation}\label{eq:bilocYang}
\gen{J}^{(1)}_\alpha =f^{\gamma \beta}{}_\alpha \sum_{1\leq j<i\leq n}\gen{J}^{(0)}_{i\beta}\,\gen{J}^{(0)}_{j\gamma}.
\end{equation}
The definition \eqref{eq:bilocYang} implies that the level-one
generators transform in the adjoint of the level-zero symmetry
\eqref{eq:Yangalg}. Note that in contrast to the local level-zero
symmetry, these bilocal generators incorporate a notion of ordered
sites. Also
note that \eqref{eq:bilocYang} singles out two ``boundary legs'' ($1$
and $n$ in this case), while in the amplitudes $\amp_n$ all legs are
on an equal footing. It was demonstrated in \cite{Drummond:2009fd}
that for $\alg{osp}(2k+2|2k)$ this definition of the Yangian is still
compatible with the cyclicity of the scattering amplitudes.
That is to say,
$\comm{\gen{J}^{(1)}_\alpha}{U}$ vanishes on the amplitudes $\amp_n$, where
$U$ is the site-shift operator.

In explicitly determining the Yangian for \osp, we follow the lines of
\cite{Drummond:2009fd}, where similar computations were performed for \psu.
To evaluate \eqref{eq:bilocYang}, we require the structure constants
$f_{\alpha\beta}{}^\gamma$ of \osp\ that can be easily read off from
the commutation relations in \Secref{sec:rep}. In order to raise or
lower their indices we also need the metric associated with the
algebra whose explicit form is given in \Appref{app:ospmetric}. That
the Yangian indeed satisfies the Serre relations \eqref{eq:serre} is
shown further below.

We want to show Yangian invariance of the four- and six-point
scattering amplitudes. In order to do so, we need to compute only one
level-one generator $\gen{J}^{(1)}_\alpha$ by means of
\eqref{eq:bilocYang}. All other level-one generators can be obtained
by commutation with level-zero generators of the \osp\ algebra
\eqref{eq:Yangalg}. Hence invariance under all other level-one
generators follows from the algebra provided we have shown invariance
under the level-zero algebra as well as under one level-one generator.
The former was done above, the latter will be demonstrated here. We
will therefore only compute the simplest generator $\gen{P}^{(1)ab}$
and show invariance of the scattering amplitudes under this generator.
As demonstrated more explicitly in \Appref{app:levelone}, the level-one generator reads
\begin{equation}\label{eq:P1}
 \gen{P}^{(1)ab}
=\half\sum_{j<i}\left(\gen{Q}_i^{(0)(aA}\gen{Q}_j^{(0)b)}{}_A-\gen{Y}_i^{(0)(a}{}_c \gen{P}_j^{(0)cb)}-(i\leftrightarrow j)\right),
\end{equation}
after we have changed the basis of generators for convenience by combining the dilatation and Lorentz generator into
\begin{equation}
 \gen{Y}^{(0)a}{}_b=\gen{L}^{(0)a}{}_b+\delta^a_b\gen{D}^{(0)}.
\label{eq:genYdef}
\end{equation}

\paragraph{Yangian Invariance of the Four-Point Amplitude.}

We now check that the four-point scattering amplitude introduced in \Secref{sec:sixpoint}
\begin{equation}\label{eq:fouragain}
\mathcal{A}_4=\delta^3(P)\delta^6(Q)f(\lambda)=\frac{\deltaPQ}{\spaa{12}\spaa{41}}=-\frac{\deltaPQ}{\spaa{23}\spaa{34}}
\end{equation}
is annihilated by the Yangian level-one generator $\gen{P}^{(1)ab}$ given in \eqref{eq:P1}. To this end we make use of
\begin{equation}
\partial_{is}\delta(Q)=\eta_i^A\frac{\partial \delta(Q)}{\partial Q^{sA}},\qquad
\partial_{is}\delta(P)=2\lambda_i^b\frac{\partial \delta(P)}{\partial P^{sb}},\qquad
\partial_{iA}\delta(Q)=\lambda_i^a\frac{\partial\delta(Q)}{\partial Q^{aA}},
\end{equation}
such that plugging in the explicit form of the generators straightforwardly yields the action of $\gen{P}^{(1)}$ on $\mathcal{A}_4$ in the following form
\begin{align}
\gen{P}^{(1)ab} \mathcal{A}_4
&= \half\sum_{j<i}\Bigbrk{\gen{Q}_i^{(0)(aR}\gen{Q}_j^{(0)b)}{}_R-\gen{Y}_i^{(0)(a}{}_r\gen{P}_j^{(0)b)r}-(i\leftrightarrow j)} \mathcal{A}_4\nonumber\\
&=\half \delta(P)\delta(Q)\sum_{j<i}\Bigbrk{-\gen{P}_{j}^{(0)r(b}\bigbrk{\eps_{rs}\eps^{st}\lambda_{i}^{a)}\partial_{it}+\half\delta^{a)}_r} f(\lambda)-(i\leftrightarrow j)}.\label{eq:P1calc}
\end{align}
Using the different expressions in \eqref{eq:fouragain} we can rewrite $f(\lambda)$ in the form of
\begin{equation}
 f(\lambda)=\frac{1}{2}\left(\frac{1}{\spaa{12}\spaa{41}}-\frac{1}{\spaa{23}\spaa{34}}\right)
\end{equation}
which yields the following derivative with respect to one of the spinors:
\begin{equation}
 \partial_{it} f(\lambda)=\eps_{st}\frac{1}{2}\left(\frac{\lambda_{i+1}^s}{\spaa{i,i+1}}-\frac{\lambda_{i-1}^s}{\spaa{i-1,i}}\right)f(\lambda).
 \label{eq:property}
\end{equation}
Now we make use of this property of the function $f(\lambda)$. First of all defining the quantity
\begin{equation}\label{eq:defU}
 U_i^{as}=\eps^{st}\lambda_i^{a}\partial_{it} f(\lambda),
\end{equation}
we find that for all $j$, the symmetric part
$U_{\mathrm{sym},i}^{as}=U_i^{(as)}$ satisfies (here $n=4$)
\begin{equation}
\sum_{i=j+1}^n
U_{\mathrm{sym},i}^{as}=\frac{1}{2}\left(\frac{\lambda_j^{(a}\lambda_{j+1}^{s)}}{\spaa{j,j+1}}-\frac{\lambda_n^{(a}\lambda_{n+1}^{s)}}{\spaa{n,n+1}}\right),
\label{eq:Ucond}
\end{equation}
where we have used momentum conservation $P^{ab}=0$.
This implies that $U_{\mathrm{sym},i}^{as}$ does not contribute to
\eqref{eq:P1calc},
\begin{equation}
 \sum_{j<i}\eps_{rs}\gen{P}_j^{r(b} U_{\mathrm{sym},i}^{a)s}=0.
\end{equation}
Hence, in \eqref{eq:P1calc} only the antisymmetric piece $U_{\mathrm{asym},i}^{as}=U_i^{[as]}$ survives and can be shown to take the form
\begin{equation}
U_{\mathrm{asym},i}^{as}=\eps^{as} f(\lambda).
\end{equation}
Thus the four-point scattering amplitude is invariant under the action of the level-one generator $\gen{P}^{(1)ab}$:
\begin{align}
 \gen{P}^{(1)ab}\amp_4&=\half \delta(P)\delta(Q)\sum_{j<i}\Bigbrk{\gen{P}_{j}^{(0)r(b}\bigbrk{\half\delta^{a)}_r-\half\delta^{a)}_r} f(\lambda)-(i\leftrightarrow j)}=0.
\end{align}
As indicated above, invariance of $\mathcal{A}_4$ under all other level-one generators follows from the algebra and hence the four-point scattering amplitude is Yangian invariant.

\paragraph{An $n$-point Invariant of $\gen{P}^{(1)}$.}

Note that the proof of $\gen{P}^{(1)}$-invariance of the four-point
scattering amplitude is based on the property \eqref{eq:property} of the
function $f(\lambda)$.
Hence we can build an $n$-point invariant of the level-one generator $\gen{P}^{(1)}$:
\begin{equation}
 \mathcal{B}_n=\delta^3(P)\delta^6(Q)f(\lambda),
\end{equation}
where the only constraint on $f(\lambda)$ is given by
\eqref{eq:property}.	
In particular, this holds for the choice
\begin{equation}
f(\lambda)=\frac{1}{\sqrt{\spaa{12}\spaa{23}\dots\spaa{n1}}}.
\end{equation}
The Gra{\ss}mann degree of $\mathcal{B}_n$, however, is too low for
being invariant under the level-zero $\alg{u}(1)$ R-symmetry
\eqref{eq:gammaweight}, and thus $\mathcal{B}_n$ cannot be an invariant of the whole Yangian.

\paragraph{Yangian Invariance of the Six-Point Amplitude.}

The six-point superamplitude was introduced in \eqref{eq:Inv6}
\begin{equation}\label{eq:A6again}
\mathcal{A}_6=\deltaPQ\bigbrk{f^+(\lambda)\,\delta^3(\alpha)+f^-(\lambda)\,\delta^3(\beta)}\,,
\end{equation}
with $f^\pm(\lambda)$ as defined in \eqref{eq:A6A6}. We will show that in fact each part of this scattering amplitude
\begin{equation}
\mathcal{A}_6^{+}=\deltaPQ f^+(\lambda)\,\delta^3(\alpha),\qquad
\mathcal{A}_6^{-}=\deltaPQ f^-(\lambda)\,\delta^3(\beta),
\end{equation}
is separately invariant under Yangian symmetry. Demonstrating this for $\mathcal{A}_6^+$, invariance of $\mathcal{A}_6^-$ follows by interchanging $+$, $-$ and $\alpha$, $\beta$ in the following calculation.

In the above paragraph we have seen that
\begin{equation}
\gen{P}^{(1)ab}\mathcal{B}_6=\gen{P}^{(1)ab}\deltaPQ\frac{1}{\sqrt{\spaa{12}\spaa{23}\dots\spaa{61}}}=0.
\end{equation}
Since $\gen{P}^{(1)}$ is a first order differential operator up to constant terms, we can factor out the invariant $\mathcal{B}_6$
in the invariance equation for the six-point amplitude in order to simplify the calculation
\begin{equation}\label{eq:P1A6+}
\gen{P}^{(1)ab}\mathcal{A}_6^+= \mathcal{B}_6\gen{\tilde P}^{(1)ab} \tilde f^+(\lambda)\,\delta^3(\alpha)+\tilde f^+(\lambda)\,\delta^3(\alpha)\gen{P}^{(1)ab}\mathcal{B}_6 .
\end{equation}
Here, of course, the second term vanishes. We have defined
\begin{equation}
 \tilde f^+(\lambda)=\sqrt{\spaa{12}\spaa{23}\dots\spaa{61}}f^+(\lambda),
\end{equation}
and have to drop constant terms in $\gen{P}^{(1)}$ since they are used up for the invariance of $\mathcal{B}_6$:
\begin{equation}
 \gen{\tilde P}^{(1)ab}=\left.\gen{P}^{(1)ab}\right |_\text{constants dropped}.
\end{equation}
Now we rewrite \eqref{eq:P1A6+} as
\begin{equation}
\mathcal{B}_6 \gen{\tilde P}^{(1)ab} \tilde f^+(\lambda)\,\delta^3(\alpha)=\half\mathcal{B}_6\sum_{j<i}\Bigbrk{\lambda_i^{(a}\lambda_j^{b)}\bigbrk{\eta_i^R\partial_{jR}-\lambda_j^r\partial_{ir}}-(i\leftrightarrow j)} \tilde f^+ \delta^3(\alpha).
 \end{equation}
 After expanding $\eta_i$ in terms of $\alpha$, $\beta$, $Q$ and $Y$ \eqref{eq:gammaexp} and using
\begin{equation}
\sum_{k=1}^6x_k^+\frac{\partial x_k^+}{\partial \lambda_{i^c}}=0,
\end{equation}
which follows from \eqref{eq:xpmdefrel}, this yields a differential equation for the function $f^+(\lambda)$ in \eqref{eq:A6again}
\begin{equation}
\gen{P}^{(1)ab}\mathcal{A}_6^+=\half\mathcal{B}_6\sum_{j<i}\Bigg[\lambda_i^{(a}\lambda_j^{b)}\Bigbrk{3x_i^-x_j^+ -3\lambda_j^rx_k^-\frac{\partial x_k^+}{\partial \lambda_i^r}-\lambda_j^r\partial_{ir} \log \tilde f^+}-(i\leftrightarrow j)\Bigg]\tilde f^+\delta^3(\alpha)\stackrel{!}{=}0.
\label{eq:P1inv}
\end{equation}
We have evaluated this equation symbolically using explicit solutions of \eqref{eq:xpmdefrel,eq:xpmnormal} for the coordinates $x^\pm$ as well as the explicit form of $f^+$ given in \eqref{eq:A6A6}. Plugging in specific numerical momentum configurations then shows that \eqref{eq:P1inv} is indeed satisfied. Hence, both summands of the six-point scattering amplitude $\mathcal{A}_6^+$ and $\mathcal{A}_6^-$ are independently invariant under the level-one generator $\gen{P}^{(1)}$ and thereby, as argued above, under the whole Yangian algebra. Note in particular that both $\mathcal{A}_6^{\pm}$ as well as \eqref{eq:P1inv} are independent of the choice of coordinates $x^{\pm}$.

\paragraph{The Serre Relations.}

In this paragraph we show that the Serre relations are indeed
satisfied for the Yangian generators defined above. We do not try to
prove the relations by brute force but first analyze their actual
content, cf.\ also \cite{Drinfeld:1985rx,MacKay:2004tc,Matsumoto:2009rf}.
This leads to helpful insights simplifying the application to the case at hand.

The Yangian algebra $Y(\alg{g})$ of some finite dimensional
semi-simple Lie algebra $\alg{g}$ (here $\alg{osp}(6|4)$) is an
associative Hopf algebra generated by the elements
$\op{J}^{(0)}_\alpha$ and $\op{J}^{(1)}_\alpha$ transforming in the
adjoint representation of $\op{J}^{(0)}$,
\begin{equation}
 \comm{\op{J}^{(0)}_\alpha}{\op{J}^{(0)}_\beta}=f_{\alpha\beta}{}^\gamma \op{J}^{(0)}_\gamma,
\qquad\qquad
 \comm{\op{J}^{(0)}_\alpha}{\op{J}^{(1)}_\beta}=f_{\alpha\beta}{}^\gamma \op{J}^{(1)}_\gamma.
\end{equation}
In all other parts of this paper we do not distinguish between the abstract
algebra elements $\op{J}$ and their representation $\gen{J}$. For the
purposes of this paragraph, however, it seems reasonable to make this
distinction. Making contact to the paragraphs above, we note that
defining a representation $\rho$ of the Yangian algebra $Y(\alg{g})$,
we have
\begin{equation}
 \rho: Y(\alg{g}) \to \mathrm{End}(V),
\qquad \rho(\op{J}^{(0)})=\gen{J}^{(0)},
\quad \rho(\op{J}^{(1)})=\gen{J}^{(1)}.
\end{equation}
The level-zero and level-one generators are promoted to tensor product
operators of $Y(\alg{g})\otimes Y(\alg{g})$ by means of the Hopf
algebra coproduct defined by
\begin{align}
 \Delta(\op{J}^{(0)}_\alpha)&=\op{J}^{(0)}_\alpha\otimes 1 +1\otimes\op{J}^{(0)}_\alpha,\label{eq:cop1}\\
 \Delta(\op{J}^{(1)}_\alpha)&=\op{J}^{(1)}_\alpha\otimes 1 +1\otimes\op{J}^{(1)}_\alpha+\frac{h}{2}f_\alpha{}^{\beta\gamma}\op{J}^{(0)}_\beta\otimes\op{J}^{(0)}_\gamma.\label{eq:cop2}
\end{align}
For consistency of the Yangian, the coproduct has to be
an algebra homomorphism, i.e.\
\begin{equation}
 \Delta(\comm{\op{X}}{\op{Y}})=\comm{\Delta(\op{X})}{\Delta(\op{Y})}
 \label{eq:homomorph}
\end{equation}
for any $\op{X}$, $\op{Y}$ in $Y(\alg{g})$. This equation trivially
holds for $\op{X}$, $\op{Y}$ being $\op{J}^{(0)}_\alpha$,
$\op{J}^{(0)}_\beta$ and for $\op{X}$, $\op{Y}$ being
$\op{J}^{(0)}_\alpha$, $\op{J}^{(1)}_\beta$. The case
\begin{equation}\label{eq:coJ1}
  \Delta(\comm{\op{J}^{(1)}_\alpha}{\op{J}^{(1)}_\beta})=\comm{\Delta\op{J}^{(1)}_\alpha}{\Delta\op{J}^{(1)}_\beta},
\end{equation}
however, is not automatically satisfied and will lead to the Serre
relations. We will now derive a rather simple criterion for
\eqref{eq:coJ1} to be satisfied by a specific representation. In
particular, this criterion will be satisfied by the Yangian
representation of $\alg{osp}(6|4)$ given above.

First of all note that both sides of \eqref{eq:coJ1} are contained in
the asymmetric part of the tensor product of the adjoint
representation with itself. We decompose this as%
\footnote{This is a standard property of all finite-dimensional
semi-simple Lie algebras.}
\begin{equation}\label{eq:decomp}
 (\mathrm{Adj}\otimes\mathrm{Adj})^{\mathrm{asym}}=\mathrm{Adj}\oplus \mathbb{X},
\end{equation}
which defines the representation $\mathbb{X}$ (not containing the
adjoint). The adjoint component of \eqref{eq:coJ1} defines
the coproduct for the level-two Yangian
generators. The Serre relations imply the vanishing
of the $\mathbb{X}$ component of the equation. For seeing this,
one can expand the right hand side of \eqref{eq:coJ1} using
\eqref{eq:cop2}, and project out the adjoint component. As shown
explicitly in \Appref{app:serre}, this yields an
equation of the form
\begin{equation}
0=\Delta(K_{\alpha\beta\gamma})-K_{\alpha\beta\gamma}\otimes 1-
1\otimes K_{\alpha\beta\gamma}.
\end{equation}
The Serre relations then are nothing but $K_{\alpha\beta\gamma}=0$, or
more explicitly
\begin{multline}
 \comm{\op{J}^{(1)}_\alpha}{\comm{\op{J}^{(1)}_\beta}{\op{J}^{(0)}_\gamma}}
+\comm{\op{J}^{(1)}_\beta}{\comm{\op{J}^{(1)}_\gamma}{\op{J}^{(0)}_\alpha}}
+\comm{\op{J}^{(1)}_\gamma}{\comm{\op{J}^{(1)}_\alpha}{\op{J}^{(0)}_\beta}}\\
=\frac{h^2}{24}
 f_{\alpha\rho}{}^{\lambda}
 f_{\beta\sigma}{}^{\mu}
 f_{\gamma\tau}{}^{\nu}
 f^{\rho\sigma\tau}
 \brc{\op{J}^{(0)}_\lambda,\op{J}^{(0)}_\mu,\op{J}^{(0)}_\nu}.
\label{eq:serrealg}
\end{multline}
It is very important to note that only the $\mathbb{X}$ component
of $\{\op{J},\op{J},\op{J}\}$ contributes to the right hand side of
these equations, cf. \Appref{app:serre}. This will be useful in the following.

It is standard knowledge (cf.\ also \Appref{app:serre}) that one can
construct a representation of the Yangian algebra starting from certain
representations of the following form:
\begin{equation}
\rho(\op{J}^{(0)})=\gen{J}^{(0)}\,,\qquad
\rho(\op{J}^{(1)})=0\,,
\label{eq:rhotilde}
\end{equation}
where $\gen{J}^{(0)}$ is a representation of the level-zero part.
The representations $\gen{J}^{(0)}$ for which this construction is
consistent with \eqref{eq:homomorph} are singled out by the Serre relations.
In the language of the present paper, $\rho$ is nothing but
\eqref{eq:jmulti,eq:bilocYang} for one site, i.e.~$n=1$.
For the
representation \eqref{eq:rhotilde}, the Serre
relations boil down to the vanishing of the right hand side of
\eqref{eq:serrealg}. As we have seen that the Serre relations are the
result of a projection onto the representation $\mathbb{X}$, this
is equivalent to
\begin{equation}
\brc{\gen{J}^{(0)}_\alpha,\gen{J}^{(0)}_\beta,\gen{J}^{(0)}_\gamma}\big|_{\mathbb{X}}=0.
\label{eq:JJJ}
\end{equation}
By repeated application of the coproduct to the generators, the
representation $\rho$ is lifted to a non-trivial representation
of the Yangian algebra. The consistency of the construction is
ensured by the homomorphicity of the coproduct \eqref{eq:coJ1}. The
form \eqref{eq:jmulti,eq:bilocYang} for generic $n$ follows from this
construction.

In the following we explicitly show that \eqref{eq:JJJ} is satisfied
for the singleton representation of $\alg{osp}(2k|2\ell)$ relevant to
this paper, cf.~\eqref{eq:genone,eq:covariangen}. Let us start with
the case $k=0$ or $\ell=0$. As demonstrated in \Appref{app:clifford}, the
representation we are using is the superanalog of the spinor
representation of $\alg{so}(2k)$ and the metaplectic representation of
$\alg{sp}(2\ell)$.%
\footnote{The treatment generalizes to $\alg{so}(2k+1)$.}

Consider the decomposition \eqref{eq:decomp} of the antisymmetric part
of the tensor product of two adjoint representations:
\begin{alignat}{2}
\alg{so}(2k):&\qquad&
\left(\autoparbox{\young{\cr\cr}}\otimes\autoparbox{\young{\cr\cr}}\right)\supup{asym}
&=\autoparbox{\young{\cr\cr}}\;\oplus\;\autoparbox{\young{&\cr\cr\cr}}\indup{g-traceless}
\nn\\[.5cm]
\alg{sp}(2\ell):&&
\left(\young{&\cr}\otimes\young{&\cr}\right)\supup{asym}
&=\young{&\cr}\;\oplus\;\autoparbox{\young{&&\cr\cr}}\indup{\Omega-traceless}
\label{eq:Xspec}
\end{alignat}
where $g$ and $\Omega$ are the relevant symmetric and symplectic form,
respectively. Note that the second contribution in these two cases
corresponds to what was called $\mathbb{X}$ above. As explained in
detail in \Appref{app:clifford}, the generators of the spinor and
metaplectic representations acting on one site take the form
\begin{equation}\label{eq:sospgens}
 T^{ij}\sim\comm{\gamma^i}{\gamma^j},\qquad
S^{ij}\sim\acomm{\xi^i}{\xi^j},
\end{equation}
respectively, where
\begin{equation}
 \acomm{\gamma^i}{\gamma^j}=g^{ij},\qquad
\comm{\xi^i}{\xi^j}=\Omega^{ij}.
\label{eq:sospform}
\end{equation}
This means that for any product of the generators \eqref{eq:sospgens}
the symmetrized $g$-traceless or antisymmetrized $\Omega$-traceless
part in two indices vanishes, respectively. Hence, in particular the
quantity $\{\gen{J},\gen{J},\gen{J}\}$ evaluated for
\eqref{eq:sospgens} cannot contain the representation $\mathbb{X}$
defined in equation \eqref{eq:Xspec}. In full detail:
\begin{alignat}{2}
\alg{so}(2k):&\qquad&
\brc{T^{ij},T^{kl},T^{mn}}&\quad\text{decomposes into}\quad
\autoparbox{\young{\cr\cr\cr\cr\cr\cr}}\;\oplus\;2\autoparbox{\young{\cr\cr}}
\nn\\[.5cm]
\alg{sp}(2\ell):&&
\brc{S^{ij},S^{kl},S^{mn}}&\quad\text{decomposes into}\quad
\young{&&&&&&\cr}\;\oplus\;2\young{&\cr}\,.
\end{alignat}
Thus the right hand side of \eqref{eq:serrealg} vanishes for these two
cases.

For the generalization to the super case $\alg{osp}(2k|2\ell)$,
notice that the two equations in \eqref{eq:Xspec} are related to each
other by flipping the tableaux. They generalize to
\begin{alignat}{2}
\alg{osp}(2k|2\ell):&\qquad&
\left(\autoparbox{\young{\cr\cr}}\otimes\autoparbox{\young{\cr\cr}}\right)\supup{asym}
&=\autoparbox{\young{\cr\cr}}\;\oplus\;\autoparbox{\young{&\cr\cr\cr}}\indup{\mathcal{G}-traceless}\,,
\end{alignat}
where in the tableaux for superalgebras, symmetrization and
antisymmetrization are graded. Symmetrization in the tableaux
by convention is defined as (anti)symmetrization in the ($\alg{sp}$)
$\alg{so}$ indices. Antisymmetrization is defined analogously.
The form $\mathcal{G}$ is composed of the metric $g$ and the symplectic form
$\Omega$, cf.~\Appref{app:clifford}. The equations
\eqref{eq:sospgens,eq:sospform} generalize to
\begin{equation}
\left[ \Theta^{\mathcal{A}},\Theta^{\mathcal{B}} \right\}=
\mathcal{G}^{\mathcal{A}\mathcal{B}}\,,
\qquad
J^{\mathcal{A}\mathcal{B}} \sim  \left\{\Theta^{\mathcal{A}} ,
\Theta^{\mathcal{B}}\right]\,.
\end{equation}
The right hand side of \eqref{eq:serrealg} generalizes to the graded
totally symmetrized product of three generators. It contains only the
representations
\begin{alignat}{2}
\alg{osp}(2k|2\ell):&\qquad&
\big\{\gen{J}^{(0)}_\alpha,\gen{J}^{(0)}_\beta,\gen{J}^{(0)}_\gamma\big]&\quad\text{decomposes into}\quad
\autoparbox{\young{\cr\cr\cr\cr\cr\cr}}\;\oplus\;2\autoparbox{\young{\cr\cr}}\,,
\end{alignat}
in particular it does not contain the representation
$\mathbb{X}$. This proves the Serre relations.

Note that only the last part of this proof used the explicit choice of
the algebra and form of the representation. Hence, adapting these last
steps might help to prove the Serre relations for different algebras
and representations.

\paragraph{Note on the Determination of Amplitudes.}

As shown in \Secref{sec:level0invariants}, all $n$-point \osp\
invariants are given by \eqref{eq:detinv}
\begin{equation}
I_{n}=\deltaPQ\sum_{k=1}^Kf_{n,k}(\lambda)F_{n,k}\,,
\label{eq:detinv2}
\end{equation}
where $\deltaQ F_{n,k}$ is a linear basis of R-symmetry
invariants, that is $F_{n,k}$ are homogeneous polynomials of degree
$3(n-4)/2$ of the Gra{\ss}mann variables
$\alpha_J,\beta_J,\ldots,\alpha_{n-4},\beta_{n-4}$ such that
\eqref{eq:Rinv} is satisfied. As is explained in \Appref{app:SO6inv},
the number $K$ of R-symmetry
invariants is given by the number of singlets in the representation
$(\rep{4}\oplus\rep{\bar4})^{\otimes(n-4)}$.

Assuming that invariance under the Yangian algebra not only holds for
the $4$- and $6$-point amplitudes, but for all tree-level amplitudes,
one can ask to what extent the amplitudes are constrained by Yangian
symmetry. Before addressing this question for the general $n$-point
case, let us summarize the cases $n=4$ and $n=6$.
After imposing Poincar\'e invariance, the $K$
functions $f_{n,k}(\lambda)$ a priori depend on $2n-6$ kinematical
invariants, cf.~\Secref{sec:kin}. Further requiring dilatation invariance reduces this
number to $2n-7$. Hence for four points, there remains only
one functional degree of freedom. Since there are no fermionic
variables $\alpha$, $\beta$ in this case,
S-invariance \eqref{eq:Sinv} is automatically satisfied. Invariance
under $\gen{P}^{(1)}$ \eqref{eq:P1} imposes one first-order
differential equation on $f(\lambda)$ and thus completely constrains the
four-point superamplitude up to an overall constant. In the case of
six points, $f^+(\lambda)$ and $f^-(\lambda)$ \eqref{eq:Inv6} depend on $2n-7=5$
parameters. Both S- and $\gen{P}^{(1)}$-invariance impose one
differential equation on each $f^+$ and $f^-$
\eqref{eq:Sinv6,eq:P1inv} without mixing the two functions. Thus
after satisfying these equations, three of the functional degrees of
freedom of $f^+$ and $f^-$ remain undetermined, and they constitute
two independent Yangian invariants.

\begin{table}
\centering
\renewcommand{\arraystretch}{1.5}
\begin{tabular}{|c
                |>{\centering }m{2.0cm}
                |>{\centering }m{2.6cm}
                |>{\centering }m{3.0cm}
                |>{\centering }m{3.0cm}
                |>{\centering\arraybackslash }m{0.9cm}|}
\hline
$n$
&$\gen{R}$-symm.\linebreak invariants
&\emph{Relevant}\linebreak $\alg{so}(n-4)$
&Irreducible rep.\linebreak of $\alg{so}(n-4)$
&Invariants $\tilde{I}_{n}$
&
\\\hline\hline
4
&$1$
&\ding{55}
&\ding{55}
&$f(\lambda)$
&\ding{51}
\\\hline
6
&$\delta(\alpha)$\linebreak$\delta(\beta)$
&$\alg{so}(2) \sim \alg{u}(1)$
&$\mathbf{+}$
\linebreak\
$\mathbf{-}$
&$f^+(\lambda)\delta(\alpha)$
\linebreak\
$f^-(\lambda)\delta(\beta)$
& \ding{51}
\\\hline
8
&$\mathcal{F}^{\tau}$\linebreak$\mathcal{F}_{[\bar{\iota} \bar{\kappa}]}^{\tau}$
&$\alg{so}(4)$
&$2\times\rep{\underline{1}}$
 \linebreak
$2 \times\rep{\underline{6}}$
& $f^{\tau}(\lambda)\, \mathcal{F}^{\tau}$
\linebreak\
$\sum_{\bar{\iota},\bar{\kappa}}f_{[\bar{\iota}
\bar{\kappa}]}^{\tau}(\lambda)\, \mathcal{F}_{[\bar{\iota}
\bar{\kappa}]}^{\tau}$
&{\large\textbf{?}}
\\\hline
10
&$\mathcal{G}_{\bar{\iota}}^{\tau}$\linebreak$\mathcal{G}_{[\bar{\iota} \bar{\kappa} \bar{\upsilon}]}^{\tau}$
&$\alg{so}(6) $
&$8\times\rep{\underline{6}}$
 \linebreak\
$6 \times\rep{\underline{20}}$
&\dots
&{\large\textbf{?}}
\\\hline
\vdots
&\quad\vdots
&\quad\vdots
&\quad\vdots
&\quad\vdots
&{\large\textbf{?}}
\\\hline
\end{tabular}
\renewcommand{\arraystretch}{1.0}
\caption{Summary of the basic R-symmetry invariants and the freedom in
the definition of the fermionic variables $\alpha_J$, $\beta_J$
\protect\eqref{eq:defalphabeta}. $n$ is the number of legs.
$\alg{so}(n-4)$ is the \emph{relevant} freedom
\protect\eqref{eq:relevant}. The Yangian invariants
$I_n=\deltaPQ\,\tilde{I}_n$ are also invariant under this
$\alg{so}(n-4)$ freedom. $\text{\emph{deg}}\times\rep{\underline{R}}$
means that the $\alg{so}(n-4)$ representation $\rep{\underline{R}}$
appears \emph{deg} times among the R-symmetry invariants.
The index $\tau$ labels this multiplicity,
the indices $\bar{\iota}, \bar{\kappa}, \bar{\upsilon}$ are
$\alg{so}(n-4)$ fundamental indices.}
\label{tab:npointinv}
\end{table}

For general number of points $n$, the S-invariance equation
\eqref{eq:Sinv} expands to
\begin{equation}
\gen{S}^A_aI_n=\deltaPQ\sum_{k=1}^K\biggbrk{
	\sum_{J=1}^{(n-4)/2}\bigbrk{\alpha_J^Ax^-_J+\beta_J^Ax^+_J}\cdot\partial_af_{n,k}(\lambda)F_{n,k}
	+f_{n,k}(\lambda)\hat{B}^A_aF_{n,k}}\,,
\label{eq:SinvExp}
\end{equation}
where $\hat{B}^A_a$ is a first-order differential operator in the
fermionic variables $\alpha_J,\beta_J$. Since the $F_{n,k}$ are
independent as functions of $\alpha_J,\beta_J$, also all elements of
$\brc{\alpha^AF_{n,k},\beta^AF_{n,k}}$ are independent (but some of
them might vanish). Thus expanding \eqref{eq:SinvExp} in the fermionic
variables yields
at most $n-4$ first-order differential
equations for each of the functions $f_{n,k}(\lambda)$. From the term
$\sum_k\hat{B}^A_aF_{n,k}$ it might yield
additional equations which only depend on the coordinates $x_J^\pm$
that define $\alpha_J,\beta_J$. Given that the $x_J^\pm$ only
parametrize a change of basis in the fermionic variables, assuming
that there exists an invariant $I_n$ already implies that these
additional equations can be solved by some choice of $x_J^\pm$.
Furthermore requiring
Yangian invariance, i.e.\ invariance under $\gen{P}^{(1)}$
\eqref{eq:P1}, yields
another first-order differential
equation for each function $f_{n,k}(\lambda)$:
\begin{equation}
\gen{P}^{(1)ab}I_n=\deltaPQ\sum_{k=1}^K\biggbrk{
	\bigbrk{\hat{C}^{ab}f_{n,k}(\lambda)}F_{n,k}
	+f_{n,k}(\lambda)\hat{D}^{ab}F_{n,k}}\,,
\end{equation}
where $\hat{C}^{ab}$ is a first-order differential operator in
$\lambda_j$, while $\hat{D}^{ab}$ is a first-order differential
operator in $\alpha_J,\beta_J$. Again, the term
$\sum_k\hat{D}^{ab}F_{n,k}$ might yield additional equations which are
solved by some $x_J^\pm$, assuming existence of an invariant.
In conclusion, there remain at least
$(2n-7)-(n-4)-1=n-4$ functional degrees of freedom for each function $f_{n,k}(\lambda)$.

While for six-point functions, the two basic R-symmetry invariants do
not mix under the S- and $\gen{P}^{(1)}$-invariance equations, for
higher number of points the mixing problem is less trivial.
Nevertheless, an analysis of the relevant freedom \eqref{eq:relevant} suggests that the
mixing should take place (at most) among the $\alg{so}(6)_{R}$
singlets contained in the same  $\alg{so}(n-4)_{\text{relevant}}$
multiplet (see \Tabref{tab:npointinv} and \Appref{app:SO6inv} for details).
This point deserves further investigations.

The above analysis shows that the
invariant \eqref{eq:detinv2} and thus the $n$-point amplitude cannot be
uniquely determined by Yangian symmetry as constructed in
\Secref{sec:yanginv}. Moreover, Yangian invariance not only leaves
constant coefficients but \emph{functional} degrees of freedom undetermined.
As in the case of \sym\
\cite{Bargheer:2009qu,Korchemsky:2009hm}, in order to fully determine
the amplitudes, symmetry constraints have to be supplemented by
further requirements. First of all, the color-ordered superamplitude
$\amp_n$ must be invariant under shifts of its arguments by two sites.
This is a strong requirement that has not been included in the
analysis above. Furthermore, one can require analyticity properties
such as the behavior of the amplitudes in
collinear or more general multiparticle factorization limits.

\section{Conclusions and Outlook}

In this paper we have determined symmetry constraints on tree-level
scattering amplitudes in \scs\ theory. Supplemented by Feynman diagram
calculations, explicit solutions to these constraints, namely the
four- and six-point superamplitudes of this theory were given. Most
notably we have shown that these scattering amplitudes are invariant
under a Yangian symmetry constructed from the level-zero \osp\
symmetry of the theory.

In order to deal with supersymmetric scattering amplitudes, we have
set up an on-shell superspace formulation for \scs\ theory.
This formulation is similar to the one for \sym\ theory, but contains
two superfields corresponding to particles and anti-particles.
Furthermore one of the superfields is fermionic. The realization of
the \osp\ algebra on superspace was used to determine constraints on
$n$-point invariants under this symmetry. In the case at hand,
introducing a new basis $\{\alpha_J,\beta_J,Y,Q\}$ for the fermionic
superspace coordinates seems very helpful in order to find
symmetry invariants. In particular it simplifies the invariance
conditions for amplitudes with few numbers of points. We have
demonstrated that the determination of symmetry invariants can be
reduced to finding $\alg{so}(6)$ singlets plus solving a set of linear
first-order differential equations.

In four dimensions, helicity is a very helpful quantum number for
classifying scattering amplitudes according to their complexity (MHV,
NMHV, etc.). In three dimensions, however, the little group of
massless particles does not allow for such a quantum number, and thus a
similar classification seems not possible. Furthermore, only the
four-point amplitude in \scs\ theory is of similar
simplicity as MHV amplitudes in \sym\ theory. The
six-point amplitude determined in this paper is already of higher degree in
the fermionic superspace coordinates than the four-point amplitude.
Its complexity is comparable with that of the six-point NMHV amplitude
in \sym\ theory. Except for the four-point case, there are
no simple (MHV-type) scattering amplitudes, but the amplitude's
complexity increases with the number of scattered particles. In terms
of complexity, the $n$-point amplitude in \scs\ theory seems
to be comparable to the most complicated, i.e.\ $\mathrm{N}^{(n-4)/2}\mathrm{MHV}$ amplitude in \sym\ theory.

We have checked that the six-point amplitudes consistently
factorize into two four-point amplitudes when the sum of three
external momenta becomes on-shell. The two-particle factorization
limit on the other hand results in a
product of scattering amplitudes with an odd number of external legs
which vanish in \scs\ theory. This is an important
difference to \sym\ theory, where the two-particle collinear
limit results in non-vanishing lower point amplitudes. In particular,
this was used to relate \sym\ scattering
amplitudes with different numbers of external legs. Symmetry plus the
collinear behavior seem to completely fix all tree-level amplitudes in
\sym\ theory \cite{Bargheer:2009qu,Korchemsky:2009hm}. Note
that similar arguments for \scs\ theory would have to make
use of a three-particle factorization or collinear limit (which are not equivalent).

\begin{figure}
\includegraphics[scale=1.2]{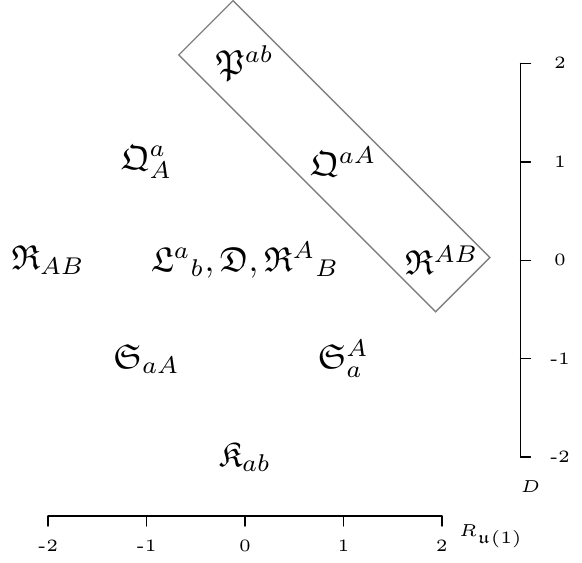}
\raisebox{2.9cm}{\quad vs \quad}
\includegraphics[scale=1.2]{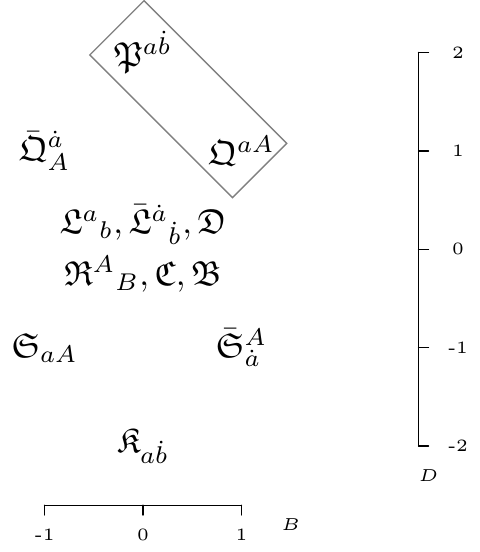}
\caption{The symmetry generators of \osp\ (lhs) and \psu\ (rhs). In \psu\ the generators can be arranged according to their hyper- and dilatation charge. Similarly, we can arrange the generators of \osp\ if we replace the hypercharge by a $\alg{u}(1)$ R-symmetry charge. In \sym\ theory, the dual or level-one Yangian generators $\gen{P}^{(1)}$ and $\gen{Q}^{(1)}$ were identified with the generators $\gen{S}^{(0)}$ and $\gen{K}^{(0)}$, respectively. The picture on the left suggests a similar dualization for \scs\ theory incorporating the R-symmetry. }
\label{fig:gens}
\label{fig:gradingen}
\end{figure}

In \cite{Bargheer:2009qu}, this relation of different \sym\ scattering
amplitudes in the collinear limit was implemented into the
representation of the \psu\ symmetry on the
scattering amplitudes. This implementation makes use of the
so-called holomorphic anomaly
\cite{Cachazo:2004by,Cachazo:2004dr,Britto:2004nj}, which originates in
the fact that four-dimensional massless momenta factorize into complex
conjugate spinors ($p_{4\mathrm{d}}=\lambda\bar\lambda$). In three
dimensions, on the other hand, massless momenta are determined by a
single real spinor ($p_{3\mathrm{d}}=\lambda\lambda$) which
does not allow for a holomorphic anomaly. Hence, a straightforward
generalization of the symmetry relation between amplitudes in the
collinear or factorization limit to \scs\ theory is not
obvious. It lacks a source for a similar anomaly as in the
four-dimensional case.

In \sym\ theory, studying the duality between scattering amplitudes
and Wilson loops revealed a dual superconformal symmetry. The presence
of this extra symmetry then lead to the finding of Yangian symmetry of
the scattering amplitudes. Even more, the dual symmetry was identified
with the level-one Yangian generators \cite{Drummond:2009fd}. Though
in \scs\ theory a similar extra symmetry is not known, there is a
straightforward way to construct level-one generators from the local
\osp\ symmetry yielding a Yangian algebra. We showed that the four-
and six-point tree-level amplitudes of \scs\ theory are indeed
invariant under this Yangian algebra, and that the Yangian generators
obey the Serre relations, which ensures that the Yangian algebra is
consistent.

The fact that \scs\ theory in the planar limit gains extra symmetries
in the form of integrability seems to be related to special properties
of the underlying symmetry algebra \osp, namely the vanishing of the
quadratic Casimir in the adjoint representation (see also
\cite{Flato:1984du,Andrianopoli:2008ea}).
It is interesting to notice that, while in the four-dimensional case
the algebra with this special property is the maximal superconformal
algebra \psu, in three dimensions it is not the maximal
superconformal algebra $\alg{osp}(8|4)$, but \osp\ that has this special
property.

Our findings point towards further investigations. Among others, one
should consider the AdS/CFT dual of \scs\ theory, since in
\sym\ theory the comparison with results from
$\mathrm{AdS}_5\times\mathrm{S}^5$ strings has been extremely useful. The dual superconformal symmetry of scattering amplitudes in \sym\ theory can be traced back to a T-self-duality of the $\mathrm{AdS}_5\times\mathrm{S}^5$ background of the dual string theory \cite{Berkovits:2008ic,Beisert:2008iq}. Such a duality seems not  to be admitted by $\mathrm{AdS}_4\times\mathbb{C}\mathrm{P}^3$, the string theory background corresponding to \scs\ theory \cite{Adam:2009kt}. Can this problem be reconsidered?

In their search for a T-dualization, the authors of \cite{Adam:2009kt}
assume that the dualization does not involve the
$\mathbb{C}\mathrm{P}^3$ coordinates.
On the other hand, the structure of the \osp\
algebra seems to call for a T-dualization of $3+3$ bosonic and $6$ fermionic coordinates
dual to the generators $\left\{ \gen{P}^{ab}, \gen{R}^{AB},\gen{Q}^{aA}\right\}$
(cf.~\Figref{fig:gradingen}). The contributions to the dilaton shift coming from bosonic and fermionic dualization
seem to cancel out. However, this \emph{formal} T-duality is not compatible
with the reality conditions of the coordinates;  still it seems
worthwhile to investigate
it further.%
\footnote{The T-duality we are proposing is very similar to another \emph{formal}
T-duality noticed in section ($3.1$) of \cite{Berkovits:2008ic}.
In that case one T-dualizes the coordinates dual to
$\big\{ \gen{P}^{a \dot a}, \gen{R}^{r r'},\gen{Q}^{a r'} ,
\gen{\bar Q}^{r \dot a}\big\}$. Here, the indices $r, r'$
correspond to the breaking $\alg{su}(4)_R \rightarrow\alg{su}(2)\times\alg{su}(2)$.
This version of the T-duality has not been used so far.}
The problem  with T-dualizing the coordinates of $\mathbb{C}\mathrm{P}^3$
appears to be connected to the lack of a definition
of $\delta(R)$ in our setup.

Other hints for rephrasing the Yangian symmetry in terms of some dual
symmetry could come form perturbative computations in \scs.
In particular, the IR divergences for scattering amplitudes could possibly be mapped to the UV divergences of some other
object (maybe a Wilson loop in higher dimensions). Any results in this
direction might also shed light on the duality between non-MHV
amplitudes and Wilson loops in \sym\ theory, since the amplitudes in
\scs\ theory are very similar to those.
A starting point for the investigation of Wilson loops in
\scs\ was set in the very recent work \cite{Henn:2010ps}.

There are many more open questions and directions for further study.
They comprise the extension of our results to higher point amplitudes,
their extension to loop-level and in particular the understanding of
corresponding quantities in the AdS/CFT dual of the three-dimensional gauge theory.
One of the most interesting problems seems to be whether one can find
a systematic way to determine (tree-level) scattering amplitudes in
\scs\ theory. An apparent ansatz would be an adaption of the
BCFW recursion relations \cite{Britto:2004ap,Britto:2005fq} of
\sym\ theory. This problem is currently under investigation.

Recently, a remarkable generating functional for \sym\ scattering
amplitudes was proposed \cite{ArkaniHamed:2009dn}. The functional takes the form of
a Gra{\ss}mannian integral that reproduces different
contributions to scattering amplitudes. These contributions
have been shown to be (cyclic by construction) Yangian invariants
\cite{ArkaniHamed:2009vw,Drummond:2010qh,Drummond:2010uq,Korchemsky:2010ut}.
It would be interesting to investigate
whether an analogous formula exists for the three-dimensional case
studied in this paper. The (S)Clifford realization presented in
\Appref{app:clifford} could play a similar role for
$\alg{osp}(2k+2|2k)$ as the twistorial realizations
plays in the case of $\alg{psu}(m|m)$.

\paragraph{Acknowledgments.}

We would like to thank Niklas Beisert, Tristan McLoughlin and Matthias
Staudacher for many helpful discussions, suggestions as well as
comments on the manuscript.
In particular, we are grateful to Tristan McLoughlin for his initial
collaboration on this project. We further thank
Lucy Gow, Yu-tin Huang, Jakob Palmkvist and Soo-Jong Rey
for discussions on various related topics.

\appendix

\section{From (S)Clifford algebra to Spinor/Metaplectic representations}
\label{app:clifford}

In this appendix we want to stress that the singleton representation
of \osp\ we are using in this paper (see \Secref{sec:rep})
is nothing but the natural generalization%
\footnote{See also \cite{Hatsuda:2008pm}
for $\alg{osp}(\mathcal N|4)$ and \cite{Aganagic:2003qj} for
$\alg{sp}(2\ell)$.}
of the familiar spinor representation of $\alg{so}(2k)$.
Moreover we will emphasize some special properties of this realization
that makes the Yangian generators defined in \Secref{sec:yanginv}
satisfy the Serre relations \eqref{eq:serre}.

Let us first review the familiar $\alg{so}(2k)$ case.
It is well-known that if one has a representation of the Clifford algebra:
\begin{equation}
\left\{ \gamma^i,\gamma^j \right\}= g^{ij}\,,
\end{equation}
for a given symmetric form $g^{ij}$, where $i,j=1, \dots 2k$,
then the objects
\begin{equation}
\label{eq:sofromclifford}
T^{ij}\sim \left[ \gamma^i,\gamma^j \right]\, ,
\end{equation}
satisfy the $\alg{so}(2k)$ algebra commutation relations
\begin{equation}
\left[ T^{ij}, T^{kl} \right]\sim g^{jk}T^{il}+\dots
\end{equation}
where the dots mean: Add three more terms such that the symmetry
properties of the indices are the same as on the right hand side.
The realization \eqref{eq:sofromclifford} still does not look
like the R-symmetry generators in \eqref{eq:genone}. To obtain \eqref{eq:genone} from
\eqref{eq:sofromclifford} one has to choose an embedding of  $\alg{u}(k)$ into
$\alg{so}(2k)$ and define creation/annihilation type fermionic variables
\begin{equation}
\eta^A \equiv \mathcal{A}^{+A}_j \gamma^j\, , \qquad \qquad
\frac{\partial}{\partial \eta^A} \equiv \mathcal{A}^{-}_{Aj} \gamma^j\,,
\end{equation}
where $A=1, \dots k$ is a $\alg{u}(k)$ index and $\mathcal{A}^{+A}_j,\mathcal{A}^{-}_{Aj}$
have to satisfy
\begin{equation}
\mathcal{A}^{+A}_i g^{ij} \mathcal{A}^{-}_{Bj}= \delta^A_B\, , \qquad \qquad
\mathcal{A}^{+A}_i g^{ij} \mathcal{A}^{+B}_j=0\,,\qquad \qquad
\mathcal{A}^{-}_{Ai} g^{ij} \mathcal{A}^{-}_{Bj}=0,
\end{equation}
in order that $\eta^A,\frac{\partial}{\partial \eta^A} $ satisfy canonical anticommutation relations.
More explicitly, the R-symmetry generators in \eqref{eq:genone} are
related to the ones in \eqref{eq:sofromclifford} via
\begin{equation}
\gen{R}^{AB} \sim \mathcal{A}^{+A}_i \mathcal{A}^{+B}_{j} T^{ij}\, , \qquad \qquad
\gen{R}^{A}_B\sim \mathcal{A}^{+A}_i \mathcal{A}^{-}_{Bj} T^{ij}, \qquad \qquad
\gen{R}_{AB} \sim \mathcal{A}^{-}_{Ai} \mathcal{A}^{-}_{Bj} T^{ij}\, .
\end{equation}
The realization one obtains in this way is not irreducible,
but splits into two irreducible representations (with opposite chirality).
Indeed, the full space of functions (necessarily polynomials) of the variables $\eta^A$ splits
into two spaces: one made of polynomials with only even powers of
$\eta^A$, the other with only
odd powers of $\eta^A$. None of the generators in \eqref{eq:genone} connects the two.

This construction works in the very same way for  $\alg{sp}(2\ell)$, the main difference is that in this case
the representation one obtains is infinite-dimensional. This representation is the direct analog of
the spinor representation and is usually called metaplectic representation.
If one starts with a representation of the algebra:
\begin{equation}
\left[ \xi^i,\xi^j \right]= \Omega^{ij}
\end{equation}
for a given anti-symmetric (non-degenerate) form $\Omega^{ij}$, then the objects
\begin{equation}
\label{eq:spfromsplifford}
S^{ij}\sim \left\{ \xi^i,\xi^j \right\}
\end{equation}
satisfy the $\alg{sp}(2\ell)$ algebra commutation relations
\begin{equation}
\left[ S^{ij}, S^{kl} \right]\sim \Omega^{jk}S^{il}+\dots
\end{equation}
where again the dots mean: Add three more terms such that the symmetry
properties of the indices are the same as on the right hand side.
As before, one has to choose an embedding of $\alg{u}(\ell)$ into
$\alg{sp}(2\ell)$ and define creation/annihilation type bosonic variables
\begin{equation}
\lambda^a \equiv \mathcal{B}^{+a}_j \xi^j\, , \qquad \qquad
\frac{\partial}{\partial \lambda^a} \equiv \mathcal{B}^{-}_{aj} \xi^j\,,
\end{equation}
where $a=1, \dots k$ is a $\alg{u}(\ell)$ index and $\mathcal{B}^{+a}_j,\mathcal{B}^{-}_{aj}$
have to satisfy
\begin{equation}
\mathcal{B}^{+a}_i \Omega^{ij} \mathcal{B}^{-}_{bj}= \delta^b_a\, , \qquad \qquad
\mathcal{B}^{+a}_i \Omega^{ij} \mathcal{B}^{+b}_j=0\,,\qquad \qquad
\mathcal{B}^{-}_{ai} \Omega^{ij} \mathcal{B}^{-}_{bj}=0
\end{equation}
in order that $\lambda^a,\frac{\partial}{\partial \lambda^a} $ satisfy canonical commutation relations.
More explicitly, the bosonic generators in \eqref{eq:genone, eq:genYdef} are
related to the ones in \eqref{eq:spfromsplifford} via
\begin{equation}
\gen{P}^{ab} \sim \mathcal{B}^{+a}_i \mathcal{B}^{+b}_{j} S^{ij}\, , \qquad \qquad
\gen{Y}^{a}_b\sim \mathcal{B}^{+a}_i \mathcal{B}^{-}_{bj} S^{ij}, \qquad \qquad
\gen{K}_{ab} \sim \mathcal{B}^{-}_{ai} \mathcal{B}^{-}_{bj} S^{ij}\, .
\end{equation}
Let us stress that at the group level spinor and metaplectic representations are representations
of $\grp{Spin}(2k)$, $\grp{Mt}(2\ell)$, respectively, which are the double
covers of $\grp{SO}(2k)$, $\grp{Sp}(2\ell)$.

All this easily generalizes  to $\alg{osp}(2k|2\ell)$ algebras. If one starts with objects satisfying
\begin{equation}
\left[ \Theta^{\mathcal{A}},\Theta^{\mathcal{B}} \right\}= \mathcal{G}^{\mathcal{A}\mathcal{B}}\, ,
\end{equation}
where $\mathcal{A},\mathcal{B}$ label the $2k+2\ell$-dimensional fundamental representation of
$\alg{osp}(2k|2\ell)$, then
\begin{equation}
\label{eq:covariangen}
J^{\mathcal{A}\mathcal{B}} \sim  \left\{\Theta^{\mathcal{A}} , \Theta^{\mathcal{B}}\right] \, ,
\end{equation}
satisfy $\alg{osp}(2k|2\ell)$ algebra commutation relations.
After choosing an embedding of $\alg{u}(k|\ell)$ into $\alg{osp}(2k|2\ell)$,
one obtains oscillator type realizations, like the one in \eqref{eq:genone}.

\section{\texorpdfstring{$\alg{so}(6)$}{so(6)} Invariants}
\label{app:SO6inv}

In this appendix we will study the problem of determining
invariants under the following realization of $\alg{so}(6)$:
\begin{equation}
\gen{R}^{AB}= \sum_{J=1}^{p}\alpha_J^{[A}\beta_J^{B]}\, ,
\end{equation}
\begin{equation}
\gen{R}_{AB}= \sum_{J=1}^{p}\frac{\partial}{\partial\alpha_J^{[A}}\frac{\partial}{\partial\beta_J^{B]}}\, ,
\end{equation}
\begin{equation}
\gen{R}^{A}_B= \sum_{J=1}^{p} \left( \alpha_J^{A}\frac{\partial}{\partial \alpha_J^B}-
\frac{\partial}{\partial \beta_J^B} \beta_J^{A} \right)\, ,
\end{equation}
where $\alpha_I^A,\beta_I^A$ are anticommuting fermionic variables
and $A,B$ are $SU(3)$ indices. $p$ is some integer, it is related to
the number $n$ of amplitude legs
as $2p=n-4$. This realization is completely
equivalent to the following one:
\begin{equation}
\gen{R}^{AB}= \sum_{\bar{\iota}=1}^{2p} \rho_{\bar{\iota}}^A\rho_{\bar{\iota}}^B\, ,
\end{equation}
\begin{equation}
\gen{R}_{AB}= \sum_{\bar{\iota}=1}^{2p}\frac{\partial}{\partial \rho_{\bar{\iota}}^A}\frac{\partial}{\partial \rho_{\bar{\iota}}^B}\, ,
\end{equation}
\begin{equation}
\gen{R}^A{}_B= \sum_{\bar{\iota}=1}^{2p}\frac{1}{2}\left( \rho_{\bar{\iota}}^A\frac{\partial}{\partial  \rho_{\bar{\iota}}^B}
-\frac{\partial}{\partial  \rho_{\bar{\iota}}^B}\rho_{\bar{\iota}}^A \right)\, ,
\end{equation}
where $\rho_{\bar{\iota}}$ are linearly related to $\alpha^A_I, \beta^A_I$, the map
between $\rho$ and $\alpha,\beta$ is parametrized by a $\grp{O}(2p)$ freedom.
Notice that this last realization makes sense also for odd $2p$.

All the generators written above are invariant under $\grp{O}(2p)$ rotations
among the family indices. In the following we will refer to this group as dual.
$\grp{O}(2p)_{\text{dual}}$ rotation symmetry is manifest in the form of the generators
written in terms of $\rho_{\bar \iota}^A$ as a rotation of the indices $\bar{\iota}$.
On the generators written in terms of $\alpha, \beta$ $\grp{O}(2p)_{\text{dual}}$ acts in the following way
\begin{align}
\alpha_I & \rightarrow \Xi_I^J \alpha_J \,,&
\beta^I & \rightarrow \left( \Xi^{-1} \right)_J^I \beta^J \,,&
\,& U(p)\,, \qquad p^2\quad \text{d.o.f.} \nn\\
\alpha_I & \rightarrow  \alpha_I  +\Omega_{IJ+} \beta^J \,,&
\beta^I & \rightarrow  \beta^I  +\Omega^{IJ}_- \alpha_J \,,&
\,& \frac{SO(2p)}{U(p)}\,, \qquad  p(p-1) \quad \text{d.o.f.} \nn\\
\alpha_I  &  \rightarrow \beta^I,,&
\beta^I  &  \rightarrow \alpha_I \,,&
\,& \mathbb{Z}_2 \sim \frac{O(2p)}{SO(2p)}\,,
\label{eq:relevant}
\end{align}
where $\Omega_\pm^{IJ}=-\Omega_\pm^{JI}$. The $\mathbb{Z}_2$ is the conjugation of $\alg{su}(p)$
(outer automorphism).
Notice that we raised the family index of $\beta$, we have to do this
in order to interpret
the family index as a $\alg{u}(p)$ index.

In the following, we will show how the $\alg{so}(6)$ invariants
can be obtained and classified.
Since the description in terms of $\alpha,\beta$ is equivalent (for integer $p$) to the one
in terms of $\rho$, we will switch between the two depending on convenience.

It is instructive to first study the case $2p=1$. This case obviously makes sense only in the
$\rho$ realization. In this case the full fermionic Fock space
is $2^3=8$-dimensional and split into $4 \oplus \bar{4}$ representations of $\alg{so}(6)$.
The two correspond to even or odd functions (just polynomials up to degree $3$)
in $\rho$, respectively.

Let us now consider the next case: $p=1$. The study of this case is particularly
transparent in terms of $\alpha,\beta$. To classify the states it is useful to introduce
an extra operator $g$
\begin{equation}
\label{eq:centralg1}
g=\alpha^A \frac{\partial}{\partial \alpha^A}- \beta^A \frac{\partial}{\partial \beta^A}\,.
\end{equation}
This operator is central with respect to $\alg{so}(6)$ and is nothing but the generator
of the previously mentioned dual $\alg{so}(2p)|_{p=1}\sim \alg{u}(1)$.
In this case  the full Fock space is $2^6=64$ dimensional,
it decomposes into irreducible representations of $\alg{so}(6)$ as
\begin{equation}
\left(4 \oplus \bar 4 \right)^2=1_3 \oplus 6_2 \oplus 15_1 \oplus  10_0  \oplus  \bar{10}_0 \oplus 15_{-1}\,,
\oplus  6_{-2} \oplus 1_{-3}
\label{eq:rsymmp1}
\end{equation}
where the subscript refers to the charge under $g$ \eqref{eq:centralg1}.
This decomposition is concretely realized by the solutions to the equation
\begin{equation}
\gen{R}_{AB}\ket{\text{State}}=\frac{\partial}{\partial\alpha^{[A}}\frac{\partial}{\partial\beta^{B]}}\ket{\text{State}}=0\, .
\end{equation}
We can be more explicit and show how these states look like in the space of
Gra{\ss}mann variables $\alpha^A, \beta^A$. For clarity we also explicitly write down the
decomposition under $SO(6)\rightarrow SU(3)$.
\begin{itemize}
\item $1_3 \rightarrow 1: \quad \epsilon_{ABC}\alpha^A \alpha^B \alpha^C$,
\item $6_2 \rightarrow \bar 3 \oplus 3: \quad \epsilon_{ABC}\alpha^B \alpha^C +\text{descendants}$,
\item $15_1 \rightarrow 3 \oplus 8 \oplus 1\oplus \bar 3: \quad  \alpha^A +\text{descendants}$,
\item $10_0 \rightarrow 1 \oplus 3 \oplus 6: \quad 1 +\text{descendants}$,
\item $\bar{10}_0 \rightarrow \bar 6 \oplus \bar 3  \oplus  1:  \quad \alpha^{(A} \beta^{B)} +\text{descendants}$,
\item $15_1 \rightarrow 3 \oplus 8 \oplus 1\oplus \bar 3: \quad  \beta^A +\text{descendants}$,
\item $6_2 \rightarrow \bar 3 \oplus 3: \quad \epsilon_{ABC}\beta^B \beta^C +\text{descendants}$,
\item $1_3 \rightarrow 1: \quad \epsilon_{ABC}\beta^A \beta^B \beta^C$,
\end{itemize}
where descendants means obtained acting with $\gen{R}^{AB}$.

We will now consider the case $p=2$, namely
\begin{equation}
\left(4 \oplus \bar 4 \right)^4\,.
\end{equation}
We can just take the expression \eqref{eq:rsymmp1} and square it.
We will not write down
the whole tensor product decomposition, but just list the singlets.
One can easily check that there are $12$ singlets coming  from $15_{\pm 1}
\otimes 15_{\pm 1}, 6_{\pm 2} \otimes 6_{\pm 2}, 1_{\pm3}\otimes1_{\pm3}$,
where the signs have to be considered independently; and two further
singlets are contained in $10_{0} \otimes \bar{10}_{0}$ ($2$ times).
For convenience we will list the explicit expressions of the singlets:
\begin{itemize}
\item $15_{\pm 1} \otimes 15_{\pm 1}$ contains 4 singlets:
\begin{equation}
\epsilon_{A B G} \epsilon_{E F C}
\alpha_1^A \alpha_1^{(B} \beta_1^{C)} \alpha_2^E \alpha_2^{(F}
\beta_2^{G)},
\qquad
(\alpha_1\leftrightarrow\beta_1)\text{ and/or }(\alpha_2\leftrightarrow\beta_2).
\end{equation}
\item $6_{\pm 2} \otimes 6_{\pm 2}$ contains 4 singlets:
\begin{equation}
\epsilon_{A B C} \epsilon^{A D E}
\alpha_1^B \alpha_1^{C}  \epsilon_{DF G} \alpha_2^F \alpha_2^{G} \epsilon_{EHI} \beta_2^H \beta_2^{I},
\qquad
(\alpha \leftrightarrow \beta)\text{ and/or }(1 \leftrightarrow 2).
\end{equation}
\item $1_{\pm 3} \otimes 1_{\pm 3}$ contains 4 singlets:
\begin{equation}
\epsilon_{ABC}\alpha_1^A \alpha_1^B \alpha_1^C\epsilon_{DEF}\alpha_2^D
\alpha_2^E \alpha_2^F,
\qquad
(\alpha_1\leftrightarrow\beta_1)\text{ and/or }(\alpha_2\leftrightarrow\beta_2).
\end{equation}
\item $10_0 \otimes \bar{10}_0$ contains $2$ singlets:
\begin{equation}
\epsilon_{ACD}\epsilon_{BEF} \alpha_1^A \beta_1^B \alpha_2^C
\alpha_2^D  \beta_2^E\beta_2^F,
\qquad
(1 \leftrightarrow 2).
\end{equation}
\end{itemize}
A question one can ask is how these singlets transform among
themselves under the $\alg{so}(2p)|_{p=2}=\alg{so}(4)\indup{dual}$ transformations.
This question can be answered noticing that the quantities
\begin{equation}
\label{centralg}
g_I=\alpha^A_I \frac{\partial}{\partial \alpha^A_I}- \beta^A_I \frac{\partial}{\partial \beta^A_I}\,.
\end{equation}
(no sum over $I$), are nothing but the Cartan generators of the
$\alg{so}(4)\indup{dual}$, and these
singlets are indeed labeled by $(g_1,g_2)$.

The $\alg{so}(4)\indup{dual}$ transformation properties of the singlets can also be obtained considering where
the singlets come from in
\begin{equation}
\label{eq:tensor22}
\left(4 \oplus \bar 4 \right)^4\,.
\end{equation}
The  $\alg{so}(4)\indup{dual}$ acts as a rotation of the four factors $(4 \oplus \bar 4)$ in the fourfold tensor product above.
Keeping in mind that a tensor product of $n_4$ fundamental with  $n_{\bar 4}$ antifundamental can contain singlets only if
$n_4-n_{\bar 4}=0 \mod 4$,
it is easy to see that singlets can only come from
\begin{itemize}
\item $4 \otimes 4 \otimes 4 \otimes 4$: $1$ singlet under $\alg{so}(6)_{\gen{R}}$, singlet also under $\alg{so}(4)_{\text{dual}}$,
\item $\bar{4} \otimes \bar{4} \otimes \bar{4} \otimes \bar{4}$: 1 singlet under $\alg{so}(6)_{\gen{R}}$, singlet also under $\alg{so}(4)_{\text{dual}}$,
\item $\bar{4} \otimes \bar{4} \otimes 4 \otimes 4$: $2$ singlets under $\alg{so}(6)_{\gen{R}}$ $\times$ $\rep{\underline{6}}$ under $\alg{so}(4)_{\text{dual}}$,
\end{itemize}
in the last line the combinatorial factor $\binom{4}{2}=\rep{\underline{6}}$ corresponding to the possible ways of choosing
two $4$ and two $\bar 4$ in \eqref{eq:tensor22}, is also the dimension
of the $\alg{so}(4)_{\text{dual}}$ representation under which
these ($\alg{so}(6)_{\gen{R}}$) singlets transform.

The cases $p=3$
\begin{equation}
\left(4 \oplus \bar 4 \right)^6\,,
\end{equation}
can be considered analogously giving
\begin{itemize}
\item $4 \otimes 4 \otimes 4 \otimes 4 \otimes 4 \otimes \bar{4} $:
$4$ singlets under $\alg{so}(6)_{\gen{R}}$ $\times$ $\rep{\underline{6}}$ under $\alg{so}(6)_{\text{dual}}$,
\item $\bar{4} \otimes \bar{4} \otimes \bar{4} \otimes \bar{4}\otimes \bar{4} \otimes 4  $:
$4$ singlets under $\alg{so}(6)_{\gen{R}}$ $\times$ $\rep{\underline{6}}$ under $\alg{so}(6)_{\text{dual}}$,
\item $\bar{4} \otimes \bar{4} \otimes \bar{4} \otimes 4 \otimes 4 \otimes 4$:
$6$ singlets under $\alg{so}(6)_{\gen{R}}$ $\times$ $\rep{\underline{20}}$ under $\alg{so}(6)_{\text{dual}}$,
\end{itemize}
where again the combinatorial factors $\binom{6}{1}=\rep{\underline{6}}$, $\binom{6}{3}=\rep{\underline{20}}$
are also the dimensions of the $\alg{so}(6)_{\text{dual}}$ representations.

The general $p>3$ cases can be studied similarly.

\section{Determinability of the Six-Point Superamplitude}
\label{app:detthing}

This appendix is devoted to the study of the invertibility of
equation \eqref{eq:A6A6}. More precisely, we will show under which conditions
one can solve \eqref{eq:A6A6}
for $f_\pm$ in terms of the component amplitudes $A_{6 \psi}$, $A_{6 \phi}$.
This is an important step, as the determination of the six-point superamplitude,
and, thus the determination of \emph{all} six-point component amplitudes
relies on it.
Let us define the following quantities
\begin{equation}
A(\pm)_{ijk} \equiv
\begin{pmatrix}
\lambda^1_i & \lambda^2_i & x^\pm_i \\
\lambda^1_j & \lambda^2_j & x^\pm_j \\
\lambda^1_k & \lambda^2_k & x^\pm_k  \\
\end{pmatrix} \, .
\end{equation}
\begin{equation}
D_\pm \equiv\det \left( A(\pm)_{ijk} \right) \, ,
\qquad \qquad
\bar{D}_\pm \equiv \det \left( A(\pm)_{\bar{i} \bar{j} \bar{k}} \right) \, ,
\end{equation}
for some fixed $i\neq j \neq k$, and let $\left\{\bar{i}, \bar{j},
\bar{k} \right\}\equiv \left\{1,\ldots,6\right\}\setminus
\left\{i,j,k \right\}$ as a set.
Equation \eqref{eq:A6A6} can be inverted iff
\begin{equation}
\label{eq:invertibcond}
D_+^3 \bar{D}_-^3-D_-^3 \bar{D}_+^3 \neq 0 \, .
\end{equation}
Using $\lambda^a \cdot \lambda^b=0$, $x^\pm \cdot \lambda^a=0$, $x^\pm \cdot x^\pm=0$,
$x^+ \cdot x^-=1$,  one can show, performing matrix multiplication, that
\begin{equation}
A^T(\pm)_{ijk} A(\pm)_{ijk} = -A^T(\pm)_{\bar{i} \bar{j} \bar{k}} A(\pm)_{\bar{i} \bar{j} \bar{k}}  \, ,
\end{equation}
\begin{equation}
A^T(\pm)_{ijk} A(\mp)_{ijk} = -A^T(\pm)_{\bar{i} \bar{j} \bar{k}} A(\mp)_{\bar{i} \bar{j} \bar{k}}
+ \Bigl( \begin{smallmatrix}
 0&0&0\\
0& 0 &0\\
0&0&1
\end{smallmatrix}  \Bigr) \, ,
\end{equation}
where $T$ means transposition.
These two equations imply respectively that
\begin{equation}
\label{eq:appendixdet0}
D_\pm^2= - \bar{D}_\pm^2 \Rightarrow  \bar{D}_\pm= i s_\pm D_\pm \, ,
\end{equation}
\begin{equation}
\label{eq:appendixdet}
D_+D_-+\bar{D}_+\bar{D}_-=\det\left((p_i+p_j+p_k)^{ab}\right)\,,
\end{equation}
where $s_\pm$ are undetermined signs. Using \eqref{eq:appendixdet0},
\eqref{eq:appendixdet} can be rewritten as
\begin{equation}
 D_+ D_- \left(1-s_+ s_-\right)
 =
 \det\left((p_i+p_j+p_k)^{ab}\right) \, .
\end{equation}
Since for generic momentum configurations
$(p_i+p_j+p_k)^2$ is not vanishing, it follows that
$s_+=s$, $s_-=-s$ for some sign $s$.
This shows that, for generic momentum configurations,
\eqref{eq:invertibcond} holds, indeed
\begin{equation}
D_+^3 \bar{D}_-^3-D_-^3 \bar{D}_+^3 = i (s_+-s_-)D_+^3 D_-^3 = 2 i s D_+^3 D_-^3 =
\frac{is}{4} \det\left((p_i+p_j+p_k)^{ab}\right)^3 \neq 0 \, .
\end{equation}
To summarize, the quantities $D_\pm$, $\bar{D}_\pm$
are not independent.
Given $(p_i+p_j+p_k)^2$, they are determined up to a sign
$s$ and a single function (which is a phase once we impose the correct reality conditions).
This freedom corresponds to the $\grp{O}(n-4)|_{n=6}=\grp{O}(2)$ \emph{relevant} freedom in the choice
of $x^\pm$ mentioned in \Secref{sec:level0invariants}. The sign is
$\grp{O}(2)/\grp{SO}(2)$,
and corresponds to exchanging $x^+$ and $x^-$; the freedom that remains,
$D^\pm \rightarrow \Xi^{\pm1}D^\pm$, corresponds to the
$\grp{SO}(2) \sim \grp{U}(1)$ freedom
of rescaling $x^\pm \rightarrow \Xi^{\pm1/3}x^\pm$.

\section{Two Component Amplitude Calculations}
\label{app:feynman}

In the following, the amplitudes between six scalars and between six
fermions are computed. As discussed in section \Secref{sec:sixpoint},
these two amplitudes uniquely determine the six-point superamplitude.
For simplicity, consider only (anti)particles
of the same flavor; set
\begin{equation}
\phi=\phi^4\,,
\quad
\bar\phi=\bar\phi_4\,,
\quad
\psi=\psi_4\,,
\quad
\bar\psi=\bar\psi^4\,.
\end{equation}
The action of $\superN=6$ superconformal Chern--Simons theory
is $S=k/4\pi\int\dd^3x\,\mathcal{L}$. Neglecting terms
that are irrelevant for the two specific amplitudes we are interested
in, the Lagrangian reads
(see e.g.\ \cite{Aharony:2008ug,Benna:2008zy,Minahan:2008hf})
\begin{equation}
\mathcal{L}=\tr\Bigsbrk{
	\varepsilon^{\mu\nu\lambda}\bigbrk{
		 A_\mu\partial_\nu A_\lambda
		+\twothird A_\mu A_\nu A_\lambda
		-\hat A_\mu\partial_\nu\hat A_\lambda
		-\twothird\hat A_\mu\hat A_\nu\hat A_\lambda}
	-\sfrac{i}{2}\bar\psi^a\slashed{D}_{ab}\psi^b
	+D_\mu\bar\phi D^\mu\phi}\,.
\end{equation}
The gauge fields $A_\mu$, $\hat A_\mu$ transform in $(\rep{ad},\rep{1})$,
$(\rep{1},\rep{ad})$ representations of the gauge group. The covariant
derivative $D_\mu$ acts on fields $\chi\in\brc{\phi,\psi}$,
$\bar\chi\in\brc{\bar\phi,\bar\psi}$ as
\begin{equation}
D_\mu \chi =\partial_\mu \chi +A_\mu \chi -\chi \hat A_\mu\,,
\quad
D_\mu\bar \chi =\partial_\mu\bar \chi +\hat A_\mu\bar \chi -\bar \chi A_\mu\,,
\quad
\slashed{D}_{ab}=\sigma^\mu_{ab}D_\mu\,.
\end{equation}
The Feynman rules can be straightforwardly derived from $\mathcal{L}$,
using the Faddeev--Popov regularization for the gauge field
propagators.

\paragraph{Six-Fermion Amplitude.}

\begin{figure}
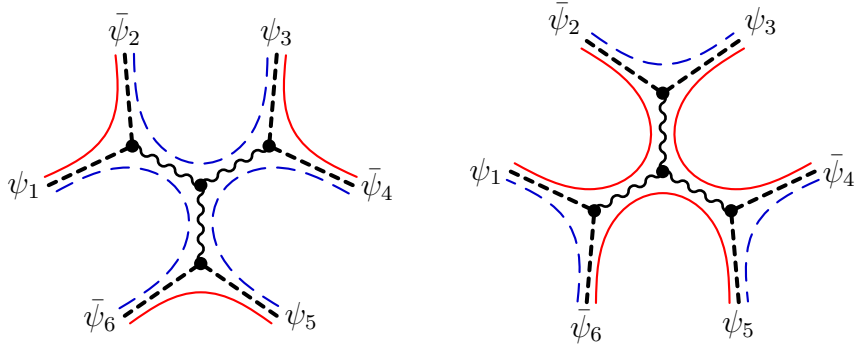
\centering
\includegraphicsbox{FigSixFerm3g1}
\qquad
\includegraphicsbox{FigSixFerm3g2}\\
\caption{Diagram A, which contributes to the six-fermion
amplitude. Blue/dashed lines represent fundamental color contractions,
red/solid lines represent antifundamental ones. When color-stripped,
the left diagram gives \protect\eqref{eq:fermA}, and the right
diagram equals the left one up to a relabeling of the external legs.}
\label{fig:fermA}
\end{figure}

\begin{figure}
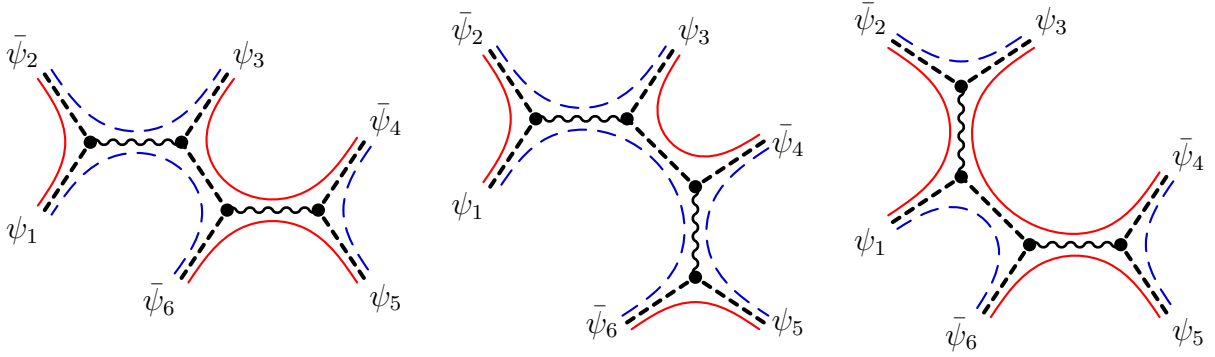
\centering
\includegraphicsbox{FigSixFermfp1}
\hfill
\includegraphicsbox{FigSixFermfp2}
\hfill
\includegraphicsbox{FigSixFermfp3}\\
\caption{Diagram B, which contributes to the six-fermion
amplitude. When color-stripped, the left diagram gives
\protect\eqref{eq:fermB}, and the other two diagrams equal the left
one up to relabelings of the external legs.}
\label{fig:fermB}
\end{figure}

The tree-level amplitude
\begin{equation}
\hat A_{6\psi}\defeq\hat A_6(\psi_1{}^{A_1}_{\bar A_1},\bar\psi_2{}^{\bar B_2}_{B_2},\psi_3{}^{A_3}_{\bar A_3},\bar\psi_4{}^{\bar B_4}_{B_4},\psi_5{}^{A_5}_{\bar A_5},\bar\psi_6{}^{\bar B_6}_{B_6})\,,
\quad
\psi_k\defeq\psi(\lambda_k)
\end{equation}
can be color-ordered \eqref{eq:colord}. The color-ordered amplitude
$A_{6\psi}(\lambda_1,\ldots,\lambda_6)$ contains all contributions in
which the fields $\psi_1,\ldots,\psi_6$ are cyclically connected by
color contractions,
\begin{equation}
\hat A_{6\psi}=
\ldots+
A_{6\psi}(\lambda)\,
\delta_{\bar A_1}^{\bar B_2}\delta_{B_2}^{A_3}\delta_{\bar A_3}^{\bar B_4}\delta_{B_4}^{A_5}\delta_{\bar A_5}^{\bar B_6}\delta_{B_6}^{A_1}
+\ldots\,,
\quad
\lambda\defeq(\lambda_1,\ldots,\lambda_6)\,.
\label{eq:colorstruct}
\end{equation}
Two kinematically different diagrams contribute to
$A_{6\psi}(\lambda)$, see \Figsref{fig:fermA,fig:fermB}.
Diagram A (left in \Figref{fig:fermA}) evaluates to%
\footnote{Define
$p_{jk}^{ab}\defeq p_j^{ab}+p_k^{ab}=\lambda_j^a\lambda_j^b+\lambda_k^a\lambda_k^b$.}
\begin{multline}
A_{6\psi,\mathrm{A}}(1,\ldots,6)
=\frac{C_6}{3}\frac{1}{\spaa{12}\spaa{34}\spaa{56}^2}\cdot\\
 \cdot\Bigsbrk{\bigbrk{\spaa{1|p_{56}|1}-\spaa{2|p_{56}|2}}\bigbrk{\spaa{5|p_3|6}-\spaa{5|p_4|6}}-\bigbrc{(1,2)\leftrightarrow(3,4)}}\,,
\label{eq:fermA}
\end{multline}
where $A(k_1,\ldots,k_6):=A(\lambda_{k_1},\ldots,\lambda_{k_6})$, and
for any momenta $q_1,\ldots,q_k$
\begin{equation}
\spaa{i|q_1|\ldots|q_k|j}\defeq\lambda_i^a\varepsilon_{ab}q_1^{bc}\varepsilon_{cd}\cdots\varepsilon_{ef}q_k^{fg}\varepsilon_{gh}\lambda_j^h\,.
\end{equation}
The overall constant $C_6$ shall be left undetermined.
Diagram B (left in \Figref{fig:fermB}) reads%
\footnote{Here, $p^2\defeq p_{ab}p^{ab}$, i.e.\
$p_{123}^2=-\spaa{12}^2-\spaa{13}^2-\spaa{23}^2$.}
\begin{equation}
A_{6\psi,\mathrm{B}}(1,\ldots,6)
=2C_6\biggsbrk{\frac{\spaa{13}\spaa{64}\spaa{1|p_{23}|4}}{\spaa{12}\spaa{45}p_{123}^2}+\brc{1\leftrightarrow2}+\brc{4\leftrightarrow5}+\brc{1\leftrightarrow2,4\leftrightarrow5}}\,.
\label{eq:fermB}
\end{equation}
The total color-ordered amplitude is a sum over all relabelings of the
diagrams in \Figsref{fig:fermA,fig:fermB} that respect
the color structure \eqref{eq:colorstruct}. The result is
\begin{align}
A_{6\psi}(1,\ldots,6)
=&+A_{6\psi,\mathrm{A}}(1,2,3,4,5,6)+A_{6\psi,\mathrm{A}}(1,6,5,4,3,2)\nn\\
 &+A_{6\psi,\mathrm{B}}(1,2,3,4,5,6)-A_{6\psi,\mathrm{B}}(6,5,4,3,2,1)\nn\\
 &+A_{6\psi,\mathrm{B}}(1,2,3,6,5,4)+A_{6\psi,\mathrm{B}}(3,2,1,4,5,6)+\brc{\text{two cyclic}}\,.
\end{align}
Here, ``two cyclic'' stands for two repetitions of all previous terms with the
relabelings $\lambda_k\to\lambda_{k+2}$, $\lambda_k\to\lambda_{k+4}$
$\mod 6$ applied. Using Schouten's identity and various relations following
from momentum conservation ($P=0$), this can be simplified to%
\footnote{$p_{j,\pm k,\ldots}\defeq p_j\pm p_k+\ldots$.}
\begin{multline}
A_{6\psi}(1,\ldots,6)=C_6\cdot\\
\cdot\biggbrk{\!\!
	\Bigbrk{
		\frac{-\third\spaa{1|p_3|p_5|1}
	          +\third\spaa{2|p_4|p_6|2}
	          -\spaa{3|p_2|p_{5,-6}|3}}{\spaa{12}\spaa{34}\spaa{56}}
		-2\frac{\spaa{2|p_{3,-4}|p_{234}|p_{5,-6}|1}}{\spaa{34}\spaa{56}p_{234}^2}
		-\brc{\text{shift by one}}\!}\\
	-2\frac{ \spaa{1|p_6|p_{6,-1,2}|p_{345}|p_{3,-4,5}|p_3|4}
	        +\spaa{1|p_2|p_{6,-1,2}|p_{345}|p_{3,-4,5}|p_5|4}}{\spaa{6|p_1|2}\spaa{3|p_4|5}p_{612}^2}}
	+\brc{\text{two cyclic}}\,,
\end{multline}
where ``shift by one'' means the relabeling
$\lambda_k\to\lambda_{k+1}\,\mod 6$.

\paragraph{Six-Scalar Amplitude.}

\begin{figure}
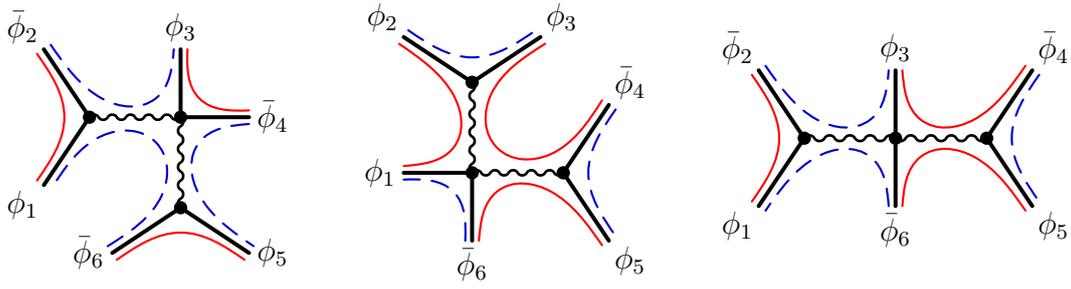
\centering
\includegraphicsbox{FigSixScalar4v1}
\qquad
\includegraphicsbox{FigSixScalar4v2}
\qquad
\includegraphicsbox{FigSixScalar4v3}
\caption{Diagram C that contributes to the
six-scalar amplitude. Again, blue/dashed lines represent fundamental
color contractions, red/solid lines represent antifundamental ones.
When color-stripped, the left diagram gives
\protect\eqref{eq:scalardiags}, and the other two diagrams equal the
left one up to relabelings of the external legs.}
\label{fig:scalardiags}
\end{figure}

Again the color-ordered amplitude
$A_{6\phi}(\lambda_1,\ldots,\lambda_6)$ contains all contributions in
which the fields $\phi_1,\ldots,\phi_6$ are cyclically connected by
color contractions,
\begin{align}
\hat A_{6\phi}
\defeq{}&\hat A_6(\phi_1{}^{A_1}_{\bar A_1},\bar\phi_2{}^{\bar B_2}_{B_2},\phi_3{}^{A_3}_{\bar A_3},\bar\phi_4{}^{\bar B_4}_{B_4},\phi_5{}^{A_5}_{\bar A_5},\bar\phi_6{}^{\bar B_6}_{B_6})
&
\phi_k&\defeq\phi(\lambda_k)\nn\\
={}& \ldots
  +A_{6\phi}(\lambda)\,
   \delta_{\bar A_1}^{\bar B_2}\delta_{B_2}^{A_3}\delta_{\bar A_3}^{\bar B_4}\delta_{B_4}^{A_5}\delta_{\bar A_5}^{\bar B_6}\delta_{B_6}^{A_1}
  +\ldots\,,
&
\lambda&\defeq(\lambda_1,\ldots,\lambda_6)\,.
\end{align}
The color-ordered amplitude receives contributions from three
kinematically different diagrams. Two of them are the diagrams A and B,
\Figsref{fig:fermA,fig:fermB}, with all fermion lines
replaced by scalar lines. The scalar version of diagram A (left in \Figref{fig:fermA}) reads
\begin{equation}
A_{6\phi,\mathrm{A}}(1,\ldots,6)=-\frac{4C_6}{3}\frac{1}{\spaa{12}\spaa{34}\spaa{56}^2}\Bigbrk{\spaa{1|p_6|2}\spaa{3|p_5|4}-\brc{5\leftrightarrow6}}\,,
\end{equation}
while the scalar version of diagram B (left in \Figref{fig:fermB}) is
\begin{equation}
A_{6\phi,\mathrm{B}}(1,\ldots,6)=8C_6\frac{\spaa{1|p_3|2}\spaa{4|p_6|5}}{\spaa{12}\spaa{45}p_{123}^2}\,.
\end{equation}
A further contribution comes from diagram C,
see \Figref{fig:scalardiags}.
It evaluates to
\begin{equation}
A_{6\phi,\mathrm{C}}(1,\ldots,6)=-2C_6\frac{\spaa{16}\spaa{25}+\spaa{15}\spaa{26}}{\spaa{12}\spaa{56}}\,.
\label{eq:scalardiags}
\end{equation}
Again, the total color-ordered amplitude is a sum over all relabelings
of these diagrams that respect the color structure. The sum of all
contributions is
\begin{align}
A_{6\phi}(1,\ldots,6)
=&+A_{6\phi,\mathrm{A}}(1,2,3,4,5,6)+A_{6\phi,\mathrm{A}}(1,6,5,4,3,2)\nn\\
 &+A_{6\phi,\mathrm{B}}(1,2,3,4,5,6)+A_{6\phi,\mathrm{B}}(6,5,4,3,2,1)\nn\\
 &-A_{6\phi,\mathrm{B}}(1,2,3,6,5,4)-A_{6\phi,\mathrm{B}}(3,2,1,4,5,6)\nn\\
 &+A_{6\phi,\mathrm{C}}(1,2,3,4,5,6)+A_{6\phi,\mathrm{C}}(3,2,1,6,5,4)\nn\\
 &-2A_{6\phi,\mathrm{C}}(1,2,3,6,5,4)+\brc{\text{two cyclic}}\,.
\end{align}
This can be simplified to
\begin{multline}
A_{6\phi}(1,\ldots,6)
=C_6\biggbrk{
	4\frac{\spaa{3|p_5|p_1|p_6|p_2|p_4|3}+\spaa{14}^2\spaa{2|p_3|p_6|p_5|2}}{\spaa{1|p_2|p_3|p_4|p_5|p_6|1}}\\
	+\Bigbrk{
		2\frac{\third\spaa{16}\spaa{35}\spaa{24}-\third\spaa{13}\spaa{56}\spaa{24}+\spaa{16}\spaa{23}\spaa{45}}{\spaa{12}\spaa{34}\spaa{56}}
		+8\frac{\spaa{5|p_1|6}\spaa{3|p_2|4}}{\spaa{34}\spaa{56}p_{234}^2}
		+\brc{\text{shift by one}}}\\
	-8\frac{\spaa{26}\spaa{35}(\spaa{16}^2\spaa{34}^2+\spaa{12}^2\spaa{45}^2)}{\spaa{2|p_1|6}\spaa{3|p_4|5}p_{612}^2}}
	+\brc{\text{two cyclic}}\,.
\end{multline}

\section{Factorization of the Six-Point Superamplitude}
\label{app:super}

Consider the quantity:
\begin{equation}
\int d^{2|3} \hat  \Lambda \, \amp_4 (\Lambda_1,\Lambda_2,\Lambda_3, \hat \Lambda)
 \frac{1}{P_{13}^2} \amp_4 (\pm i \hat \Lambda,\Lambda_4,\Lambda_5,\Lambda_6)\,,
\end{equation}
where $\Lambda=(\lambda^a,\eta^A)$ and the result doesn't depend on
the choice of sign $\pm$ .
The integration can be trivially performed because of the delta functions using:
\begin{equation}
\int d^3 \hat \eta \, \delta^6 (Q_1^{A a} +\hat \eta^A \mu^a) \, \delta^6 (Q_2^{A a} -\hat \eta^A \mu^a)= \delta^6 (Q_1^{A a} +Q_2^{A a}) \, \delta^3(\epsilon_{ab}Q_1^{A a}\mu^b)
\end{equation}
and
\begin{equation}
\int d^2 \hat \lambda \, \delta^3 (P_1^{a b} + \hat \lambda^a \hat \lambda^b) \,
 \delta^3 (P_2^{a b} - \hat \lambda^a \hat \lambda^b) F(\hat \lambda)= \delta^3 (P_1^{ ab} +P_2^{ab}) \,
 \delta(P_1^2)\left(F(\hat \lambda) +F(-\hat \lambda) \right)
\end{equation}
where on the right hand side $\hat \lambda$ is the solution to the equation $\hat \lambda^a\hat \lambda^b= P_2^{ab} $.
Reminding that \eqref{eq:4point}
\begin{equation}
\amp_4 (1,2,3,4)= \deltaPQ f(\lambda_1,\lambda_2,\lambda_3,\lambda_4)\,,
\end{equation}
and using the properties of $f(\lambda)$, we obtain:
\begin{equation}
\frac{1}{P_{13}^2} \delta(P_{13}^2) \,
\delta^3(P) \, \delta^3(Q) \, \delta^3({\epsilon_{ab}Q_{13}^{A a}\hat \lambda^b})
f(\lambda_1,\lambda_2,\lambda_3,\hat \lambda)f(\pm i \hat
\lambda,\lambda_4,\lambda_5,\lambda_6)\,.
\end{equation}
This can be rewritten as
\begin{equation}
\frac{1}{P_{13}^2} \delta(P_{13}^2)\,
\deltaPQ\delta^3(\alpha)f^+(\lambda)\,,
\end{equation}
which equals $\amp_6$ in the limit $P_{13}^2\to0$,
cf.~\Secref{sec:sixpoint}.

\section{The Metric of \texorpdfstring{\osp}{osp(6|4)}}
\label{app:ospmetric}

Introducing matrices $(E^\mathbb{A}{}_\mathbb{B})^i{}_j=\delta^{\mathbb{A}i} \delta_{\mathbb{B}j}$ with $\mathbb{A}, \mathbb{B} =a, b, A, B, \dots$, the fundamental representation $M$ of \osp\ consisting of $(4|6)\times (4|6)$ matrices can be written as
\begin{equation}
\text{\footnotesize
$\displaystyle
M\left[\left(
\begin{array}{cc|cc}
 \gen{L}^a{}_b&\gen{P}^{ab}&\gen{Q}^a{}_A&\gen{Q}^{aA}\\
\gen{K}_{ab}&\gen{L}^a{}_b&\gen{S}_{aA}&\gen{S}_a{}^A
\\\hline
\gen{S}_a{}^A&\gen{Q}^{aA}&\gen{R}^A{}_B&\gen{R}^{AB}\\
\gen{S}_{aA}&\gen{Q}^a{}_A&\gen{R}_{AB}&\gen{R}^A{}_B
\end{array}
\right)\right]
=
\left(
\begin{array}{cc|cc}
E^a{}_b-\half \delta^a_b\mathbb{I}&E^{ab}+E^{ba}&E^a{}_A&-E^a{}_A\\
E_{ab}+E_{ba}&E_b{}^a+\half\delta^a_b \mathbb{I}&E_{aA}&-E_{aA}
\\\hline
E_A{}^a&-E_{Ac}&E^A{}_B & E_B{}^A-E^A{}_B\\
-E_A{}^a&E_{Ac}&E_B{}^A-E^A{}_B&E_B{}^A
\end{array}
\right).
$}
\end{equation}
For the Lorentz generator for instance, this equation is to be understood as
\begin{equation}
 M[\gen{L}^a{}_b]=
\left( \begin{array}{cc|cc}
E^a{}_b-\half \delta^a_b\mathbb{I}&0&0&0\\
0&E_b{}^a+\half\delta^a_b \mathbb{I}&0&0
\\\hline
0&0&0&0\\
0&0&0&0
\end{array}\right),
\end{equation}
where we raise and lower Lorentz indices with $\eps^{ab}$, $\eps_{ab}$ and have
\begin{equation}
E_B{}^A=(E^A{}_B)^T,\quad
E_A{}^a=(E^a{}_A)^T,\quad
E_{Ac}=(\eps_{ac} E^c{}_A)^T.
\end{equation}
Furthermore, the dilatation generator is defined by
\begin{equation}
M[\gen{D}]=
\left( \begin{array}{cc|cc}
\half \mathbb{I}&0&0&0\\
0&-\half\mathbb{I}&0&0
\\\hline
0&0&0&0\\
0&0&0&0
\end{array}\right).
\end{equation}
The Killing form of \osp\ vanishes. We compute the metric defined by
\begin{equation}
 g_{\alpha \beta}=g(\gen{J}_\alpha,\gen{J}_\beta)=\sTr{M[\gen{J}_\alpha]M[\gen{J}_\beta]},
\end{equation}
which obeys
\begin{equation}
 g_{\alpha\beta}=(-1)^{\grade{\alpha}} g_{\beta\alpha},\qquad g_{\alpha \beta}=0\,\,\text{if}\,\,\grade{\alpha}\neq\grade{\beta}.
\end{equation}
Here, $\abs{\alpha}$ denotes the Gra{\ss}mann degree of the generator $\gen{J}_\alpha$.
We change the basis of generators and introduce
\begin{equation}
 \gen{Y}^a{}_b=\gen{L}^a{}_b+\delta^a_b\gen{D}.
\end{equation}
Then the metric has the following non-vanishing components
\begin{align}
 g(\gen{Y}^a{}_b,\gen{Y}^c{}_d)&=2\delta^a_d \delta^c_b,
\nonumber\\
g(\gen{P}^{ab},\gen{K}_{cd})&=g(\gen{K}_{cd},\gen{P}^{ab})=-2 \delta^a_c\delta^b_d-2\delta^a_d\delta^b_c,
\nonumber\\
g(\gen{Q}^{aA},\gen{S}_{bB})&=-g(\gen{S}_{bB},\gen{Q}^{aA})=2\delta^A_B\delta^a_b,
\nonumber\\
g(\gen{Q}^a{}_A,\gen{S}_b{}^B)&=-g(\gen{S}_b{}^B,\gen{Q}^a{}_A)=2\delta^B_A\delta^a_b,
\nonumber\\
g(\gen{R}^A{}_B,\gen{R}^C{}_D)&=g(\gen{R}^C{}_D,\gen{R}^A{}_B)=-2\delta^A_D\delta^C_B,
\nonumber\\
g(\gen{R}^{AB},\gen{R}_{CD})&=g(\gen{R}_{CD},\gen{R}^{AB})=2\delta^A_C\delta^B_D-2\delta^A_D\delta^B_C.
\end{align}
The inverse metric $g^{\alpha\beta}=g^{-1}(\gen{J}_\alpha,\gen{J}_\beta)$ satisfies
\begin{equation}
 g_{\alpha\beta}g^{\beta\gamma}=\delta^\gamma_\alpha=g^{\gamma\beta}g_{\beta\alpha}.
\end{equation}
Its non-zero components are
\begin{align}
 g^{-1}(\gen{Y}^a{}_b,\gen{Y}^c{}_d)&=\half\delta^a_d \delta^c_b,
\nonumber\\
g^{-1}(\gen{P}^{ab},\gen{K}_{cd})&=g^{-1}(\gen{K}_{cd},\gen{P}^{ab})=-\sfrac{1}{8} \delta^a_c\delta^b_d-\sfrac{1}{8}\delta^a_d\delta^b_c,
\nonumber\\
g^{-1}(\gen{Q}^{aA},\gen{S}_{bB})&=-g^{-1}(\gen{S}_{bB},\gen{Q}^{aA})=-\half\delta^A_B\delta^a_b,
\nonumber\\
g^{-1}(\gen{Q}^a{}_A,\gen{S}_b{}^B)&=-g^{-1}(\gen{S}_b{}^B,\gen{Q}^a{}_A)=-\half\delta^B_A\delta^a_b,
\nonumber\\
g^{-1}(\gen{R}^A{}_B,\gen{R}^C{}_D)&=g^{-1}(\gen{R}^C{}_D,\gen{R}^A{}_B)=-\half\delta^A_D\delta^C_B,
\nonumber\\
g^{-1}(\gen{R}^{AB},\gen{R}_{CD})&=g^{-1}(\gen{R}_{CD},\gen{R}^{AB})=\sfrac{1}{8}\delta^A_C\delta^B_D-\sfrac{1}{8}\delta^A_D\delta^B_C.
\end{align}

\section{The Level-One Generators \texorpdfstring{$\gen{P}^{(1)ab}$}{P1} and \texorpdfstring{$\gen{Q}^{(1)aB}$}{Q1}}
\label{app:levelone}
We can use the metric and read off the structure constants from the commutation relations of \osp\ to compute the Yangian level-one generators $ \gen{P}^{(1)ab}$ and $\gen{Q}^{(1)aA}$.
According to \eqref{eq:bilocYang} we have
\begin{align}
\gen{P}^{(1)ab}={}&f^{\gamma\beta}{}_{\gen{P}^{ab}}\sum_{j<i}\gen{J}^{(0)}_{i\beta}\gen{J}^{(0)}_{j\gamma}\nonumber\\
={}&f_{\tilde\gamma\tilde \beta}{}^{\gen{K}_{cd}}g_{\gen{K}_{cd}\gen{P}^{ab}}g^{\tilde \beta \beta}g^{\tilde \gamma \gamma}\sum_{j<i}\gen{J}^{(0)}_{i\beta}\gen{J}^{(0)}_{j\gamma}\nonumber\\
={}&-2(\delta^a_c\delta^b_d+\delta^a_d\delta^b_c)\sum_{j<i}\Big(f_{ \gen{K}_{ef}\gen{Y}^l{}_m}{}^{\gen{K}_{cd}}g^{\gen{Y}^l{}_m\gen{Y}^g{}_h}g^{\gen{K}_{ef}\gen{P}^{rs}}\gen{Y}_i^{(0)g}{}_h\gen{P}_j^{(0)rs}\nonumber\\
&\qquad\qquad\qquad\qquad+f_{\gen{S}_e{}^E\gen{S}_{fF}}{}^{\gen{K}_{cd}}g^{\gen{S}_{fF}\gen{Q}^{gG}}g^{\gen{S}_e{}^E\gen{Q}^h{}_H}\gen{Q}_i^{(0)gG}\gen{Q}_j^{(0)h}{}_H\nonumber\\
&\qquad\qquad\qquad\qquad-(i\leftrightarrow j)\Big)\nonumber\\
={}&\half\sum_{j<i}\left(\gen{Q}_i^{(0)(aA}\gen{Q}_j^{(0)b)}{}_A-\gen{Y}_i^{(0)(a}{}_c \gen{P}_j^{(0)cb)}-(i\leftrightarrow j)\right).
\end{align}
In order to check consistency, we also determine $\gen{Q}^{(1)aA}$ :
\begin{align}
\gen{Q}^{(1)aA}={}&f^{\gamma\beta}{}_{\gen{Q}^{aA}}\sum_{j<i}\gen{J}^{(0)}_{j\beta}\gen{J}^{(0)}_{j\gamma}\nonumber\\
={}&f_{\tilde\gamma\tilde\beta}{}^{\gen{S}_{bB}}g_{\gen{S}_{bB}\gen{Q}^{aA}}g^{\tilde\beta\beta}g^{\tilde\gamma\gamma}\sum_{j<i}\gen{J}^{(0)}_{j\beta}\gen{J}^{(0)}_{j\gamma}\nonumber\\
={}&-2 \delta^A_B\delta^a_b\sum_{j<i}\Big(f_{\gen{K}_{cd}\gen{Q}^e{}_E}{}^{\gen{S}_{bB}}g^{\gen{K}_{cd}\gen{P}^{fg}}g^{\gen{Q}^e{}_E\gen{S}_h{}^H}\gen{S}^{(0)}_{ih}{}^H\gen{P}_j^{(0)fg}\nonumber\\
&\qquad+f_{\gen{Y}^c{}_d \gen{S}_{eE}}{}^{\gen{S}_{bB}}g^{\gen{Y}^c{}_d\gen{Y}^f{}_g} g^{\gen{S}_{eE}\gen{Q}^{hH}}\gen{Q}_i^{(0)hH}\gen{Y}_j^{(0)f}{}_g\nonumber\\
&\qquad+f_{\gen{R}^C{}_D\gen{S}_{eE}}{}^{\gen{S}_{bB}}g^{\gen{R}^C{}_D\gen{R}^{F}{}_G}g^{\gen{S}_{eE}\gen{Q}^{hH}}\gen{Q}_i^{(0)hH}\gen{R}_j^{(0)F}{}_G\nonumber\\
&\qquad+f_{\gen{R}_{CD}\gen{S}_e{}^E}{}^{\gen{S}_{bB}} g^{\gen{R}_{CD}\gen{R}^{FG}}g^{\gen{S}_e{}^E\gen{Q}^h{}_H}{\gen{Q}_i^{(0)h}}_H\gen{R}_j^{(0)FG}\nonumber\\
&\qquad-(i\leftrightarrow j)\Big)\nonumber\\
={}&\half\sum_{j<i}\Big(\gen{Q}_i^{(0)bA}{\gen{Y}_j^{(0)a}}_b+\gen{Q}^{(0)a}_i{}_B \gen{R}_j^{(0)BA}-\gen{Q}_i^{(0)aB}{\gen{R}_j^{(0)A}}_B-\gen{S}^{(0)}_{ib}{}^A\gen{P}_j^{(0)ba}-(i\leftrightarrow j)\Big)
\end{align}
One can easily convince oneself that consistently
\begin{equation}
 \acomm{\gen{Q}^{(1)aA}}{\gen{Q}^b{}_B}=\delta^A_B\gen{P}^{(1)^{ab}}.
\end{equation}

\section{The Serre Relations}
\label{app:serre}

In the following, we will show how the homomorphicity condition
\eqref{eq:coJ1} of the coproduct \eqref{eq:cop1,eq:cop2} leads to the
Serre relations \eqref{eq:serrealg}.
First, we multiply \eqref{eq:coJ1} by the algebra structure constants
and take cyclic permutations to find
\begin{equation}\label{eq:coJ1sym}
 f_{\beta\delta}{}^\gamma\Delta(\comm{\op{J}^{(1)}_\alpha}{\op{J}^{(1)}_\gamma})+\mathrm{cyclic}(\alpha,\beta,\delta)=f_{\beta\delta}{}^\gamma\comm{\Delta(\op{J}^{(1)}_\alpha)}{\Delta(\op{J}^{(1)}_\gamma)}+\mathrm{cyclic}(\alpha,\beta,\delta).
\end{equation}
It is obvious that \eqref{eq:coJ1sym} follows from \eqref{eq:coJ1};
how about the other direction? The answer is that \eqref{eq:coJ1sym}
equals the $\mathbb{X}$ component of \eqref{eq:coJ1} while the adjoint
component is projected out. The reason for this is rather simple:
Equation \eqref{eq:coJ1} can be written in the form
$f_{\alpha\beta}{}^\delta Z_\delta+X_{\alpha\beta}=0$, where
$X_{\alpha\beta}\in\mathbb{X}$ and $Z_\delta\in \mathrm{Adj}$ (cf.
\eqref{eq:decomp}). Now showing that \eqref{eq:coJ1sym} does not
contain the adjoint boils down to using the Jacobi identity in the
form
\begin{equation}
 f_{\beta\delta}{}^\gamma f_{\alpha\gamma}{}^\epsilon +\mathrm{cyclic}(\alpha,\beta,\delta)=0.
\end{equation}
Furthermore that only the adjoint and nothing else is projected out in
going from \eqref{eq:coJ1} to \eqref{eq:coJ1sym} follows from
\begin{equation}
 f_{\alpha}{}^{\beta\gamma} u_{\beta\gamma}=0 \qquad \Rightarrow\quad u_{\alpha\beta}=f_{\alpha\beta}{}^\gamma v_\gamma,
\end{equation}
for some $v_\gamma$ (or equivalently that the second cohomology of
$\alg{g}$ vanishes). Since $\mathbb{X}$ does not contain the adjoint,
we have separately $X_{\alpha\beta}=0$ and $f_{\alpha\beta}{}^\delta
Z_\delta=0$. The first equation will lead to the Serre relations. The
second equation represents the definition of the coproduct for the
level-two generators.

In order to derive the Serre relations we rewrite the right hand side
of \eqref{eq:coJ1} as (cf. \cite{Matsumoto:2009rf})
\begin{align}
\comm{\Delta(\op{J}^{(1)}_\alpha)}{\Delta(\op{J}^{(1)}_\beta)}
=&\comm{\op{J}^{(1)}_\alpha}{\op{J}^{(1)}_\beta}\otimes 1+1\otimes\comm{\op{J}^{(1)}_\alpha}{\op{J}^{(1)}_\beta}\nonumber\\
&+\frac{h}{2}\bigbrk{f_\alpha{}^{\gamma\delta}\comm{\op{J}^{(0)}_\gamma\otimes\op{J}^{(0)}_\delta}{\op{J}^{(1)}_\beta\otimes
1+1\otimes \op{J}^{(1)}_\beta}-(\alpha\leftrightarrow\beta)}\nonumber\\
&+\frac{h^2}{4} f_\alpha{}^{\gamma\delta} f_\beta{}^{\rho\epsilon}\comm{\op{J}^{(0)}_\gamma\otimes\op{J}^{(0)}_\delta}{\op{J}^{(0)}_\rho\otimes\op{J}^{(0)}_\epsilon}.
\end{align}
It is rather straightforward to rewrite the last two lines in this
equation in the form of
\begin{align}
 &\frac{h}{2} f_{\alpha\beta}{}^\rho f_\rho{}^{\gamma\delta}\bigbrk{\op{J}^{(1)}_\gamma\otimes\op{J}^{(0)}_\delta-\op{J}^{(0)}_\delta \otimes\op{J}^{(1)}_\gamma},\label{eq:eqone} \\
&\frac{h^2}{4}f_\alpha{}^{\epsilon\rho}f_\beta{}^{\gamma\mu}f_{\epsilon\gamma}{}^\kappa\bigbrk{\op{J}^{(0)}_\kappa\otimes\op{J}^{(0)}_\mu \op{J}^{(0)}_\rho+\op{J}^{(0)}_\mu\op{J}^{(0)}_\rho\otimes \op{J}^{(0)}_\kappa}\label{eq:eqtwo}.
\end{align}
Now it is easy to see that \eqref{eq:eqone} vanishes due to the Jacobi
identity when plugged into the right hand side of \eqref{eq:coJ1sym}.
Using the Jacobi identity twice, the contribution to
\eqref{eq:coJ1sym} coming from the second piece \eqref{eq:eqtwo} reads
\begin{equation}
 \frac{h^2}{8}f_{\alpha\rho}{}^\lambda f_{\beta\delta}{}^\mu f_{\gamma\kappa}^\nu f^{\kappa\delta\rho}\bigbrk{\acomm{\op{J}^{(0)}_\lambda}{\op{J}^{(0)}_\mu}\otimes \op{J}^{(0)}_\nu+\op{J}^{(0)}_\nu\otimes\acomm{\op{J}^{(0)}_\lambda}{\op{J}^{(0)}_\mu}}+\mathrm{cyclic}(\alpha,\beta,\gamma).
\end{equation}
Since the coproduct on $\op{J}^{(0)}$ has the trivial form
\eqref{eq:cop1} one can rewrite this as%
\footnote{We thank Lucy Gow for discussions on this point and sharing
some of her notes with us.}
\begin{equation}
\Delta(S_{\alpha\beta\gamma})-S_{\alpha\beta\gamma}\otimes 1- 1\otimes S_{\alpha\beta\gamma},
\end{equation}
where
\begin{equation}
 S_{\alpha\beta\gamma}=\frac{h^2}{24}f_{\alpha}{}^{\rho\lambda} f_{\beta}{}^{\delta\mu} f_{\gamma}{}^{\kappa\nu} f^{\kappa\delta\rho}\{\op{J}^{(0)}_\rho,\op{J}^{(0)}_\delta,\op{J}^{(0)}_\kappa\}.
\end{equation}
Putting everything together \eqref{eq:coJ1sym} becomes
\begin{equation}\label{eq:0K}
0=\Delta(K_{\alpha\beta\gamma})-K_{\alpha\beta\gamma}\otimes 1- 1\otimes K_{\alpha\beta\gamma},
\end{equation}
where now
\begin{equation}
 K_{\alpha\beta\gamma}= S_{\alpha\beta\gamma}-\bigbrk{f_{\alpha\beta}{}^{\delta} \comm{\op{J}^{(1)}_\gamma}{\op{J}^{(1)}_\delta}+\mathrm{cyclic}(\alpha,\beta,\gamma)}.
\end{equation}
A sufficient condition for \eqref{eq:0K} to be satisfied is
$K_{\alpha\beta\gamma}=0$ which, rewriting
$f_{\alpha\beta}{}^\delta\op{J}^{(1)}_\delta=\comm{\op{J}^{(0)}_\alpha}{\op{J}^{(1)}_\beta}$,
are the well known Serre relations \eqref{eq:serrealg}. One of the
reasons for rederiving the Serre relations here is to convince the
reader and ourselves that only the $\mathbb{X}$ component of
$\{\op{J},\op{J},\op{J}\}$ contributes to the right hand side of
\eqref{eq:serrealg}. As we have seen in \Secref{sec:yanginv}, this is
very useful for proving the Serre relations for specific
representations.

In order to show that the Serre relations are indeed satisfied for a
certain representation, one can start with the case $n=1$, i.e.\ a
representation acting on only one vector space and define
\begin{align}
 \rho|_{n=1}(\op{J}^{(0)}_\alpha)&=\gen{J}^{(0)}_\alpha,\nonumber\\
 \rho|_{n=1}(\op{J}^{(1)}_\alpha)&=0.
\label{eq:onerep}
\end{align}
The left hand side of \eqref{eq:serrealg} vanishes for the one-site
representation $\rho|_{n=1}$. Assuming that also the right
hand side of this equation vanishes for the one-site representation,
one can promote \eqref{eq:serrealg} from one to $n$
sites. The point is that the coproduct preserves the Serre relations,
that is if $\op{J}^{(0)}$ and $\op{J}^{(1)}$ satisfy the Serre
relations then also $\Delta(\op{J}^{(0)})$ and $\Delta(\op{J}^{(1)})$
do. The reason behind this is an inductive argument. Assuming the Serre
relations to be satisfied for $n$ sites implies the coproduct to be a
homomorphism \eqref{eq:homomorph} for $n+1$ sites. Acting with
$\Delta$ on \eqref{eq:serrealg} thus yields the Serre relations for
$n+1$ sites which in turn implies \eqref{eq:homomorph} for $n+2$
sites. This means that the Serre relations will be automatically
satisfied by the choice \eqref{eq:onerep} promoted to $n$ vector
spaces by successive application of the coproduct. To be explicit, the
action on two sites is given by
\begin{align}
 \rho|_{n=2}(\Delta\op{J}^{(0)}_\alpha)&=1\otimes\gen{J}^{(0)}_\alpha+\gen{J}^{(0)}_\alpha\otimes 1=\sum_{i=1}^2\gen{J}^{(0)}_{i\alpha},\nonumber\\
 \rho|_{n=2}(\Delta\op{J}^{(1)}_\alpha)&=f^{\beta\gamma}{}_\alpha \gen{J}^{(0)}_\gamma\otimes\gen{J}^{(0)}_\beta=f^{\beta\gamma}{}_\alpha\sum_{1\leq j<i\leq 2}\gen{J}^{(0)}_{i\gamma}\gen{J}^{(0)}_{j\beta},
\end{align}
where we recover the original bilocal form of the level-one generators \eqref{eq:bilocYang}. Here,
\begin{equation}
\rho|_{n=2}(A\otimes B)=(\rho|_{n=1} A)\otimes(\rho|_{n=1} B).
\end{equation}

Note that the above analysis is completely independent of the explicit
representation $\rho$. The criterion
for any representation to obey the Serre relations is thus the
vanishing of the right hand side of
\eqref{eq:serrealg} for that specific representation. For showing this,
it is crucial that the right hand side of \eqref{eq:serrealg}
transforms in the representation $\mathbb{X}$ as shown above.

\section{Conventions and Identities}

Throughout the article, the spacetime metric is fixed to
$\eta^{\mu\nu}=\eta_{\mu\nu}=\mathrm{diag}(-++)$.
The totally antisymmetric tensor $\varepsilon^{\mu\nu\rho}$ is defined
such that $\varepsilon_{012}=-\varepsilon^{012}=1$.
\begin{equation}
\varepsilon_{12}=-\varepsilon^{12}=1.
\end{equation}
The relation between spacetime vectors and bispinors is given by
\begin{equation}
p^{ab}=(\sigma^\mu)^{ab}p_\mu\,,
\qquad
p^\mu=-\half(\sigma^\mu)_{ab}p^{ab}\,,
\end{equation}
where a convenient choice for the matrices $(\sigma^\mu)^{ab}$ is
\begin{equation}
\brk{\sigma^0}^{ab}=\matr{rr}{-1&0\\0&-1}\,,\quad
\brk{\sigma^1}^{ab}=\matr{rr}{-1&0\\0&1}\,,\quad
\brk{\sigma^2}^{ab}=\matr{rr}{0&1\\1&0}\,.
\label{eq:sigmaupup}
\end{equation}
They obey the following relations:
\begin{equation}
\sigma^\mu_{ab}\sigma^{\nu ab}=-2\eta^{\mu\nu}\,,
\end{equation}
\begin{equation}
\sigma^\mu_{ab}\sigma_{\mu cd}=-\varepsilon_{ac}\varepsilon_{bd}-\varepsilon_{ad}\varepsilon_{bc}\,,
\end{equation}
\begin{multline}
\varepsilon_{\mu\nu\rho}\brk{\sigma^\mu}_{ab}\brk{\sigma^\nu}_{cd}\brk{\sigma^\rho}_{ef}
=\half\brk{
  \varepsilon_{ac}\varepsilon_{be}\varepsilon_{df}
 +\varepsilon_{ac}\varepsilon_{bf}\varepsilon_{de}
 +\varepsilon_{ad}\varepsilon_{be}\varepsilon_{cf}
 +\varepsilon_{ad}\varepsilon_{bf}\varepsilon_{ce}\\
 +\varepsilon_{ae}\varepsilon_{bc}\varepsilon_{df}
 +\varepsilon_{ae}\varepsilon_{bd}\varepsilon_{cf}
 +\varepsilon_{af}\varepsilon_{bc}\varepsilon_{de}
 +\varepsilon_{af}\varepsilon_{bd}\varepsilon_{ce}
 }\,.
\end{multline}
The matrices
$(\sigma^\mu)^a{}_b=\varepsilon_{bc}(\sigma^\mu)^{ac}$ obey the
algebra
\begin{equation}
(\sigma^\mu)^a{}_b(\sigma^\nu)^b{}_c
=g^{\mu\nu}\delta^a{}_b+\varepsilon^{\mu\nu\rho}(\sigma_\rho)^a{}_c\,.
\end{equation}
We use $(..)$ and $[..]$ for symmetrization or antisymmetrization of indices, respectively, i.e.\
\begin{equation}
 X_{(ab)}=X_{ab}+X_{ba},\qquad\quad
 X_{[ab]}=X_{ab}-X_{ba}.
\end{equation}

\bibliographystyle{nb}
\bibliography{n6scs}

\end{document}